\documentclass[%
superscriptaddress,
nofootinbib,
 amsmath,amssymb,
 aps,
 twocolumn,
 longbibliography,
]{revtex4-2}

\usepackage{graphicx}
\usepackage{dcolumn}
\usepackage{bm}

\usepackage{amsmath,amssymb,amsthm,amsfonts,latexsym,mathtools,mathdots,graphicx,float,bbold}
\usepackage{caption,subcaption,ellipsis,xcolor,textcomp,thmtools,thm-restate,cmap,mleftright}
\usepackage[colorlinks,citecolor=blue]{hyperref}
\usepackage[nameinlink]{cleveref}
\usepackage[letterpaper,margin=1in]{geometry}
\usepackage{enumitem}

\usepackage{bold-extra}

\usepackage{algorithm}
\usepackage{algpseudocode}
\usepackage[nottoc,numbib]{tocbibind}
\usepackage[normalem]{ulem}

\usepackage{microtype}

\declaretheorem[numberwithin=]{theorem}
\declaretheorem[sibling=theorem]{lemma}

\declaretheorem[sibling=theorem]{proposition}

\theoremstyle{definition}
\declaretheorem[sibling=theorem]{definition}
\declaretheorem[sibling=theorem]{fact}

\declaretheorem[sibling=theorem]{notation}

\declaretheorem[sibling=theorem]{example}

\crefname{fact}{fact}{facts}
\crefname{notation}{notation}{notations}
\crefname{ineq}{inequality}{inequalities}
\creflabelformat{ineq}{#2{\upshape(#1)}#3}

\newcommand{\ignore}[1]{}
\let\Pr\relax
\DeclareMathOperator*{\Pr}{\mathbf{Pr}}
\DeclareMathOperator*{\E}{\mathbf{E}}

\DeclareMathOperator*{\Var}{\mathbf{Var}}
\DeclareMathOperator*{\Cov}{\mathbf{Cov}}

\DeclareMathOperator*{\avg}{avg}

\DeclarePairedDelimiter{\bra}{\langle}{\rvert}%
\DeclarePairedDelimiter{\ket}{\lvert}{\rangle}%
\DeclarePairedDelimiterX\innerp[2]{\langle}{\rangle}{#1\delimsize\vert\mathopen{}#2}%
\DeclarePairedDelimiterX\braket[2]{\langle}{\rangle}{#1\delimsize\vert\mathopen{}#2}%
\DeclarePairedDelimiterX\braketOP[3]{\langle}{\rangle}{#1\,\delimsize\vert\,\mathopen{}#2\,\delimsize\vert\,\mathopen{}#3}%
\DeclarePairedDelimiterX\ketbra[2]{\lvert}{\rvert}{#1\delimsize\rangle\!\delimsize\langle#2}%
\DeclarePairedDelimiterX\outerp[2]{\lvert}{\rvert}{#1\delimsize\rangle\!\delimsize\langle#2}%
\DeclarePairedDelimiterX\projector[1]{\lvert}{\rvert}{#1\delimsize\rangle\!\delimsize\langle#1}%

\DeclareMathOperator{\poly}{poly}
\DeclareMathOperator{\polylog}{polylog}

\newcommand{\R}{\mathbb{R}}
\newcommand{\C}{\mathbb{C}}
\newcommand{\N}{\mathbb{N}}

\newcommand{\F}{\mathbb{F}}

\newcommand{\eps}{\epsilon}

\newcommand{\wt}[1]{\widetilde{#1}}

\renewcommand{\ol}[1]{\overline{#1}}

\newcommand{\transp}{^\intercal}

\newcommand{\symmdiff}{\triangle} \newcommand{\symdiff}{\symmdiff}
\newcommand{\half}{\tfrac{1}{2}}
\newcommand{\bbone}{\mathbb{1}}
\newcommand{\Id}{\bbone}

\newcommand{\calA}{\mathcal{A}}
\newcommand{\calB}{\mathcal{B}}
\newcommand{\calC}{\mathcal{C}}

\newcommand{\calF}{\mathcal{F}}
\newcommand{\calG}{\mathcal{G}}
\newcommand{\calH}{\mathcal{H}}
\newcommand{\calI}{\mathcal{I}}

\newcommand{\calK}{\mathcal{K}}

\newcommand{\calN}{\mathcal{N}}

\newcommand{\calP}{\mathcal{P}}

\newcommand{\calR}{\mathcal{R}}
\newcommand{\calS}{\mathcal{S}}

\newcommand{\calX}{\mathcal{X}}
\newcommand{\calY}{\mathcal{Y}}
\newcommand{\calZ}{\mathcal{Z}}

\newcommand{\boldeta}{\boldsymbol{\eta}}

\newcommand{\bphi}{\boldsymbol{\phi}}
\newcommand{\bPsi}{\boldsymbol{\Psi}}
\newcommand{\bPhi}{\boldsymbol{\Phi}}

\newcommand{\bGamma}{\boldsymbol{\Gamma}}
\newcommand{\bPi}{\boldsymbol{\Pi}}

\newcommand{\ba}{\boldsymbol{a}}
\newcommand{\bb}{\boldsymbol{b}}

\newcommand{\bj}{\boldsymbol{j}}

\renewcommand{\bm}{\boldsymbol{m}}

\newcommand{\bv}{\boldsymbol{v}}

\newcommand{\bz}{\boldsymbol{z}}
\newcommand{\bA}{\boldsymbol{A}}

\newcommand{\bF}{\boldsymbol{F}}

\newcommand{\bH}{\boldsymbol{H}}

\newcommand{\bS}{\boldsymbol{S}}

\newcommand{\bcalA}{\boldsymbol{\calA}} 
\newcommand{\bcalB}{\boldsymbol{\calB}} 
\newcommand{\bcalG}{\boldsymbol{\calG}} 
\newcommand{\bcalI}{\boldsymbol{\calI}} 
\newcommand{\bcalF}{\boldsymbol{\calF}} 
 
\newcommand{\bcalH}{\boldsymbol{\calH}}

\DeclarePairedDelimiter\parens{\lparen}{\rparen}
\DeclarePairedDelimiter\abs{\lvert}{\rvert}
\DeclarePairedDelimiter\norm{\lVert}{\rVert}

\DeclarePairedDelimiter\ceil{\lceil}{\rceil}
\DeclarePairedDelimiter\braces{\lbrace}{\rbrace}
\DeclarePairedDelimiter\bracks{\lbrack}{\rbrack}

\newcommand{\Kik}{\calK}

\newcommand{\Opt}{\mathrm{Opt}}
\renewcommand{\adv}{\mathrm{adv}}
\newcommand{\lambdamax}[1]{\lambda_{\mathrm{max}}(#1)}
\newcommand{\opnorm}[1]{\norm{#1}}
\newcommand{\maxnorm}[1]{\norm{#1}_{\mathrm{max}}}
\newcommand{\Part}{\mathrm{Part}}

\renewcommand{\O}{\widetilde{O}}
\renewcommand{\i}{\mathrm{i}}

\newenvironment{widequotation}
    {\begin{list}{}{\setlength{\leftmargin}{0.2cm} \setlength{\rightmargin}{0.2cm}}%
    \item[]}
    {\end{list}}

\newcommand{\T}{\boldsymbol{T}}
\newcommand{\G}{\boldsymbol{G}}

\declaretheorem[sibling=theorem]{problem}
\crefname{problem}{Problem}{Problems}
\Crefname{problem}{Problem}{Problems}

\begin{document}


\title{Quartic quantum speedups for planted inference}%
\author{Alexander Schmidhuber}
\affiliation{
Google Quantum AI, Venice, CA
}\affiliation{Center for Theoretical Physics, Massachusetts Institute of Technology, Cambridge, MA}
\author{Ryan O'Donnell}%
\affiliation{%
Computer Science Department, Carnegie Mellon University, Pittsburgh, PA}%
\author{Robin Kothari}\author{Ryan Babbush}
\affiliation{
Google Quantum AI, Venice, CA
}%

\date{\today}

\begin{abstract}
    We describe a quantum algorithm for the Planted~Noisy~$k$XOR problem (also known as sparse Learning Parity with Noise) that achieves a nearly \emph{quartic} ($4$th power) speedup over the best known classical algorithm while using exponentially less space. Our work generalizes and simplifies prior work of Hastings~\cite{Has20}, by building on his quantum algorithm for the Tensor Principal Component Analysis (PCA) problem. We achieve our quantum speedup using a general framework based on the Kikuchi Method (recovering the quartic speedup for Tensor PCA), and we anticipate it will yield similar speedups for further planted inference problems. These speedups rely on the fact that planted inference problems naturally instantiate the Guided Sparse Hamiltonian problem. Since the Planted Noisy $k$XOR problem has been used as a component of certain cryptographic constructions, our work suggests that some of these 
     are susceptible to super-quadratic quantum attacks. 
\end{abstract}

\maketitle

\section{Introduction}
Existing quantum algorithms generally provide either exponential speedups over classical algorithms, such as Shor's algorithm for integer factorization \cite{shor1999polynomial}, or quadratic (or smaller) speedups, such as Grover's algorithm for unstructured search \cite{grover1996fast}. 
While exponential speedups represent a dramatic advantage, they remain exceedingly rare and are limited to only a handful of celebrated examples. In contrast, quadratic Grover-like speedups are much more prevalent, but recent studies suggest that they are unlikely to lead to real-world quantum advantage in practice due to the overheads associated with quantum error correction \cite{BMN+21, hoefler2023disentangling}. Interestingly, both these studies highlight that an intermediate regime -- super-quadratic polynomial quantum speedups -- may be far more promising for practical use, with Ref.~\cite{BMN+21} specifically highlighting that ``quartic ($4$th power) speedups look significantly more practical.''

In this work, we describe a quantum algorithm that achieves
a quartic speedup over the best known classical algorithm for the Planted~Noisy~$k$XOR problem (sometimes referred to as ``Sparse Learning Parity with Noise''). Moreover, the number of qubits used by the quantum algorithm is exponentially smaller than the classical algorithm's space complexity. Our work generalizes and simplifies prior work of Hastings~\cite{Has20}, by building on his quartic quantum speedup for the Tensor Principal Component Analysis (PCA) problem. We achieve our quantum speedup
using a general framework based on the Kikuchi Method (recovering the quartic speedup for Tensor PCA), and we anticipate it will yield similar speedups for further planted inference problems.

The Planted Noisy $k$XOR problem is at the core of the field of average-case analysis for constraint satisfaction problems, with a history dating back at least forty years~\cite{Has84}.
An additional motivation for studying it comes from cryptography, where the average-case hardness of Planted Noisy $k$XOR has long been known to imply Public Key Encryption~\cite{ABW10}, and its variants have recently been used to obtain multiparty homomorphic secret sharing~\cite{DIJL23}, lossy trapdoor functions, and collision-resistant hash functions~\cite{DJ24}. Our results suggest that some of these cryptographic primitives based on problems closely related to the Planted Noisy $k$XOR problem might need to increase their security parameters to sufficiently protect against future quantum attacks.

The Planted Noisy $k$XOR problem is an example of a larger class of problems called \emph{planted inference}, which concern the detection and recovery of a signal hidden in random noise. This class includes, among others, Planted Clique \cite{alon1998finding}, Community Detection \cite{JMLR:v18:16-480,abbe2015detection}, refutation of $k$-colorability of graphs \cite{bandeira2021spectral}, Group Synchronization \cite{singer2011angular,perry2018message}, and Tensor PCA \cite{richard2014statistical}. Planted problems find applications in machine learning, signal processing, and combinatorial optimization, and moreover serve as a testbed for the complexity-theoretic study of average-case hardness and statistical-computational gaps.

Planted inference problems have long been known to exhibit close connections to statistical physics \cite{mezard1987spin, hastings2006community, moore2017computer}. Indeed, the best known classical method for solving the Planted Noisy $k$XOR problem -- the so-called Kikuchi method \cite{WAM19} -- is motivated by a certain marginal approximation to Gibbs free energy, called \emph{Kikuchi free energy} \cite{PhysRev.81.988}. Here, we reinterpret the Kikuchi method as mapping planted inference problems to the groundstate energy estimation problem of a certain sparse, implicitly defined mean-field Hamiltonian. We show that this mean-field Hamiltonian admits an efficiently preparable \emph{guiding state} that has improved overlap with the ground state space. As in Hastings's work \cite{Has20}, this guiding state -- combined with Amplitude Amplification -- is the source of our quartic quantum speedup. 

As such, our quantum algorithms can be viewed as mapping planted inference problems (such as Planted Noisy $k$XOR and Tensor PCA) to concrete instantiations of the Guided Sparse Hamiltonian problem, where both the Hamiltonian and the guiding state in the problem are constructed from the planted inference problem. The Guided Sparse Hamiltonian problem was previously studied in the special case where the Hamiltonian is local and the guiding state has exponentially increased overlap over a random sate, in which case the problem is BQP-complete \cite{GL22}. However, finding problems where such exponentially improved guiding states are efficiently preparable remains an outstanding challenge.  Our work highlights that guiding states with only polynomially increased overlap can appear naturally, as in the context of the planted inference problems studied here, resulting in significant polynomial quantum speedups. 

Importantly, we show that our quantum algorithms not merely achieve a nearly quartic speedup over the current best classical algorithm based on the Kikuchi method, but even over hypothetical improved classical algorithms that still use the same framework.
Our approach can be viewed more as a method of upgrading a classical algorithm of this Kikuchi form to a quantum algorithm, rather than a quantum algorithm for a particular problem. 
Since our proofs use very few specific properties of the problems ($k$XOR or Tensor PCA) themselves, we expect our approach to generalize to other planted inference problems that can be solved using the Kikuchi method. 

\bigskip
The remainder of this paper is structured as follows. In \Cref{sec:bg}, we define the Planted Noisy $k$XOR problem and give a brief review of the Kikuchi method that is used to solve it. Our main results and new techniques are presented in \Cref{sec:main}. Finally, \Cref{sec:quantum_algo} summarizes the quantum algorithm that achieves our (nearly) quartic speedup.

\Cref{sec:Kikuchi_properties} contains a comprehensive technical description of the Planted Noisy $k$XOR problem and the Kikuchi method. In \Cref{sec:kik_motiv}, we outline an interpretation of the Kikuchi method in terms of mean-field theory and monogamy of entanglement. Our main technical contributions are in \Cref{sec:proof2,sec:proof3}, which prove correctness of our quantum algorithms for Planted Noisy $k$XOR and Tensor PCA, respectively. The final section, \Cref{sec:quantumalgorithm}, describes the implementation of the quantum algorithm and establishes its efficiency.  

\section{Technical background}
\label{sec:bg}
We first review the Planted Noisy $k$XOR problem and then use it to introduce the Kikuchi method.
\subsection{Planted Noisy $k$XOR}
\label{sec:bg_kXOR}
The Planted Noisy $k$XOR problem is at the core of the field of average-case analysis for constraint satisfaction problems, with a history dating back at least forty years~\cite{Has84}. It encompasses several other well-studied problems, such as Maximum Cut for random graphs with a planted partition, as special cases. An additional motivation for studying it comes from cryptography, where for some time now there has been a focus on ``new'' hardness assumptions. 
One reason for this focus is, of course, the necessity of developing post-quantum cryptography, given that Shor's algorithm breaks all cryptography based on the hardness of factoring or discrete logarithm. 
Another reason is that alternative assumptions --- such as the hardness of Learning With Errors (LWE), Learning Parity with Noise (LPN), and breaking pseudorandom generators (PRGs) in $\mathsf{NC}^0$ --- have been crucial for the construction of advanced cryptographic primitives such as Indistinguishability Obfuscation~\cite{JLS21,JLS22}.

Here, we focus on the very closely related cryptographic assumption known as \emph{Sparse LPN}~\cite{Ale03} --- which precisely concerns the average-case hardness of Planted Noisy $k$XOR. 
Under certain parameter settings, Sparse LPN is known to imply Public Key Encryption~\cite{ABW10}, multiparty homomorphic secret sharing~\cite{DIJL23}, lossy trapdoor functions, and collision-resistant hash functions~\cite{DJ24}.

\subsubsection{Problem definition}
In the $k$XOR constraint satisfaction problem (with $k \geq 2$ a constant) the input consists of~$m$ linear equations in $n$~variables $x_1, \dots, x_n$ over~$\{\pm 1\}^n$, in which each ``left-hand side'' is ``$k$-sparse''; i.e., it contains exactly $k$ out of~$n$ variables. 
Formally, a \emph{$k$XOR instance} $\calI$ over variables indexed by $[n]$ consists of $m$ \emph{constraints} $\calC = (S,b)$, where each \emph{scope} $S \subseteq [n]$ has cardinality~$k$ and each \emph{right-hand side}~$b$ is in $\{\pm 1\}$.  Each constraint  specifies a $k$XOR equations via 
\begin{equation}
    x^S  = b \in \{\pm 1\} \qquad \text{(where $x^S \coloneqq \prod_{i \in S} x_i$)}.
\end{equation} Since we represent the Boolean variables $x_1, \dots, x_n$ as elements of $\{\pm 1\}^n$ instead of $\{0, 1\}^n$, the linear constraints take the form of multiplicative equations.

In the Planted Noisy version, a ``secret'' variable assignment $z \in \F_2^n$ is (randomly) chosen, then $m$ random $k$-sparse equations consistent with~$z$ are chosen, and finally each ``right-hand side'' is flipped with probability equal to some ``noise rate''~$\eta \in (0,1/2]$\footnote{The problem is easy with noise rate $\eta=0$, since it amounts to solving a system of linear equations over a finite field.}. We will more commonly use the parameter $\rho = 1 - 2 \eta \in [0,1)$, called the ``planted advantage''.

Given this input, there are  several (closely related) natural algorithmic tasks: to find~$z$; to find an assignment having good correlation with~$z$; or, to distinguish the input equations 
from ones drawn from the ``purely random'' distribution (equivalent to planted advantage $\rho = 0$, in which case the distribution has no dependence on~$z$).
For simplicity, we focus our discussion on this last, ``decision'' variant of the problem, where the goal is to distinguish between a draw from the planted noisy distribution and a draw from the purely random distribution. More formally, we define the Planted Noisy $k$XOR problem as follows.

We let $\calR_{n,k}(m)$ (for \textbf{r}andom) denote the distribution of uniformly random $k$XOR instances in which $m$ scopes $S_1, \dots, S_m \in \binom{n}{k}$ are chosen uniformly at random, and the associated right-hand sides $b_1, \dots, b_m$ are i.i.d.\ Rademacher random variables (either $+1$ or $-1$, with equal probability). We also use the notation $\bcalI \sim \wt{\calR}_{n,k}(m)$ to denote the Poissonized version, in which first $\bm \sim \mathrm{Poi}(m)$ is drawn, and then (independently) $\bcalI \sim \calR_{n,k}(\bm)$. 

In the planted case, the right-hand sides are instead correlated with a ``secret'' assignment $z \in \{\pm 1\}^n$.  For $0 \leq \rho \leq 1$, we let  $\calP^z_{n,k}(m,\rho)$ (for \textbf{p}lanted) denote the distribution of planted noisy $k$XOR instances in which $m$ scopes $S_1, \dots, S_m \in \binom{n}{k}$ are chosen uniformly at random, and the associated right-hand sides are $\boldeta_1 z^{S_1}, \cdots, \boldeta_m z^{S_m}$, where $\boldeta_1, \dots, \boldeta_m$ are i.i.d.\ $\{\pm 1\}$-valued random variables satisfying $\E[\boldeta_i] = \rho$. When $\rho = 1$ we write simply $\calP^z_{n,k}(m)$, and when $\rho = 0$ the distribution is simply $\calR_{n,k}(m)$ (independent of~$z$). We continue to use $\wt{\calP}^z_{n,k}(m,\rho)$ to denote the Poissonized version. 

If $m \gg n$, with high probability no assignment satisfies significantly more than half of the constraints of a uniformly random $k$XOR instance, whereas for a planted $k$XOR instance the assignment $z$ satisfies $(1+\rho)/2$ of the constraints in expectation. The Planted Noisy $k$XOR problem is to decide whether a given $k$XOR instance is uniformly random or planted.  
\begin{problem}[Planted Noisy $k$XOR]
\label{prob:planted_noisy_kXOR_informal}
    For a quantity $0 < \rho \leq 1$, an algorithm that takes as input a $k$XOR instance $\calI$ and outputs a bit $r(\calI) \in \{0,1\}$ is said to solve the Planted Noisy $k$XOR problem at noise rate $\frac{1}{2}-\frac{\rho}{2}$ if 
    \begin{multline} 
    \qquad\quad \Pr_{\bcalI \sim \wt{\calR}_{n,k}(m)}\left[ r(\bcalI) = 1 \right] = 1- o(1)
\\ \text{ and } 
    \\ \forall z \in \{\pm 1\}^n:
    \Pr_{\bcalI \sim \wt{\calP}^z_{n,k}(m,\rho)}\left[ r(\bcalI) = 0\right] = 1 - o(1). \end{multline}  
    Here, the probability is over the randomness in the input and the internal randomness of the algorithm. The notation $o(1)$ denotes a term that vanishes in the limit $n \to \infty.$
\end{problem}

We have defined the task such that the algorithm should correctly identify instances drawn from the planted distribution (up to $o(1)$ failure probability) for \emph{every} $z$. One could also define the problem to be that of distinguishing a random instance from a planted instance for $z$ drawn uniformly at random. While the latter problem seems slightly easier, it has the same complexity due to a worst-case to average-case reduction over $z$.
\subsubsection{The computational threshold}
The main challenge in the Planted Noisy $k$XOR problem is to understand the computational complexity as a function of the ``constraint density'' $\Delta = m/n$, which can also be thought of as a ``signal-to-noise ratio'' (SNR). This question (and the very closely related one for Planted $k$SAT) has seen intensive study over the last $20^{+}$ years. 
When $\Delta \leq 1$, the two distributions are not statistically distinguishable, but once $\Delta > 1$, the two distributions can be distinguished, although not necessarily in polynomial time. 
We now know (see, e.g.,~\cite{Fei02,GK01,GJ02,CGL07,CCF10,AOW15,BM22,DT23}) that for  constant~$k$, provided $\Delta \geq C n^{(k-2)/2}$ (for a certain~$C$), there is a $\poly(n)$-time algorithm for the Planted Noisy $k$XOR problem.  
Moreover, there is reasonable evidence (see, e.g.,~\cite{AR01,Sch08,OW14,MW16,KMOW17}) that no $\poly(n)$-time algorithm is possible when $\Delta \ll n^{(k-2)/2}$.

For cryptographic applications, one seeks to refine this evidence so as to understand the tradeoff between the extent to which $\Delta < n^{(k-2)/2}$ and the (presumed) superpolynomial running time required for the task.
The reasons this is important are twofold. 
First, the cryptographic applications become stronger for larger $\Delta$; e.g., for $\mathsf{NC}^0$ PRGs based on Noisy $k$XOR, $\Delta$ directly corresponds to the stretch of the PRG.  
Second, for concrete cryptographic security one typically wishes to precisely quantify the (usually exponential) running time required for a cryptographic break. 
For example, more precisely understanding when the problem becomes hard allows us to choose key sizes in cryptographic protocols more aggressively.

Along these lines, work of Raghavendra, Rao, and Schramm~\cite{RRS17} showed that the degree-$\ell$ Sum-of-Squares (SoS) method --- which runs in $n^{O(\ell)}$ time --- suffices to solve the Planted Noisy $k$XOR problem provided $\Delta$ is at least a quantity of the form $(\frac{n}{\ell})^{(k-2)/2} \cdot \log^{c(k)}(n)$.
(See also~\cite{Ahn20}.)
The upper bound in~\cite{RRS17} was later made sharper, faster, and simpler by the development of the ``Kikuchi Method'' of Wein, Alaoui, and Moore~\cite{WAM19} (independently discovered by Hastings~\cite{Has20}), which improved the bound on $\Delta$ to a quantity of the form $(\frac{n}{\ell})^{(k-2)/2} \cdot \log n$.

The lower bounds in \cite{KMOW17}  matched this up to polylog factors, showing that indeed degree-$\ell$ SoS \emph{cannot} solve the problem unless $\Delta \gg (\frac{n}{\ell \log(n/\ell)})^{(k-2)/2}$. Since it would be a major algorithmic breakthrough if any classical algorithm outperformed the SoS bound for this problem, the algorithm based on the Kikuchi method is believed to be close to optimal (up to the gap between the upper and lower bounds above).

\subsection{Kikuchi method}
\label{sec:bg_kikuchi}
The Kikuchi Method can be interpreted as a way of transforming a degree-$k$ polynomial optimization problem into a degree-$2$ polynomial optimization problem (e.g., a $k$XOR instance into a $2$XOR instance, or a hypergraph problem into a graph problem, etc.). This is desirable because degree-$2$ problems may be modeled with matrices, allowing linear algebraic methods to be used. This section gives a concise review of the method. We refer the reader to \Cref{sec:Kikuchi_properties} for a comprehensive technical description, and to \Cref{sec:kik_motiv} for an interpretation in terms of statistical mechanics and monogamy of entanglement.  
\subsubsection{Baseline: $2$XOR} 
Consider a $k$XOR instance $\mathcal{I}$ with $m = \Delta n$ equations in the variables $x_1, \dots, x_n \in \{\pm 1\}$. If $k = 2$, then there is a very natural way to approach it.  Since $k =2$, each equation is of the form $x_i x_j = b_{ij} \in \{\pm 1\}$, and we can arrange the right-hand sides $b_{ij}$ into a matrix \footnote{Strictly speaking, the input might have several equations with the same left-hand side $x_i x_j$; in that case, $B_{ij}$ should equal the sum of the associated right-hand sides. For simplicity, we ignore the possibility of any left-hand side appearing more than once in this explanation, but incorporate it in \Cref{sec:Kikuchi_properties}.}~$B$ (with $B_{ij} = 0$ when no constraint on $x_i x_j$ is present). Then $\adv(x) \coloneqq \frac1{2m} x\transp B x$ represents the fraction of  satisfied constraints minus the fraction of violated constraints. The advantage of an assignment $x$ thus corresponds to its expected energy under the Hamiltonian $B$, which describes an Ising model without external field.

Note that in the Planted Noisy $k$XOR problem,  the planted assignment $z \in \{\pm 1\}^n$ achieves $\adv(z) = \rho$ in expectation, and we would generally expect (if $\Delta$ is large enough) that $\rho \approx \adv(z) \approx \mathrm{Opt} \coloneqq \max_x \{\adv(x)\}$.
So if we were trying to distinguish purely random $2$XOR instances from ones with planted advantage $\rho = 0.9$ (say), we could hypothetically succeed by detecting whether $\mathrm{Opt} \approx 0$ or  $\mathrm{Opt} \approx 0.9$.
Of course, approximating $\mathrm{Opt}$ is NP-hard, but for $2$XOR we can compute the continuous relaxation $\lambdamax{B}$, which satisfies
\begin{equation}
    \lambdamax{B} = \max_{y \in \R^n} \frac{y\transp By}{y\transp y} \geq \frac{2m}{n} \cdot \mathrm{Opt} = \ol{d} \cdot \mathrm{Opt},
\end{equation}
where $\ol{d} \coloneqq 2\Delta$ is the \emph{average degree} of the (edge-signed) graph associated to~$B$. In other words, we compute the largest eigenvalue (or ground state energy after the transformation $B \mapsto -B$) of the Hamiltonian $B$.

In the other direction, for purely random ($\rho = 0$) instances, $B$ is a randomly edge-signed random graph of average degree~$\ol{d}$, or equivalently, a random Ising Hamiltonian. In this case, one can try a variety of proof methods of varying sophistication~\footnote{E.g., Matrix Chernoff bounds as in~\cite{WAM19} and this paper, the Trace Method, nonbacktracking matrices, or even ``quantum field theory techniques'' as in~\cite{Has20}.} to prove upper-bounds on  $\lambdamax{B}$ of the form $C \cdot \ol{d}^{1/2}$ (for $C = O(\sqrt{\log n})$ or $C = O(1)$ or $C \approx 2$).
These allow the prover to confidently solve the Planted Noisy $2$XOR problem provided $C \ol{d}^{1/2}$ is a bit smaller than $\rho \cdot \ol{d}$; i.e., provided $2\Delta = \ol{d}$ is a bit bigger than $C^2/\rho^2$. 
So roughly speaking, we can solve the Planted Noisy $k$XOR problem once the average degree is slightly large, like $\Omega(1)$ or $\Omega(\log n)$, by estimating the largest eigenvalue of $B$.


\subsubsection{Reducing $k$XOR to $2$XOR}
For $k> 2$, the constraints are described by a $k$-Tensor instead of a matrix, and it becomes less obvious how to approach the Planted Noisy $k$XOR problem. A general idea, employed in almost all previous works on Planted Noisy $k$XOR, is to try to reduce to the above $2$XOR strategy. The Kikuchi Method, introduced by \cite{WAM19} and independently discovered by \cite{Has20}, seems to be both the simplest and most effective method for doing this. 

It starts (at least, for even~$k$) by introducing many new variables $y_T$ indexed by subsets $T \subset [n]$ with $\abs{T} = \ell$ for some fixed parameter~$\ell \geq k/2$.
There are then $\binom{n}{\ell} \approx n^\ell$ new variables \footnote{The approximation $\binom{n}{\ell} \approx n^\ell$ holds only if $\ell$ is considered to be a constant.  In fact, it makes sense even to allow  $\ell$ which is polynomially related to~$n$, but for the sake of intuition in this section, we assume $\ell$ is constant.}.
Each constraint $x^S = b$ in the original $k$XOR instance gets converted to all possible $\binom{n-k}{\ell - k/2}\binom{k}{k/2}$ equations of the form $y_T y_U = b$, conditioned on $T,U \in \binom{[n]}{\ell}$ satisfying $T \symdiff U = S$. 
This means we convert a $k$XOR instance with constraint density $\Delta$ into a $2$XOR instance -- the ``Kikuchi graph'' or ''Kikuchi Hamiltonian'' $\calK_\ell$ --  with average degree
\begin{equation} \label{eqn:heur}
    \ol{d}_{\calK_{\ell}} = \frac{\Delta n \cdot \binom{n-k}{\ell - k/2}\binom{k}{k/2}}{\binom{n}{\ell}} \sim \Delta \cdot \binom{k}{k/2} \cdot \ell \cdot \parens*{\frac{\ell}{n}}^{(k-2)/2}.
\end{equation}

As before, we aspire that the largest eigenvalue of the Kikuchi Hamiltonian reveals whether the $k$XOR instance is planted or random. By design, if $z \in \{\pm 1\}^n$ achieves advantage~$\alpha$ for the original $k$XOR instance, then $z^{\odot \ell}$ (that is, the vector in $\{\pm 1\}^{\binom{[n]}{\ell}}$ whose entries are indexed by sets of length $\ell$, and the $T$-coordinate is $z^T$) achieves advantage~$\alpha$ in this new Kikuchi $2$XOR instance.  In particular, starting from a Planted Noisy $k$XOR with planted advantage~$\rho$, the Kikuchi instance $\calK_\ell$ will have $\Opt \gtrapprox \rho$ and hence $\lambdamax{\calK_\ell} \gtrapprox \rho \cdot \ol{d}_{\calK_\ell}$.

On the other hand, if we start from a truly random $k$XOR instance, the resulting Kikuchi instance is \emph{some kind of} randomly edge-signed random graph of average degree~$\ol{d}_{\calK_\ell}$.  
So we might heuristically hope that still $\lambdamax{\calK_\ell} \leq C \cdot \ol{d}_{\calK_\ell}^{1/2}$ for some relatively small factor $C$ such as $C = O\Bigl(\sqrt{\log \binom{n}{\ell}}\Bigr)$.
If this is indeed proven true, we would have the same tradeoff as in our basic $2$XOR discussion, meaning that we could successfully solve the Planted Noisy $k$XOR problem as soon as $\ol{d}_{\calK_\ell}$ is a bit bigger than~$C^2/\rho^2$ (a quantity which is presumably/hopefully $O(\log n)$). By \Cref{eqn:heur}, this corresponds to the original constraint density~$\Delta$ being a bit bigger than $\frac{\log n}{\rho^2}\binom{k}{k/2}^{-1} \cdot \parens*{\frac{n}{\ell}}^{(k-2)/2}$.
This matches the original \cite{RRS17} result, provided one can prove the suggested upper bound for~$C$. In \Cref{sec:kik_motiv}, we discuss an interpretation of the Kikuchi matrix as an $\ell$-particle mean-field Hamiltonian.

\subsubsection{Example}
It will help us to introduce classical Alice and quantum Bob, in summarizing the above discussion. Alice uses a classical computer to solve the Planted Noisy $k$XOR problem. In the next section, we explain how Bob solves the same problem as Alice quartically faster on a quantum computer. To say that Alice uses ``\emph{Kikuchi Method with parameter~$\ell$}'' to solve the Planted Noisy $k$XOR problem (even~$k$) means:
\begin{itemize}[left=0.1cm]
    \item Alice proves a theorem of the following form:

    \textbf{Alice Theorem.} \emph{Assume $\Delta \geq C_\kappa \cdot (n/\ell)^{(k-2)/2}$.
    Then given a truly random $n$-variate, $m$-constraint $k$XOR instance, it holds whp \footnote{With high probability, by which we mean except with probability at most $o(1)$.
    } that the associated Kikuchi graph satisfies}
    $
        \lambdamax{\calK_\ell} \leq \kappa \cdot \ol{d}_{\calK_\ell}
    $.
    
    Here we allow that $C_\kappa$ might not strictly be a constant, but, say, $\polylog(n)$.
    \item (Alice recognizes the simple fact that in the noisy $k$XOR case with planted advantage~$\rho$, for any constant $\rho' < \rho$ we have whp
    $\lambdamax{\calK_\ell} \geq \rho' \cdot \ol{d}_{\calK_\ell}$.)
    \item Given a noisy $k$XOR instance $\bcalI$ of constraint density~$\Delta$ which is either truly random or has planted advantage~$\rho$, Alice instantiates her Theorem with $\kappa = \hat{\rho} < \rho' < \rho$ (but still $\hat{\rho} \approx .99\rho$, say).
    She then selects
    \begin{equation}
        \ell \geq n(C_{\hat{\rho}}/\Delta)^{2/(k-2)}
    \end{equation}
    to conclude that whp, she can solve the Planted Noisy $k$XOR problem by deciding whether $\lambdamax{\calK_\ell} \leq \hat{\rho} \cdot \ol{d}_{\calK_\ell}$ or $\lambdamax{\calK_\ell} \geq {\rho'} \cdot \ol{d}_{\calK_\ell}$.
    \item Finally, Alice decides this whp in $\wt{O}(\binom{n}{\ell}) \approx \wt{O}(n^\ell)$ time using, say, the Power Method on $\calK_\ell$. (Note that simply writing down one vector in the dimension of $\calK_\ell$ takes $\Omega(n^\ell)$ time.) 
\end{itemize}

In \Cref{sec:proof2}, we prove an example explicit ``Alice Theorem'', \Cref{thm:alice}, using the Matrix Chernoff bound (as do~\cite{WAM19}). In the case of $k = 4$ and $\rho = 0.25$, it implies that if $\Delta \geq (\frac{8}{\ell} \ln n) \cdot n$, then whp $\lambdamax{\calK_\ell}\leq .24 \ol{d}$.  It follows by taking $\ell = 32$ that Alice may solve the Planted Noisy $4$XOR problem (whp) at constraint density $\Delta \sim  (\tfrac{8}{32} \ln n) \cdot n = (.25 \ln n) \cdot n$  in time $\wt{O}(n^{32})$.

\section{Main results}
\label{sec:main}
\subsection{Planted Noisy $k$XOR}
\noindent Our main result, informally stated, is:
\begin{widequotation}
    Suppose Alice uses the Kikuchi Method with parameter~$\ell$ to solve the Planted Noisy $k$XOR problem for constraint density~$\Delta$ in time $\wt{O}(n^\ell)$. 
    Then Bob can use the same ``Alice Theorem'' to prove correctness of a quantum algorithm for the Planted Noisy $k$XOR problem, at the same constraint density~$\Delta$, with gate complexity $n^{\ell/4} \cdot \poly(n)$ and only $\O(\log n)$ qubits.
\end{widequotation}
More precisely, so long as $\ell$ is a multiple \footnote{Even if $\ell$ is not a multiple of $k$, we can increase $\ell$ to the next multiple of $k$, which increases the runtime of our quantum algorithm by at most a factor of $O(n^{k/4})$ and thus yields a total runtime of $n^{\ell/4} \cdot {\textstyle\widetilde{O}}(n^{3k/4})$.} 
of~$k$,  we show the quantum algorithm has complexity $n^{\ell/4} \cdot \wt{O}(n^{k/2})$.
Observe that as $\ell$ becomes large, this tends to a quartic speedup over the best known classical algorithm.  
For example, as a consequence of the Alice Theorem discussed in the previous section, we can prove that there is a quantum algorithm solving (whp) the Planted Noisy $4$XOR problem  with $\rho = 0.25$ in time $n^{8} \cdot \wt{O}(n^2)$.
(Here the improvement in the exponent, $\frac{32}{10} = 3.2$, is not quite~$4$ because $\ell$ is ``merely''~$32$.) 

\textbf{Caution.} Please note that we are not conjecturing that the fastest classical algorithm for Planted Noisy $4$XOR with $\rho = 0.25$ at constraint density $\Delta \sim (.25 \ln n) \cdot n$ is $\wt{O}(n^{32})$ (while also showing our quantum algorithm is $\wt{O}(n^{10})$ time). Rather, our result is that 
whenever Alice uses the Kikuchi method (as described above) to achieve time $\wt{O}(n^\ell)$ for a certain density~$\Delta$, Bob can use Alice's Theorem to achieve quantum complexity $n^{\ell/4} \cdot \poly(n)$. Importantly, we do not merely get a nearly quartic speedup over the current best classical algorithm based on the Kikuchi method, but even over hypothetical improved classical algorithms that still use the same framework.

Our quantum method for achieving quartic speedup on the Planted Noisy $k$XOR problem works by using an improved algorithm in the \emph{planted} case. The quartic speedup arises by combining two quadratic speedups, one due to Amplitude Amplification and one due to a better guiding state, which we describe in \Cref{sec:techniques}.


\subsection{Tensor PCA and Planted Inference}

We show that our quantum algorithm also achieves a (nearly) quartic quantum speedup for Tensor PCA, which is the problem originally studied by Hastings~\cite{Has20}.
We give a technical description in \Cref{sec:TensorPCA}, but informally, in this problem we are given a tensor of order $k$ in $n$ dimensions of the form $\T = \beta z^{\otimes k} + \G$, 
where $\G$ is a random tensor with each entry drawn i.i.d.\ from a standard Gaussian distribution, and $z$ is a fixed planted spike of norm $\sqrt{n}$. 
The task is to decide if $\beta=0$ (the random case) or $\beta=\beta^* > 0$ (the planted case). Spiked Noisy Tensor PCA can be viewed as a degree-$k$ polynomial optimization problem, and hence is also amenable to the Kikuchi Method. We show in \Cref{sec:tensor_guide} how our approach also solves this problem with a nearly quartic speedup.

Planted Noisy $k$XOR and Tensor PCA are examples of a larger class of problems called planted inference, which generally concern the detection of a signal or planted structure hidden in random noise. Our proofs use very few specific properties of the problems ($k$XOR or Tensor PCA) themselves and we expect our approach to generalize to further planted inference problems that can be solved using the Kikuchi method. A famous planted inference problem is Planted Clique \cite{alon1998finding}, the task of recovering a large planted clique hidden in a random Erdös-Renyi graph. However, Planted Clique is not amenable to the Kikuchi method as it is already a degree-$2$ optimization problem. Higher-degree examples where our approach may be effective include finding optimal solutions to random constraint satisfaction problems (CSPs) with planted assignments \cite{RRS17,WAM19}, community detection
in the stochastic block model \cite{JMLR:v18:16-480,abbe2015detection}, certification of an upper bound on the maximum k-cut of random (hyper)graphs \cite{bandeira2021spectral}, 
refutation of $k$-colorability of (hyper)graphs \cite{bandeira2021spectral}, and Group Synchronization \cite{singer2011angular,perry2018message}.

\subsection{Techniques}
\label{sec:techniques}
The weighted adjacency matrix of the Kikuchi graph is a sparse Hamiltonian of dimension $\binom{n}{\ell}$, which can be represented using (approximately) $\ell \log n$ qubits. As discussed in \Cref{sec:bg_kikuchi}, estimating the largest eigenvalue of the Kikuchi Hamiltonian solves the Planted Noisy $k$XOR problem. 
Specifically, in the case of a Planted $k$XOR instance with planted advantage~$\rho$, our algorithm would like to confirm that $\lambdamax{\calK_\ell} \geq \rho' \cdot \ol{d}_{\calK_\ell}$ (for some $\rho'$ slightly less than~$\rho$).
We will use the phrase ``cutoff eigenspace'' to mean the span of eigenvectors of $\calK_\ell$ of eigenvalue at least $\rho' \cdot \overline{d}_{\calK_\ell}$; thus the algorithm would like to confirm that the cutoff eigenspace is nonempty.
Approximating the largest eigenvalue of a sparse Hamiltonian --- or equivalently the ground state energy --- is a classic problem in quantum computing.
It is well studied, especially for local Hamiltonians, a special type of sparse Hamiltonian.
Indeed, the task of deciding whether the largest eigenvalue of a local Hamiltonian is above or below two nearby thresholds is the canonical QMA-complete problem~\cite{KSV02,Boo14}. 

The problem does get easier if we also have the ability to prepare a \emph{guiding state}, a quantum state having nontrivial \emph{overlap} \footnote{We say the ``overlap'' between two pure quantum states $\ket{\psi}, \ket{\phi}$ is $\abs{\braket{\psi}{\phi}}^2$.} with the largest eigenvector, or more generally with a vector in the cutoff eigenspace.  
For local Hamiltonians, this problem is called the \emph{Guided Local Hamiltonian problem} and has been studied recently~\cite{GL22,CFG+23}. We describe how to solve the more general \emph{Guided Sparse Hamiltonian problem} in \Cref{sec:guidedsparseHamSim}, but for now let us briefly say that if the sparse Hamiltonian's nonzero entries are efficiently computable, then the Guided Sparse Hamiltonian problem has complexity $\poly(n)/{\gamma}$, 
where $\gamma^2$ is the overlap 
of the guiding state with the cutoff eigenspace. 
The gist of the algorithm is to combine (known) algorithms for Hamiltonian Simulation, Phase Estimation, and Amplitude Estimation.

\subsubsection{Quadratic speedup: Amplitude Amplification}\label{sec:quadratic}

One possible choice for the guiding state is a completely random unit vector.
When we're in the planted case, so that the cutoff eigenspace is nonempty (but presumably of small dimension), such a random guiding state will (whp) have overlap inversely proportional to the dimension, meaning $\gamma^2=\Theta(1/n^{\ell})$ in our setting. This already yields a quantum algorithm with complexity $n^{\ell/2}\cdot \poly(n)$, a (nearly) quadratic speedup over the classical algorithm. The algorithm also only requires $\O(\log n)$ qubits. This quadratic quantum speedup is simply due to Amplitude Amplification.
\subsubsection{Quartic speedup: Guiding state}\label{sec:quartic}
As in \cite{Has20}, the key to improving to a (nearly) quartic speedup is obtaining a better guiding state. 
In principle, if $z$ is the planted assignment, 
a great choice for the guiding state would (hypothetically) be the normalized vector  $\ket{z^{\odot \ell}}$ corresponding to $z^{\odot \ell}$.
This is because $z^{\odot \ell}$ is the natural vector that ``certifies'' nonemptiness of the cutoff eigenspace, by virtue of
\begin{equation}
    \lambdamax{\calK_\ell} \geq \braketOP{z^{\odot \ell}}{\calK_\ell}{z^{\odot \ell}} = \adv(z^{\odot \ell}) \cdot \ol{d}_{\calK_\ell}  \geq \rho' \cdot \ol{d}_{\calK_\ell} 
\end{equation}
(whp).
We do have to be additionally concerned about lower-bounding the \emph{overlap} of $\ket{z^{\odot \ell}}$ with the cutoff eigenspace, but the above inequality --- together with an easy \emph{upper} bound of $O(\ell \log n)$ for $\lambdamax{\calK_\ell}$ ---ensures that the overlap is quite huge: $\gamma^2 = \Omega(\frac{1}{\ell \log n})$.
Thus \emph{if} a quantum algorithm could actually prepare the guiding state $\ket{z^{\odot \ell}}$, it could certify being in the planted case extremely efficiently: in $\poly(n) \cdot \ell$ time.

An additional virtue of using $\ket{z^{\odot \ell}}$ as a guiding state is that it is ``essentially'' of the form $\ket{z}^{\otimes \ell}$.\footnote{We put ``essentially'' in quotes because $\ket{z^{\odot \ell}}$ is actually a symmetrized version of $\ket{z}^{\otimes \ell}$. We prove in \Cref{sec:proof2} that this does not cause any major complications.} 
This crucially means that even though $\ket{z^{\odot \ell}}$ is of dimension $\binom{n}{\ell}$ (a running time we don't want to suffer), we could prepare it simply by preparing~$\ell$ unentangled copies of~$\ket{z}$, a state of dimension merely~$n$.

Unfortunately, this plan is not feasible, since preparing $\ket{z}$ amounts to finding the planted vector! The idea does inspire our actual solution, though.

Our guiding state is based on the observation that the very input to the problem gives a state that is somewhat correlated with $z^{\odot k}$. Specifically, given a $k$XOR instance with $m$ constraints of the form $x^{S_i} = b_i$, we prepare a vector $\psi \in \{\pm 1\}^{\binom{[n]}{k}}$, which has $b_U$ as its $U$th entry whenever $x^U = b_U$ is in the input, and has $0$ as its $U$th entry otherwise.
\begin{equation}
\label{eq:small_guide_main}
    \psi_U = \begin{cases}  
        b_{U} \quad \text{ if } U = S_i \text{ for some } 1\leq i\leq m
        \\ 0 \text{ else }. 
    \end{cases}
\end{equation}
Its normalized version $\ket{\psi}$ can be prepared in $O(m\log n)$ time using standard methods. Then by definition, assuming the $k$XOR instance does not containt the same constraint multiple times\footnote{We handle this case formally in \Cref{sec:proof2}.},
\begin{align}
    \mu^2 &\coloneqq \abs{\braket{\psi}{z^{\odot k}}}^2 = \frac{1}{m \binom{n}{k}}\left( \sum_i b_i z^{S_i} \right)^2 \\&= \adv(z)^2 \frac{m}{\binom{n}{k}}
= \rho^2 \cdot \frac{m}{\binom{n}{k}},
\end{align} meaning $\mu^2 = \wt{\Theta}(n^{-k/2})$ for the value of~$m = \wt{\Theta}(n^{k/2})$ arising in any natural Alice Theorem.  In turn, if we define the guiding state $\ket{\Psi} \coloneqq \ket{\psi}^{\otimes \ell/k}$ (which is hardly more complex to prepare than $\ket{\psi}$), it is reasonable to expect overlap 
\begin{align}\label{eq:quadraticoverlap}
    \abs{\braket{\Psi}
    {z^{\odot \ell}}}^2 
    &\approx   \abs{\braket{\psi}
    {z^{\odot k}}}^{2(\ell/k)} 
    \approx (\mu^2)^{\ell/k}\\  &\approx \wt{\Theta}(n^{-k/2})^{\ell/k} = \wt{\Theta}(n^{-\ell/2}).
\end{align}
We prove an overlap statement of this from in \Cref{sec:proof2}, which we state informally here. 
\begin{theorem}[Overlap proof: informal version]
\label{thm:overlap_informal}
 Let $\bcalI$ be a planted $k$XOR instance and $\Kik_\ell$ its Kikuchi Hamiltonian. Given just a description of $\bcalI$, one can efficiently construct a \emph{guiding state} $\ket{\Psi}$ such that there exists a unit vector $\ket{\bv}$ in the cutoff eigenspace of $\Kik_\ell$ with 
    \begin{equation}
        \braket{\bv}{\Psi}^2 \geq \O\left( \left(\frac{m}{\tbinom{n}{k}}\right)^{\ell/k} \right).
    \end{equation}
\end{theorem}
The formal version and proof of this theorem is given in \Cref{thm:kXOR_analysis}. For the natural parameter setting $m = \wt{\Theta}(n^{k/2})$ of the Planted Noisy $k$XOR Problem, this overlap tends towards $\O(n^{\ell/2})$. 
The inverse-square-root of this is the source of our final quantum complexity, $n^{\ell/4} \cdot \poly(n)$. Importantly, as explained more in \Cref{sec:putting_all_together}, we show that this statement holds independently of the specific ``Alice Theorem'' and the corresponding value of $\ell$. 

\subsection{Complications}  There are several oversimplifications in the above outline.  One is to do with the conflation of $\ket{z^{\odot k}}^{\otimes \ell/k}$ and $\ket{z^{\odot \ell}}$; and additionally, $\ket{\Psi}$ with our actual guiding state, which is a symmetrized version of $\ket{\psi}^{\otimes \ell/k}$. Another difficulty is verifying that $\abs{\braket{\Psi}{z^{\odot \ell}}}^2$ is large not just in expectation but with high probability. 

The crucial difficulty is that we have only discussed the overlap of the guiding state $\ket{\Psi}$ with $\ket{z^{\odot \ell}}$, rather than with the cutoff eigenspace (as required by our quantum algorithm for the Guided Sparse Hamiltonian problem). Now earlier we argued that $\ket{z^{\odot \ell}}$ was likely to have huge overlap with the cutoff eigenspace, and so it is plausible to hope that we can put these two statements together to conclude that the guiding state still has overlap roughly $n^{-\ell/2}$ with the cutoff eigenspace. Unfortunately, proving this rigorously is technically challenging. The ``large'' overlap between the guiding state $\ket{\Psi}$ and $\ket{z^{\odot \ell}}$ of order $\approx n^{-\ell/2}$ is still small enough that a straight forward composition of the two overlap statements is not sufficient. This complication is further enhanced by the fact that the two overlap statements are not independent: the Kikuchi matrix~$\calK_\ell$ and the guiding state $\ket{\Psi}$ are both defined from the same random input, and hence depend on each other.

For the Tensor PCA problem, Hastings faces similar difficulties, and he uses a somewhat intricate strategy to deal with it.  
He first creates a noisy version of the input data, and from this he creates a Hamiltonian similar to the Kikuchi matrix as well as several slightly  different noisy guiding states. 
Then, repeatedly adding additional noise to each guiding state and using several special properties of Gaussian random variables and low-degree polynomials (e.g., the Carbery--Wright theorem on anticoncentration), he is able to show that with high probability at least one of the guiding states has overlap $\approx n^{-\ell/2}$ with the Kikuchi matrix's high eigenspace. This method exploits powerful properties of Gaussian random variables and does not have an obvious analogue in the Planted Noisy $k$XOR problem.

We employ a simpler strategy in our work, which also generalizes to Tensor PCA.  Essentially, we separate the $m$ input $k$XOR constraints into two independent batches, one of size $(1-\zeta)m$ and the other of size $\zeta m$.  (Think of $\zeta = .001$, say.)  
We construct the Kikuchi matrix from the first batch, and since $(1-\zeta) m \approx m$, the ``Alice Theorem'' is hardly impacted and we still essentially have $\lambdamax{\calK_\ell} \leq \hat{\rho} \cdot \ol{d}_{\calK_\ell}$ in the truly random case.
From the second batch, we construct the guiding state, and although it has just $\zeta m$ constraints, this only reduces the anticipated overlap of $\wt{\Omega}(n^{-\ell/2})$ by a factor of $\zeta^\ell$ (and thus increases the final quantum running time by a factor of $2^{O(\ell)}$ which is negligible compared to $n^{O(\ell)}$). The details of this procedure are described in \Cref{sec:proof2}.

Once independence has been established, we project the guiding state and the Kikuchi matrix into the ``homogeneous subspace'' of $[n]^\ell$ spanned by $\ell$-tuples with no repeating entries. In this subspace, a second-moment method suffices to establish that our guiding state has the desired overlap with the cutoff eigenspace of the Kikuchi matrix.

At the end of this procedure, we have arrived at an instance of the Guided Sparse Hamiltonian problem, which can be solved in the stated complexity if we can prepare the guiding state efficiently and compute nonzero entries of the Kikuchi Hamiltonian efficiently. We describe the corresponding quantum algorithm in \Cref{sec:quantum_algo} and prove its efficiency in \Cref{sec:quantumalgorithm}.

\subsection{Intuition for quantum advantage} The quartic speedup arises by combining two quadratic speedups: one from the (Grover-like) Amplitude Estimation subroutine, and the other from having a state with quadratically improved (vs.~a random vector) overlap with the cutoff eigenspace. 
In the previous section, we have discussed the origin of both these speedups. We now provide some intuition for why having a state with improved overlap does not help the classical algorithm. 

Consider the Guided Sparse Hamiltonian problem on $N$ qubits, so the Hamiltonian is of size $2^N \times 2^N$. The problem is QMA-complete if we are given no guiding state, or equivalently, if we are given a guiding state with overlap $2^{-N}$, since a random vector has this overlap. On the other hand, Gharibian and Le Gall~\cite{GL22} show that the problem becomes BQP-complete if the guiding state has $1/\poly(N)$ overlap. As the overlap changes from inverse exponential to inverse polynomial, the quantum complexity of the problem changes drastically: it goes from exponential (assuming QMA-complete problems take exponential time to solve on a quantum computer) to polynomial. On the other hand, assuming classical algorithms require exponential time for both BQP-complete and QMA-complete problems, the classical complexity  remains exponential. This suggests that classical algorithms are unable to exploit good guiding states, and hence the ability to prepare states with improved overlap only helps the quantum algorithm. To give a concrete example, consider the Power Method for approximating the spectral norm of a Hamiltonian. While a good guiding state reduces the number of iterations required for convergence, the dominating cost is the cost of each single iteration. And this single-iteration cost is already at least $2^N$, since the algorithm has to multiply a $2^N$-dimensional vector with a $2^N \times 2^N$ dimensional matrix. 

Super-quadratic quantum speedups have been demonstrated or suggested in prior works, including \cite{hastings2018short, Dalzell2023mind, Boulebnane2022,chakrabarti2024generalized,kapit2023approximability,zhou2024statistical}. Some of these approaches share similarities with ours; for instance, both \cite{hastings2018short} and \cite{Dalzell2023mind} employ a guiding state in combination with quantum phase estimation. Hence the underlying intuition for super-quadratic speedups developed in those works can also apply here. We note that the quartic speedup achieved in our work is substantially larger than the speedups obtained in these earlier results.

\section{Quantum algorithm}
\label{sec:quantum_algo}
In this section, we present an overview of our quantum algorithm that achieves a (nearly) quartic quantum speedup for the Planted Noisy $k$XOR problem. The input to our algorithm is a $k$XOR instance $\calI = \{S_i,b_i \}_{i = 1}^m$ on $n$ variables, and an ``Alice Theorem'' as described in \Cref{sec:bg_kikuchi}.

\begin{center}
    \textbf{ Algorithm}
\end{center}
\noindent\hrulefill

\noindent\textbf{Preprocessing:} Choose a sufficiently large $\ell$ and a threshold $\lambda_{th}$ in accordance with the ``Alice Theorem'' at constraint density $\Delta  = \frac{m}{n}$. That is, the ``Alice Theorem'' guarantees that if $\calI$ is a purely random instance, then whp the spectral norm  $||K_\ell||$ of the Kikuchi Hamiltonian is less than $\lambda_{th}$. \\
\textbf{Quantum algorithm:}
Encode the following in Amplitude Amplification and repeat $\O(n^\frac{\ell}{4})$ times:
\begin{itemize}[left = 0.2cm]
        \item Prepare $\ell/k$ unentangled copies of the guiding state $\ket{\psi}$ defined in \Cref{eq:small_guide_main}. Symmetrize the resulting state to obtain a guiding state $\ket{\Psi}$. 
        \item Perform Quantum Phase Estimation with the sparse Hamiltonian $K_\ell$ on the initial state $\ket{\Psi}$.
        \item Measure the eigenvalue register and record whether an eigenvalue above the threshold $\lambda_{th}$ was sampled.
\end{itemize}
\textbf{Output:} If during any of the repetitions, an eigenvalue above $\lambda_{th}$ was found, return ``Planted''. Otherwise, return ``Random''. 

\noindent\hrulefill

We give a detailed technical description of the quantum algorithm in \Cref{sec:quantumalgorithm}, where we prove efficiency of the above procedure. Preparing the guiding state is efficient because it essentially amounts to preparing $\ell/k$ copies of the $m$-sparse state $\ket{\psi}$, which can be done in time that scales linearly in $m$. We prove this fact in \Cref{sec:guidingstateprep} by building on state preparation routines described in \Cref{sec:state_prep_prelim}. Because $\Kik_\ell$ is sparse, the second step involving Quantum Phase Estimation is also efficient, which we prove in \Cref{sec:KikuchiHamSim} utilizing sparse Hamiltonian simulation techniques described in \Cref{sec:guidedsparseHamSim}. The gate complexity of this quantum algorithm is thus dominated by the number of repetitions of Amplitude Amplification, and of the form $n^{\ell/4} \cdot \poly(n)$. Finally, we prove in \Cref{sec:proof2} that -- with high probability over the randomness of the input and the internal randomness of the algorithm -- our quantum algorithm correctly detects whether the input instance $\calI$ is random or planted.

What kind of quantum resources are required to execute this algorithm? While we leave a full resource analysis for future work, preliminary implementations of the algorithm and its subroutines in Qualtran are provided in \cite{qualtranPR1348, harrigan2024expressing}. Based on these implementations, we estimate that classically intractable instances would require on the order of $10^{15}$ to $10^{16}$ Toffoli gates. The cost of block-encoding the Kikuchi matrix is modest (on the order of $10^6$ Toffoli gates), as expected, while the dominant cost arises from repeated applications of Quantum Phase Estimation and Amplitude Amplification. The resource requirements of our algorithm are thus akin to early estimates for problems in quantum chemistry, such as simulating FeMoco \cite{reiher2017elucidating}. Since our estimates arise from a coarse first pass, further optimizations are almost certainly possible. For comparison, the resource estimates for FeMoco have since been reduced by more than a factor of a million \cite{gunther2025phase}.
\section{Conclusion}
We have shown how to ``quantize'' the classical algorithms for Planted Noisy $k$XOR and Tensor PCA to obtain quantum algorithms with nearly quartic quantum speedups. 
Importantly, we do not merely get a nearly quartic speedup over the current best classical algorithm based on the Kikuchi method, but even over hypothetical improved classical algorithms that still use the same framework.
Our approach can be viewed more as a method of upgrading a classical algorithm of this Kikuchi form to a quantum algorithm, rather than a quantum algorithm tailored to a particular problem. 

Our quartic speedups exploit the fact that planted inference problems naturally instantiate the ``Guided Sparse Hamiltonian problem''. Specifically, the Kikuchi method maps planted inference problems to the estimation of the ground-state energy (largest eigenvalue) of a large, implicitly defined, sparse Hamiltonian (the Kikuchi matrix). We show that this matrix admits a \emph{guiding state} which has polynomially increased (over a random state) overlap with its leading eigenspace, and which is moreover efficiently preparable given just the input of the planted inference problem. The Guided Sparse Hamiltonian problem was previously studied in the special case where the Hamiltonian is local and the guiding state has exponentially increased overlap, in which case the problem is BQP-complete \cite{GL22}. However, finding problems where such exponentially improved guiding states are efficiently preparable remains an outstanding challenge.  Our work highlights that guiding states with only polynomially increased overlap can appear naturally, as in the context of the planted inference problems studied here, resulting in significant polynomial quantum speedups. 

Our work further underscores that techniques such as the guided local Hamiltonian problem -- originally developed with applications in quantum chemistry in mind -- can extend to unexpected domains within theoretical computer science. In the present setting, this connections emerges through the interplay between planted inference problems and statistical physics, in particular mean-field theory. To establish further complexity-theoretic evidence for the quartic speedup developed here, a key open research direction is to investigate the fine-grained complexity of the Guided Sparse Hamiltonian problem, in the setting where the guiding state offers only polynomial, rather than exponential, overlap enhancement relative to a random state.

Since the proofs of our quartic quantum speedups use very few specific properties of the problems ($k$XOR or Tensor PCA) themselves, we expect our approach to generalize to other planted inference problems that can be solved using the Kikuchi method. As such, our core techniques can serve as a new framework for obtaining large polynomial quantum speedups for problems of practical interest. In particular, problems closely related to the Planted Noisy $k$XOR problem have been used as components of certain cryptographic constructions, for instance the 6-ary TSA predicate studied in \cite{OW14}, where in each clause the first four variables act as a 4XOR constraint and the remaining two variables encode the noise. While we do not claim to have a specific instantiation of a cryptosystem that is broken by our algorithm, the generality of our approach suggests that some of these cryptosystems might need to increase their security parameters in order to sufficiently protect against our attack.


\begin{acknowledgments}
We would like to thank Stephen Jordan for comments on an earlier draft, and Edward Farhi, Aram Harrow, Bill Huggins, Tanuj Khattar, Anurudh Peduri, Rolando Somma, and Prashant Nalini Vasudevan for helpful discussions. Alexander Schmidhuber was partially funded by NSF grant PHY-2325080. Ryan O'Donnell's work on this project was funded by a grant from Google Quantum AI.
\end{acknowledgments}

\clearpage
\newpage
\appendix
\onecolumngrid
\section{Properties of the Kikuchi method}
\label{sec:Kikuchi_properties}
We first recall the definition of the Planted Noisy $k$XOR problem and then use this problem to formally introduce the Kikuchi method and to discuss its properties.
\subsection{Planted Noisy $k$XOR problem}
\label{sec:problem_statement}

\begin{definition}      \label{def:I}
    A \emph{$k$XOR instance} $\calI$ over variables indexed by $[n]$ consists of a multiset of \emph{constraints} $\calC = (S,b)$, where each \emph{scope} $S \subseteq [n]$ has cardinality~$k$ and each \emph{right-hand side}~$b$ is in $\{\pm 1\}$.  
\end{definition}

A natural associated task is to either find an $n$-bit string that satisfies as many constraints $(S,b)$ as possible or compute the maximum number of constraints that can be satisfied.
\begin{definition}
    Given a $k$XOR instance $\calI$ as above, 
    we say that assignment $x \in \{\pm 1\}^n$ \emph{satisfies} constraint $\calC = (S,b)$ if $x^S = b$, where $x^S$ denotes $\prod_{i \in S} x_i$.
    The \emph{advantage} of assignment $x$ for~$\calI$ is 
    \begin{equation}
        \adv_{\calI}(x) = \avg_{(S,b) \in \calI} \braces{b \cdot x^S} \in [-1,+1].
    \end{equation}
    Note that $\half + \half \adv_{\calG}(x)$ represents the fraction of constraints satisfied by~$x$.
    Finally, we write $\Opt(\calI) = \max_{x} \braces{\adv_{\calI}(x)}$.
\end{definition}

In this work, we study the average-case problem of distinguishing uniformly random from planted $k$XOR instances. We now define the random instances we consider.

\begin{notation}    \label{not:R}
    We let $\calR_{n,k}(m)$ (for \textbf{r}andom) denote the distribution of uniformly random $k$XOR instances in which $m$ scopes $S_1, \dots, S_m \in \binom{n}{k}$ are chosen uniformly at random, and the associated right-hand sides $b_1, \dots, b_m$ are i.i.d.\ Rademacher random variables (either $+1$ or $-1$, with equal probability). We also use the notation $\bcalI \sim \wt{\calR}_{n,k}(m)$ to denote the Poissonized version, in which first $\bm \sim \mathrm{Poi}(m)$ is drawn, and then (independently) $\bcalI \sim \calR_{n,k}(\bm)$.
\end{notation}


In the planted case, the right-hand sides are instead correlated with a ``secret'' assignment. 

\begin{notation}    \label{not:P}
    Fix $z \in \{\pm 1\}^n$ and $0 \leq \rho \leq 1$.
    We let  $\calP^z_{n,k}(m,\rho)$ (for \textbf{p}lanted) denote the distribution of planted noisy $k$XOR instances in which $m$ scopes $S_1, \dots, S_m \in \binom{n}{k}$ are chosen uniformly at random, and the associated right-hand sides are $\boldeta_1 z^{S_1}, \cdots, \boldeta_m z^{S_m}$, where $\boldeta_1, \dots, \boldeta_m$ are i.i.d.\ $\{\pm 1\}$-valued random variables satisfying $\E[\boldeta_i] = \rho$. 
    When $\rho = 1$ we write simply $\calP^z_{n,k}(m)$, and when $\rho = 0$ the distribution is simply $\calR_{n,k}(m)$ (independent of~$z$). We continue to use $\wt{\calP}^z_{n,k}(m,\rho)$ to denote the Poissonized version. 
\end{notation}

If $m \gg n$, with high probability no assignment satisfies significantly more than half of the constraints of a uniformly random $k$XOR instance, whereas the optimal advantage of planted $k$XOR instance is evidently related to $\rho$:

\begin{fact}
    For $\bcalI \sim \calP^z_{n,k}(m,\rho)$ we have  $\adv_{\bcalI}(z) = \avg\{\boldeta_i\}$.  
    Thus in general $\E[\adv_{\bcalI}(z)] = \rho$, and for $\bcalI \sim \calP^z_{n,k}(m)$ we have $\adv_{\bcalI}(z) = 1$ with certainty.
\end{fact}

The Planted Noisy $k$XOR problem is to decide whether a given $k$XOR instance is uniformly random or planted.  

\begin{problem}[Planted Noisy $k$XOR]
\label{prob:planted_noisy_kXOR}
    For a quantity $0 < \rho \leq 1$, an algorithm that takes as input a $k$XOR instance $\calI$ and outputs a bit $r(\calI) \in \{0,1\}$ is said to solve the Planted Noisy $k$XOR problem at noise rate $\frac{1}{2}-\frac{\rho}{2}$ if 
    \begin{equation} 
    \Pr_{\bcalI \sim \wt{\calR}_{n,k}(m)}\left[ r(\bcalI) = 1 \right] = 1- o(1) 
    \quad \text{ and } \quad \forall z \in \{\pm 1\}^n:
    \Pr_{\bcalI \sim \wt{\calP}^z_{n,k}(m,\rho)}\left[ r(\bcalI) = 0\right] = 1 - o(1).
    \end{equation}
\end{problem}

We have defined the task such that the algorithm should correctly identify instances drawn from the planted distribution (up to $o(1)$ failure probability) for \emph{every} $z$. One could also define the problem to be that of distinguishing a random instance from a planted instance for $z$ drawn uniformly at random. While the latter problem seems slightly easier, it has the same complexity due to a worst-case to average-case reduction over $z$.

Here we define our problem formally to be about distinguishing a draw from $\wt{\calR}_{n,k}(m)$ vs.\ $\wt{\calP}^z_{n,k}(m)$ 
instead of distinguishing an instance drawn from ${\calR}_{n,k}(m)$ vs.\ ${\calP}^z_{n,k}(m)$, because the former setup is more natural to reason about. But note that the two problems have almost the same complexity since we can reduce from one to the other by changing the value of $m$ to $(1- o(1))m$ in this reduction. 
For example, if we want to reduce the Poissonized problem to the non-Poissonized problem, when $\bm \sim \mathrm{Poi}(m)$ is drawn, we have that $\bm \geq (1-o(1))m$ with overwhelming probability, which gives us an instance of the non-Poissonized problem for a slightly smaller value of $m$.

\subsection{{$k$}{k}XOR and 2XOR}
\label{sec:xor-defs}
It will be fruitful to think of the special case of $2$XOR, in which each constraint specifies the parity of two variables, as being about edge-signed graphs.

\begin{definition}      \label{def:G}
    By an $n$-vertex \emph{edge-signed graph}, we mean an undirected graph~$\calG$ on vertex set~$[n]$, with parallel edges allowed, and with each edge being labeled by a sign~$\pm 1$.
    We identify $\calG$ with a multiset of ``signed edges'' $(S,b)$, where $S \subseteq [n]$ has cardinality~$2$ and~$b \in \{\pm 1\}$; in this way, $\calG$ may also be thought of as a $2$XOR instance.
    Finally, we will also identify $\calG$ with its (symmetric) \emph{adjacency matrix}: 
    \begin{equation}
        {\calG =}  \sum_{(\{i,j\}, b) \in \calG} b\cdot (\ketbra{i}{j} + \ketbra{j}{i}) = \sum_{\{i,j\} \in \binom{[n]}{2}} B_{\calG}(\{i,j\})\cdot (\ketbra{i}{j} + \ketbra{j}{i}).
    \end{equation}
\end{definition}  

We remark that the ``Max-Cut'' problem on graph~$G$ corresponds to the edge-signing ($2$XOR instance) in which all edges are labeled~$-1$.

\begin{notation}
    Given an $n$-vertex graph $\calG$ (possibly edge-signed), we write $D(\calG)$ for its maximum degree and $\ol{d}(\calG) = \frac{2\abs{\calG}}{n}$ for its average degree.
    We remark that $D(\calG) \geq \norm*{\calG}_{1 \mapsto 1}$ (the maximum $\ell_1$ norm of matrix columns).
\end{notation}

This allows us to write $\adv_{\calG}(x)$ as a normalized bilinear form of the adjacency matrix $\calG$.
\begin{notation}
    Suppose we consider the $n$-vertex edge-signed graph $\calG$ to be a $2$XOR instance.  
    Then, writing $\ket{x}$ for the unit vector in the direction of $x \in \{\pm 1\}^n$, namely $\ket{x} = \frac{1}{\sqrt{n}} x$, we have
    \begin{equation}
        \adv_{\calG}(x) = \avg_{(S,b) \in \calG} \braces{b \cdot x^S}  
        = \frac{n}{2\abs{\calG}} \braketOP{x}{\calG}{x} = \frac{\braketOP{x}{\calG}{x} }{\overline{d}(\calG)}.
    \end{equation}
\end{notation}

The following immediate spectral bounds relate the spectrum of $\calG$ to the advantage of the associated 2XOR instance:
\begin{fact}    \label{fact:1}
    For an $n$-vertex edge-signed graph~$\calG$:
    \begin{gather}
        \lambdamax{\calG} \leq 
        \opnorm{\calG} \leq \|\calG\|_{1 \mapsto 1} \leq D(\calG); \label[ineq]{ineq:dd}\\
        \adv_{\calG}(x) 
        = \frac{\braketOP{x}{\calG}{x}}{\ol{d}(\calG)}  
        \leq \frac{\lambdamax{\calG}}{\ol{d}(\calG)}
        \qquad\implies\qquad
        \Opt(\calG) \leq \frac{\lambdamax{\calG}}{\ol{d}(\calG)}.
    \end{gather}
\end{fact}
Here $\opnorm{M}$ denotes the operator norm of a matrix $M$, i.e., its maximum singular value. 

\subsection{Random signed matchings}
\label{sec:random_matchings}
A fruitful strategy for attacking a $k$XOR optimization problem is to reduce it to a $2$XOR optimization problem, which can then be analyzed via the spectral bounds outlined above. A particular effective method for doing so is the Kikuchi method, which we describe in the next two subsections. It will be convenient to introduce this method in terms of multisets of signed matchings, which generalize edge-signed graphs. 
\begin{definition}
    A \emph{signed matching on $[N]$} is a pair $(F,b)$, where $F$ is a matching (nonempty graph of degree at most~$1$) on vertex set~$[N]$, and $b \in \{\pm 1\}$ is a sign.
    If $\calF$ is a multiset of signed matchings, we identify it with its (symmetric) \emph{adjacency matrix}
    \begin{equation} \label{eqn:bM}
        \calF = \sum_{(F,b) \in \calF} b \cdot F,
    \end{equation}
    where $F$ is identified with its  adjacency matrix (when regarded as an unsigned graph).  
    In this way, $\calF$ may be regarded as an edge-signed graph.
    (Our original \Cref{def:G} is just the special case of this in which all matchings have cardinality~$1$; i.e., they are single edges.)
\end{definition}

As we are concerned with distinguishing uniformly random $k$XOR from planted noisy $k$XOR, we will be considering random and semi-random signed matchings. In order to apply the spectral relations in \Cref{fact:1}, we are interested in bounding the operator norm of the resulting edge-signed graphs. Let us first consider the case where the ``right-hand sides'' $\bb_i$ of the $k$XOR equations are random. 
\begin{proposition} \label{prop:mc}
    Let $\calF = (F_1, \dots, F_m)$ be a multiset of matchings on~$[N]$, and suppose $\bcalF$ is formed from it by associating signs $\bb_1, \dots, \bb_m$ that are independent and uniformly random.
    Then for any $\eps > 0$, except with probability at most $2N^{-\eps}$, 
    we have
    \begin{equation}
        \lambdamax{\bcalF} \leq \opnorm{\bcalF} \leq \sqrt{D(\calF)} \cdot \sqrt{2(1+\eps) \ln N}. 
    \end{equation}
\end{proposition}
\begin{proof}
    In \Cref{eqn:bM}, the signs are independent Rademacher random variables.
    Thus ``matrix Chernoff (Khintchine) bounds''
    (see, e.g.,~\cite{Tro12}) tell us that
    \begin{equation}
        \Pr[\opnorm{\bcalG} \geq \lambda_0] \leq 
        2N \cdot \exp\parens*{-\frac{\lambda_0^2}{2\opnorm{\Sigma^2}}},
    \qquad\text{where}\qquad
        \Sigma^2 \coloneqq \sum_{i=1}^m F_i^2.
    \end{equation}
    The matrix $F_i^2$ is diagonal, with its diagonal being the $0$-$1$ indicator of the vertices in the matching.  
    Hence $\Sigma^2$ is diagonal, with $i$th diagonal entry equal to the degree of vertex~$i$ in $\calF = F_1 \sqcup \cdots \sqcup F_m$.
    Thus $\opnorm{\Sigma^2} = D(\calF)$, and the result follows.
\end{proof}
In the Planted Noisy $k$XOR problem, the scopes or ``left-hand sides'' of the $k$XOR equations are also chosen randomly. This means that $\calF$ will usually be randomly chosen from a larger, fixed collection of matchings:
\begin{notation}    \label{not:matchy}
    Suppose $\calK$ is a multiset of matchings on~$[N]$.
    We will write $\bcalF \sim \calP_{\calK}(m)$ to denote that $\bcalF$ is the multiset of matchings $(\bF_1, \dots, \bF_m)$ formed by choosing each $\bF_i$ independently and uniformly at random from~$\calK$.
    We continue to use the notation $\bcalG \sim \wt{\calP}_{\calK}(m)$ to denote the Poissonized version, in which first $\bm \sim \mathrm{Poi}(m)$ is drawn, and then (independently) $\bcalG \sim \calP_{\calK}(\bm)$.
\end{notation}

\begin{definition}
    We say a collection $\calK$ of matchings on~$[N]$ is \emph{$\delta$-bounded} if, for all $i \in [N]$, the fraction of $F \in \calK$ that touch~$i$ is at most~$\delta$. 
    If, moreover, all $F \in \calK$ touch exactly $\delta N$ vertices, we call $\calK$ \emph{$\delta$-balanced}.
    \label{def:balanced}
\end{definition}

Standard Chernoff+union bound arguments yield the following:
\begin{proposition} \label{prop:md}
    Let $\calK$ be a $\delta$-bounded collection of matchings on~$[N]$. 
    Let  $\bcalF$ be drawn from either $\calP_{\calK}(m)$ or $\wt{\calP}_{\calK}(m)$.
    Then for any $\kappa > 0$, and writing $d = \delta m$, we have $D(\bcalF) \leq (1+\kappa)d$ except with probability at most $N \cdot \exp(-\frac{\kappa^2}{2+\kappa} d)$.
    
\end{proposition}

Combining \Cref{prop:mc,prop:md}, we arrive at an upper bound on the spectral norm of a random multiset of matchings:
\begin{proposition} \label{prop:rand0}
    Let $\calK$ be a $\delta$-bounded collection of matchings on~$[n]$
    and let $\bcalF$ be formed by first drawing from either $\calP_{\calK}(m)$ or $\wt{\calP}_{\calK}(m)$ and then associating independent and uniformly random signs.
    Then writing $d = \delta m$, we have
    \begin{equation}
        \lambdamax{\bcalF} \leq \opnorm{\bcalF} \leq \sqrt{2(1+\eps)(1+\kappa)\ln N} \cdot \sqrt{d} 
    \end{equation}
    except with probability at most $
        2N^{-\eps} + N \cdot \exp(-\tfrac{\kappa^2}{2+\kappa} d)$. 
    In particular,
    \begin{equation}
        d \geq \frac{2(1+\eps)(1+\kappa)}{\kappa^2} \cdot \ln N
    \quad\implies\quad
        \opnorm{\bcalF} \leq \kappa d
    \end{equation}
    except with probability at most $2N^{-\eps} + N^{-\frac{\kappa}{2+\kappa}}$.
\end{proposition}


\subsection{Kikuchi matchings}
\label{sec:Kikuchi_matchings}
Let us now focus our attention on the particular family of matchings that appears in the Kikuchi method. 
\begin{definition}
    Let $\ell, n, k \in \N^+$ with $k$ even and $n \geq \ell \geq k/2$.
    The associated collection of \emph{Kikuchi matchings} is
    \begin{equation}
        \Kik^{\ell, n, k} = \braces*{K^{\ell, n, k}_S : S \in \binom{[n]}{k}},
    \end{equation}
    where $K^{\ell,n,k}_S$ is the matching on vertex set $\binom{[n]}{\ell}$ in which $T$, $U$ are matched iff $T \symdiff U = S$.
    Note that $\Kik^{\ell,n,k}$ is $\delta_{\ell,n,k}$-balanced for 
    \begin{equation}    \label{eqn:del}
        \delta_{\ell,n,k} \coloneqq \tbinom{k}{k/2}\frac{\binom{n-k}{\ell-k/2}}{\binom{n}{\ell}} = (1-o(1))\tbinom{k}{k/2}\cdot (\ell/n)^{k/2},
    \end{equation}
    where we assume that $\ell = o(n/k)$.
    (Indeed, we typically think of $k$ as constant, $n \to \infty$, and $\ell$ as possibly large but certainly at most $n^{1-\Omega(1)}$.)
    \label{def:Kikuchi_matching}
\end{definition}
The particular form of Kikuchi matchings is motivated\footnote{For a more detailed discussion of the Kikuchi method, see \Cref{sec:kik_motiv}.} by the fact that for any assignment $x \in \{\pm 1\}^n$, and sets $T,U \in \binom{[n]}{\ell}$ with $T \Delta U = S$, we have $x^T x^U = x^S$. This relates the maximal advantage of a $k$XOR instance to the spectral properties of its associated Kikuchi graph. The adjacency matrix of the Kikuchi graph is also called \emph{Kikuchi matrix}. 
\begin{definition}  \label{def:corresp}
    There is a $1$-$1$ \emph{correspondence} between: a $k$XOR instance $\calI$ over~$[n]$ with constraints $\{(S_1,b_1), (S_2, b_2), \dots\}$; and, its associated (edge-signed) \emph{$\ell$-Kikuchi graph} $\calK_\ell(\calI)$, formed from the signed matchings $\{(K_{S_1}^{\ell,n,k}, b_1), (K_{S_2}^{\ell,n,k}, b_2), \dots\}$.
\end{definition}
\begin{notation}
    For $x \in \{\pm 1\}^n$ and $\ell \in \N^+$, we write $x^{\odot \ell} \in \{\pm 1\}^{\binom{[n]}{\ell}}$ for the assignment in which the $T$-coordinate is~$x^T$.
\end{notation}
\begin{fact}    \label{fact:xL}
    For any particular single Kikuchi matching $K = K^{\ell,n,k}_S$, we have $\adv_K(x^{\odot \ell}) = x^S \in \{\pm 1\}$ for any assignment $x \in \{\pm 1\}^n$.
    Thus for any $k$XOR instance $\calI$, we have $\adv_{\calI}(x) = \adv_{\mathcal{K}_\ell(\mathcal{I})}(x^{\odot \ell})$.
\end{fact}

We can now apply our results from \Cref{sec:random_matchings} to obtain bounds on the operator norm of the Kikuchi graph $\calK_\ell(\calI)$ for such random instances $\calI$. Via the correspondence in \Cref{def:corresp},  \Cref{prop:rand0} implies the following (using $\ln \binom{n}{\ell} \leq \ell \ln n$ and \Cref{eqn:del}).\footnote{The reciprocal of the $1-o(1)$ in \Cref{eqn:del} should enter into the definition of $C$ from \Cref{ineq:C}.  However, this can be covered up by the fact that for any fixed $\kappa > 0$, the expression $\exp(-\frac{\kappa^2}{2+\kappa})$ appearing in \Cref{prop:rand0} can be improved to the strictly smaller constant $\exp(\kappa)/(1+\kappa)^{1+\kappa}$.}\textsuperscript{,}\footnote{We don't claim any novelty in this theorem, as very similar theorems were proven by \cite{WAM19,Has20}. We state it merely to give an example of the constants that can be achieved if one follows \cite{WAM19}'s method carefully.} 
\begin{theorem}
\label{thm:alice}
    Let $\ell,n,k \in \N^+$ with $k$  even, $\ell \geq k/2$, and $n \gg k \ell$.  
    Let $\kappa \leq 1$ and $0 < \eps \leq \frac{\kappa}{2+\kappa}$, and assume that $m$ satisfies
    \begin{equation}    \label[ineq]{ineq:C}
        \Delta \coloneqq \frac{m}{n} \geq C_\kappa \cdot (n/\ell)^{(k-2)/2}, \quad \text{where } C_\kappa = \frac{2(1+\eps)(1+\kappa)}{\kappa^2} \binom{k}{k/2}^{-1} \cdot \ln n.
    \end{equation}
    Then for $\bcalI$ drawn  from either $\calR_{n,k}(m)$ or $\wt{\calR}_{n,k}(m)$, except with probability at most $3n^{-\eps \ell}$ we have
    \begin{equation} \label[ineq]{ineq:Cnext}
        \lambdamax{\calK_{\ell}(\bcalI)}\leq \kappa d, \quad \text{where } d = \delta_{\ell,n,k} \cdot m.
    \end{equation}
\end{theorem}
The above is an example of an ``Alice Theorem'' that may be used for a ``Kikuchi-style'' algorithm for solving Planted Noisy $k$XOR.
We give an illustrative example in the next section, after discussing planted random instances. The particular ``Alice Theorem'' we state here specifies some (potentially not optimal) value of $C_\kappa$, but note that even future hypothetical improvements of this ``Alice Theorem'' cannot achieve arbitrary small $C_\kappa$. Indeed, solving the Planted Noisy kXOR problem via the Kikuchi method requires the Kikuchi graph to have degree at least one, which implies $C_\kappa  \geq \frac{1}{\ell} \binom{k}{k/2}^{-1}$.  

\subsection{Planted noisy instances}\label{sec:noisy-planted}

In the previous section, we developed \emph{upper} bounds on the largest eigenvalue of the $\ell$-Kikuchi graph associated with a uniformly random $k$XOR instance. We now show \emph{lower} bounds in the case where the $k$XOR instance is planted. Both statements together imply that estimating the largest eigenvalue of the Kikuchi graph solves the Planted Noisy $k$XOR problem for a suitable choice of $\ell$. 

Recall from \Cref{not:P} the distribution $\calP^z_{n,k}(m,\rho)$ of $k$XOR instances with planted advantage $\rho$. 
We continue to use tildes to denote Poissonization of the parameter~$m$. Note that a poissonized planted noisy $k$XOR instance with planted advantage $\rho$ can be viewed as the union of a uniformly random $k$XOR instance and a planted $k$XOR instance with planted advantage 1. 
\begin{fact} \label{fact:barb}
    By Poisson splitting, $\bcalI \sim \wt{\calP}^z_{n,k}(m, \rho)$ is distributed as $\bcalI_0 \sqcup \bcalI_1$, where $\bcalI_0 \sim \wt{\calR}_{n,k}((1-\rho)m)$ and $\bcalI_1 \sim \wt{\calP}^z_{n,k}(\rho m)$ are independent.
\end{fact}
A planted noisy $k$XOR instance is a special case of an $n$-variate 
degree-$k$ polynomial. We introduce this more general notion because it also encompasses the Tensor PCA problem we study in \Cref{sec:tensorPCA}. 

\begin{definition}
    By an \emph{$n$-variate degree-$k$ polynomial} $A$, we will always mean a homogeneous multilinear polynomial of the form 
    \begin{equation}
        A(X) = A(X_1, \dots, X_n)  = \sum_{S \in \binom{[n]}{k}} a_S X^S,
    \end{equation}
    for real \emph{coefficients} $a_S$ and indeterminates $X_1, \dots, X_n$, where $X^S$ denotes $\prod_{i \in S} X_i$.
\end{definition}
\begin{notation}\label{not:kXOR} 
    Given a $k$XOR instance $\calI$ over $[n]$, and $S \in \binom{[n]}{k}$, we use the notation $B_\calI(S) = \sum_{(S,b) \in \calI} b$. 
    With this notation, we may say that a $k$XOR instance $\calI$ corresponds to an $n$-variate degree-$k$ polynomial $\sum_S B_{\calI}(S) X^S$. Moreover, the advantage $\adv_{\calI}(x)$ equals  $\adv_{\calI}(x)= \frac{1}{\abs{\calI}} \sum_S B_{\calI}(S) x^S$. 
\end{notation}    

\begin{definition}  \label{def:nono}
    Let $\calH$ denote a probability distribution on~$\R$ and let $z \in \R^n$.  
    We say that $\bcalA$ is an \emph{$\calH$-noisy $z$-planted} random degree-$k$ polynomial if its coefficients $\ba_S$ are i.i.d.\ with $\ba_S z^S$ distributed as~$\calH$.
\end{definition}


\begin{example} \label{eg:skell}
    When viewed as a degree-$k$ polynomial, $\bcalI \sim \wt{\calP}^z_{n,k}(m, \rho)$ is $\calH$-noisy $z$-planted for $\calH$ being the so-called \emph{Skellam} distribution $\mathrm{Skel}((\frac12 + \frac12\rho)q,(\frac12 - \frac12\rho)q)$, with $q = m/\binom{n}{k}$.
    (Recall that $\mathrm{Skel}(\mu_0,\mu_1)$ is the distribution of the difference of independent $\mathrm{Poi}(\mu_0)$ and $\mathrm{Poi}(\mu_1)$ random variables.  It has mean $\mu_0 - \mu_1$ and variance $\mu_0 + \mu_1$.) 
\end{example}
\begin{example}
\label{ex:tensorPCA}
    Fix $\beta \geq 0$.
    Then a (symmetric) Spiked Noisy $k$-Tensor~$\bcalA$, with Boolean spike $z \in \{\pm 1\}^n$ and signal-to-noise ratio~$\lambda$ (cf. \Cref{def:tensor_spiked}), is an $\calH$-noisy $z$-planted polynomial with $\calH$ being the translated standard Gaussian distribution $\calN(\lambda, 1)$.
\end{example}
The relationship between the maximal advantage of a $k$XOR instance $\bcalI$ and the operator norm of the corresponding Kikuchi graph (\Cref{fact:1}) in particular implies that the largest eigenvalue of $\calK_\ell(\bcalI)$ is lower bounded by the advantage of any assignment to $\bcalI$.
\begin{fact}    \label{fact:barb2}
    For $\bcalI \sim \wt{\calP}^z_{n,k}(m,\rho)$, let $\bA$ be the difference between the number of constraints in~$\bcalI$ satisfied and violated by~$z$, then $\bA \sim \mathrm{Skel}((\frac12 + \frac12 \rho) m, (\frac12 - \frac12 \rho) m)$.
    Thus (cf.\ \Cref{fact:xL}) when $\calK_\ell(\bcalI)$ is viewed as a $2$XOR instance, the difference between satisfied and unsatisfied constraints for assignment $z^{\odot \ell}$ is distributed $\tfrac12 \delta_{\ell,n,k} \cdot \binom{n}{\ell} \cdot \bA$.
    Finally, using \Cref{fact:1}, we may conclude that $\lambdamax{\calK_\ell(\bcalI)} \geq \braketOP{z^{\odot \ell}}{\calK_\ell(\bcalI)}{z^{\odot \ell}} \geq \delta_{\ell,n,k} \cdot \bA$.
\end{fact}
Bounding the tail of this random variable is a standard probability exercise (perhaps already well known):
\begin{proposition} \label{prop:fun}
    For $\bA \sim \mathrm{Skel}((\frac12 + \frac12 \rho) m, (\frac12 - \frac12 \rho) m)$  and $0 < \gamma <1$, we have 
    \begin{equation}
        \Pr[\bA \leq (1-\gamma) \rho m] \leq \exp(-\tfrac{\gamma^2 \rho^2}{2} m).
    \end{equation}
\end{proposition}
\begin{proof}
    Write $\mu_j = (\half + (-1)^j \half \rho)\cdot m$ for $j = 0,1$, and also $a=(1-\gamma) \rho m$.
    Then for any $t < 0$ the probability is upper-bounded by 
    \begin{equation}
        \E[\exp(t\bA)] \exp(-ta) = \exp(\mu_0(e^t - 1) + \mu_1(e^{-t}-1) - ta)
    \end{equation}    
    where we used the moment-generating function of Poisson random variables. 
    Selecting $t = \ln(1-\gamma \rho)$ yields the bound
    \begin{equation}
        \left[(1-\gamma\rho)^{-(1-\gamma)\rho}\exp\parens*{-\frac{\gamma \rho^2 (2 - \gamma(1+\rho))}{2(1-\gamma \rho)}}\right]^m \leq \exp(-\tfrac{\gamma^2 \rho^2}{2})^m
    \end{equation}
    as needed.
\end{proof}    
Combining \Cref{fact:barb2} and \Cref{prop:fun} we finally obtain the desired lower bound:
\begin{proposition} \label{prop:simple}
    For $\bcalI \sim \wt{\calP}^z_{n,k}(m, \rho)$ and any $0 < \gamma < 1$, except with probability at most $\exp(-\tfrac{\gamma^2 \rho^2}{2} m)$ we have
    \begin{equation}
        \lambdamax{\calK_\ell(\bcalI)} \geq \braketOP{z^{\odot \ell}}{\calK_\ell(\bcalI)}{z^{\odot \ell}} \geq (1-\gamma) \rho d, \quad \text{where } d = \delta_{\ell,n,k} \cdot m.
    \end{equation}
\end{proposition}

\textbf{Discussion.} 
With the simple \Cref{prop:simple} in hand, we illustrate how ``Alice'' can use \Cref{thm:alice} to obtain a Kikuchi-style algorithm for distinguishing between $\wt{\calR}_{n,4}(m)$ (i.e., purely random $4$XOR instances) and $\wt{\calP}^z_{n,4}(m,0.25)$  (i.e., random $4$XOR instances with planted advantage $\rho = 0.25$).
In \Cref{thm:alice} Alice might select $\kappa = 0.24$, $\eps = 0.1$, hence $C \approx 7.9$ in \Cref{thm:alice}, concluding: 
\begin{equation}
    \text{Provided } \Delta \geq (8/\ell) \cdot n \ln n, \quad \lambdamax{\calK_\ell(\bcalI)} \leq .24 \cdot \delta_{\ell,n,4} \cdot m \quad \text{except with probability at most } 3n^{-.1\ell}.
\end{equation}
On the other hand, taking $\gamma = .02$ in \Cref{prop:simple}, Alice knows that 
\begin{equation}
    \lambdamax{\calK_\ell(\bcalI)} \geq .245 \cdot \delta_{\ell,n,4} \cdot m \quad \text{except with probability at most } \exp(-m/80000) \ll n^{-.1\ell}.
\end{equation}
So if indeed $\Delta \sim (8/\ell) \cdot n \ln n$, then  with high probability (at least $1-4n^{-.1\ell}$), Alice can succeed by using the Power Method to distinguish $\lambdamax{\calK_\ell(\bcalI)} \leq .24 d$ vs.\ $\geq .245 d$, where $d \coloneqq \delta_{\ell,n,4} \cdot m = \Theta(\ell \log n)$ is the (expected) average degree of the Kikuchi graph.
In this regime, the Power Method takes time $\wt{O}(\binom{n}{\ell})$.
So we have, for example, the following:
\begin{quotation}
    By virtue of \Cref{thm:alice}, if $\Delta \sim .25 \cdot n \ln n$ then Alice can use the $\ell = 32$ Kikuchi Method to distinguish truly random $4$XOR instances from ones with planted advantage $\rho = 0.25$ in $\wt{O}(n^{32})$ time, $\wt{O}(n^{32})$ space, and failure probability $O(n^{-3.2})$.
\end{quotation}    

\section{Mean-field theory and planted inference}
\label{sec:kik_motiv}  \label{sec:motivation}
In this section, we describe an interpretation of the Kikuchi method in terms of statistical physics. The Kikuchi Method is a simple way of transforming a degree-$k$ optimization problem into a related degree-$2$ optimization problem, which can then be attacked via standard spectral methods. In recent years, it has become a powerful tool for addressing problems in a wide variety of areas (see, e.g., \cite{WAM19,guruswami2022algorithms,hsieh2023simple,alrabiah2023near,hsieh2024small}). 

This method was first introduced by Wein et al.~\cite{WAM19} and independently discovered by Hastings \cite{Has20}, although there are some differences, which we discuss in more detail in \Cref{sec:has_vs_kik}. Roughly speaking, the matrix constructed in \cite{WAM19} is a projection of the matrix in \cite{Has20} into the symmetric subspace, which reduces the size of the matrix and simplifies the analysis at the expense of losing rotational symmetry and locality. 
While the constructions in \cite{WAM19} and \cite{Has20} are closely related, their motivation is different. Wein et al.\ were motivated by free energy considerations,
and they obtain the Kikuchi matrix described here as a submatrix of the Hessian associated with the so-called Kikuchi free energy, a certain marginal approximation to the Gibbs free energy. On the other hand, Hastings's motivation stems from the theory of mean-field approximations to quantum many-body systems~\cite{Has20}. We now describe this second motivation in more detail and discuss a related motivation based on the monogamy of entanglement. 

\subsection{Degree-$k$ optimization.}
Consider the task of maximizing a homogeneous $n$-variate degree-$k$ polynomial \begin{equation}
  \Opt (\calP):= \max_{x} p(x) := \max_{x}   \sum_{S \in [n]^k} a_S  x^S
    \label{eq:poly_opt}
\end{equation} over assignments $x \in \{\pm1\}^n$ and with coefficients $a_S$ in some suitable field such as $\C$. This task captures many different combinatorial optimization problems simultaneously. For example, \cref{eq:poly_opt} corresponds to MAX-CUT if the coefficients $a_S$ are the (negated) edge weights of a graph or to $k$XOR if each $a_S$ represents the right-hand side of a scope $S$. By arranging the coefficients $a_S$ into a $k$-tensor $T$, the goal corresponds to finding its best ``rank-one approximation'' \begin{equation}
    \Opt(\calP) = \max_{x \in \{\pm1\}^n} \langle T, x^{\otimes k} \rangle := \max_{x \in \{\pm1\}^n}   \sum_{S \in [n]^k} a_S  x^S.
\end{equation}
\subsection{Degree-2 relaxation.}
In order to solve this problem, it is fruitful to relax it to a degree-2 optimization problem. The most straightforward approach (for even $k$) is to flatten the coefficient tensor $T$ 
to a $n^{k/2} \times n^{k/2}$ matrix $H$ and compute\begin{equation}
    \max_{v \in \C^{n^{k/2}}, ||v|| = 1} \bra{v} H \ket{v},
\end{equation}  which is efficient using linear algebraic methods. 
While this relaxation always upper bounds $\Opt(\calP)$, the bound is typically loose because the spectral norm of $H$ is maximized over arbitrary unit-norm states, whereas only product states of the form \begin{equation}
    v = x^{\otimes k/2}
\end{equation} for Boolean $x \in \{\pm1\}^n$ correspond to valid values of $p(x)$. The Kikuchi Method remedies this by replacing $H$ with a larger Kikuchi matrix $\Kik_\ell$ defined on $\C^{\binom{n}{\ell}}$, for which the leading eigenvector approaches the product form $v = z^{\odot \ell}$ as $\ell$ increases. From the perspective of statistical physics, this phenomenon is described by mean-field theory. 
\subsection{Mean-field theory.} 
The key idea is to recognize $H$ as the Hamiltonian of a $k/2$-particle system, and to increase the number of particles to $\ell \gg k/2$ until the system is accurately modeled by a mean-field theory. 
To illustrate this, let us fix $k = 4$. 

In this case, the flattened version $H$ of the coefficient tensor $T$ maps states on $\C^{n} \otimes \C^{n}$ to $\C^{n} \otimes \C^n$, that is, it describes the interaction of two particles where each particle has local dimension $n$. In this language, our degree-4 optimization problem corresponds to the maximal energy of the interaction Hamiltonian $H$ over \emph{product states},  \begin{equation}
    \max_{x \in \{\pm1\}^{n}} \bra{x^{\otimes 2}} H \ket{x^{\otimes 2}}.
    \label{eq:tens_opt}
\end{equation} 
However, the leading eigenvector of $H$ is typically an entangled state, hence the spectral norm of $H$ is not close to our optimum. This can be remedied by increasing the number of particles from $2$ to $\ell$ while maintaining the same interaction $H$ between any pair of two particles. The corresponding Hamiltonian is (up to symmetrization, see \Cref{sec:has_vs_kik}) the adjacency matrix of the Kikuchi graph  \begin{equation}
 \label{eq:app_kik_def}
     \Kik_{\ell;\{S,V\} }= \begin{cases}
a_{S\Delta V} \quad \text{   if } |S\Delta V|= k, \\
0 \quad \text{ else }
\end{cases}
\end{equation} defined on the $\binom{n}{\ell}$-dimensional space indexed by subsets $S,V \subset [n]$ of length $\ell$, and where $S\Delta V = S\cup V - S\cap V$ denotes the symmetric difference of two $\ell$-sets. It is folklore in condensed matter physics that if the number of particles $\ell$ is large, the system is well approximated by a mean-field theory. That is, the movement of each particle is well approximated by the action of a global mean-field, or in other words, the ground state of this problem (in our context, the top eigenvector of $\Kik_\ell$) is well approximated by a product state. Hence the spectral norm $||\Kik_\ell||$ is a good upper bound for the maximum of $\bra{v^{\odot \ell}} \Kik_\ell \ket{v^{\odot \ell}}$ over product states of the form $\ket{v^{\odot \ell}}$ for unit-norm vectors $v \in \C^n$. Unlike the original degree-$k$ optimization problem in \cref{eq:poly_opt}, this spectral norm can be computed using simple linear algebraic techniques in time that is polynomial in the dimension of the Hamiltonian (which, however, is now $\binom{n}{\ell} \approx n^{\ell}$, instead of the original input size $n^k$). 

In order to correspond to a valid value of the original degree-4 polynomial $p(x)$, $v$ should moreover be close to a Boolean vector $x \in \{\pm1\}^n$. However, unit-norm vectors can be rounded to Boolean vectors by essentially following the Goemans--Williamson procedure, without degrading the approximation ratio by too much. 

For rotationally symmetric systems, the above mean-field intuition can be made rigorous via the quantum de Finetti theorem \cite{christandl2007one}, which in general requires large $\ell$. Hastings \cite{Has20} established that a relatively small number $\ell \sim \log(n)$ of particles already suffices if the coefficient tensor $T$ carries the specific structure of the Planted Noisy Tensor PCA problem studied in \Cref{sec:tensorPCA}.
\subsection{Monogamy of entanglement.} 
Another way to understand the above relaxation is through monogamy of entanglement, which is closely related to mean-field theory. Roughly speaking, monogamy of entanglement states that entanglement cannot be strongly shared between many parties simultaneously: If one particle is maximally entangled with another particle, it cannot have any entanglement with a third system. More generally, for $\ell$ particles, the Osborne and Verstraete inequality puts a limit on the sum of pairwise entanglement between the first particle and the rest: 
\begin{equation}
\tau\left(\rho_{1 2}\right)+\tau\left(\rho_{1 3}\right)+\cdots+\tau\left(\rho_{1 \ell}\right) \leq \tau\left(\rho_{1 ; \left(2 3 \ldots \ell\right)}\right).
\end{equation}
Here, $\tau(\rho_{AB})$ is the ``tangle'', a measure of the bipartite entanglement between subsystem $A$ and $B$. The Hamiltonian $\Kik_\ell$ is by construction invariant under permutations of the $\ell$ particles, and any entanglement is forced to spread out equally between all possible pairs of particles. The entanglement between any pair of particles thus decays inversely with the number of particles, i.e., $\tau\left(\rho_{i j}\right) \sim  1/\ell$.
As $\ell$ increases, the top eigenvector approaches an unentangled (= product) state, which together with the permutation invariance implies that the leading eigenvector of $\Kik_\ell$ is approximately of the form $v^{\odot \ell}$ for unit-norm $v \in \C^n$.

\subsection{Comparison with Hastings' construction}
\label{sec:has_vs_kik}

In the previous section, we have motivated the Kikuchi matrix as describing a system of $\ell$ particles with interactions between any $k/2$-subsets of particles governed by the entries of a $k$-Tensor $T$. The natural Hamiltonian associated with this system is 
\begin{equation}
H_\ell(T)=\frac{1}{2} \sum_{i_1, \ldots, i_{k/2}}\left(\sum_{\mu_1, \ldots, \mu_k} T_{\mu_1, \mu_2, \ldots, \mu_k}\left|\mu_1\right\rangle_{i_1}\left\langle\mu_{1+k / 2}|\otimes| \mu_2\right\rangle_{i_2}\left\langle\mu_{2+k / 2}|\otimes \ldots \otimes| \mu_{k / 2}\right\rangle_{i_{k / 2}}\left\langle\mu_k\right|+\text {h.c.}\right),
\label{eq:app_hastings_def}
\end{equation} 
where the first sum is over distinct $1 \leq i_1, i_2, \dots ,i_{k/2} \leq \ell$ which pick out $k/2$ of the $\ell$ particles, and the second sum is over (not necessarily distinct) indices $ 1 \leq \mu_1, \ldots, \mu_k \leq n$ of the $k$-tensor $T$, which index the possible states of the particles. Here ``+h.c.'' means ``plus the Hermitian conjugate'' which means we add the Hermitian conjugates of all the terms in the sum, making the 
right-hand side a Hermitian matrix.

In the simplest case, when $\ell=k/2$, the Hamiltonian $H_\ell(T)$ is just a flattened version of the tensor $T$, i.e., a matrix of dimension $n^{k/2} \times n^{k/2}$. As another example, consider the case of $k=4$, and let $H$ be the matrix $H_2(T)$, which is of size $n^2 \times n^2$. $H$ is the Hamiltonian of a system of 2 particles, each of which is a qudit of dimension $n$. Then $H_3(T)$ can be viewed as the Hamiltonian of 3 particles, where each pair of paricles has the Hamiltonian $H$. In other words $H_3(T)=H_{12}\otimes\Id_3 + H_{13}\otimes\Id_2 + H_{23}\otimes\Id_1$, where $H_{ij}$ means the Hamiltonian $H$ acts on particles $i$ and $j$.

This Hamiltonian is the matrix studied by Hastings in his quantum algorithm for Tensor PCA. 
It is defined on the whole space 
$\left(\mathbb{C}^{n}\right)^{\otimes \ell}$ of $\ell$ qudits with internal dimension $n$. Comparing \cref{eq:app_hastings_def} with \cref{eq:app_kik_def} shows that the Kikuchi matrix $\Kik_\ell(T)$ arises by projecting $H_\ell(T)$ first into the ``homogeneous'' subspace where all the $\mu_1, \dots, \mu_k$ are distinct, and additionally working in the symmetric subspace where the order of the indices $i_1, \dots, i_{k/2}$ is irrelevant (i.e., the tuple $(i_1, \dots, i_{k/2})$ is identified with the set $\{i_1, \dots, i_{k/2}\}$). 

These projections come with certain advantages and disadvantages. First, note that $H_\ell(T)$ is $\frac{k}{2}$-local, whereas $\Kik_\ell(T)$ is sparse but not local. The dimension of $\Kik_\ell(T)$ is smaller than the dimension of $H_\ell(T)$, and working in the symmetric and homogeneous subspace simplifies the analysis of the Kikuchi method. However, the specific form of the Kikuchi matrix is particularly well-suited for the analysis of Boolean spikes, and it is somewhat tedious to prove detection and recovery for spike priors that are not Boolean (see \Cref{sec:other_priors}). On the other hand, Hastings's construction $H_\ell(T)$ is rotationally invariant, which makes it a natural choice for Gaussian spikes. The rotational invariance also seems to make it easier to prove a simpler and improved randomized guarantee of recovery (at least for Tensor PCA, see \cite{Has20}).

\section{Correctness of the quantum algorithm for Planted Noisy $k$XOR}
\label{sec:proof2}
We now prove correctness of our quantum algorithm for the Planted Noisy $k$XOR problem. The main technical challenge is to establish that our guiding state has improved overlap with the cutoff eigenspace of the Kikuchi Hamiltonian, as informally presented in \Cref{thm:overlap_informal}.
\subsection{Cutoff eigenspace}
Before describing our guiding state, we record here a slight refinement of \Cref{prop:simple}'s statement that $\lambdamax{\calK_\ell(\bcalI)}$ is large by virtue of the planted solution~$x$:
\begin{proposition} \label{prop:simple2}
    In the setting of \Cref{prop:simple}, let $\bPi_{\geq}$ denote the projector onto the eigenspaces of $\calK_{\ell}(\bcalI)$ of eigenvalue at least $(1-\gamma) \rho d$.
    Then for any $0 < \hat{\gamma} < \gamma-\eps$, 
    we have
    \begin{equation}
        \braketOP{z^{\odot \ell}}{\bPi_{\geq}}{z^{\odot \ell}} \geq \frac{\rho \eps}{49\ell \ln n}.
    \end{equation}
    except with probability at most
    $
        \mathrm{\textsc{Fail}}_1 \coloneqq \exp(-\frac{\hat{\gamma}^2 \rho^2}{2} m) + n^{-\ell (46d-1)}. $
    Moreover, $\mathrm{\textsc{Fail}}_1  \leq \exp(-\frac{\hat{\gamma}^2 \rho^2}{2} m) + n^{-45\ell}$ whenever a  bound of the form \Cref{ineq:Cnext} is in effect. 
\end{proposition}
\begin{proof}
    We have the loose upper bound $\lambdamax{\calK_\ell(\bcalI)} \leq D(\calK_\ell(\bcalI)) \leq (1+48\ell \ln n) d$ except with probability at most $\binom{n}{\ell} \cdot \exp(-46.08 \cdot d \cdot \ell \ln n) \leq n^{-\ell (46d-1)}$, which uses \Cref{ineq:dd} and \Cref{prop:md} (with $\kappa = 48 \ell \ln n$).
    The first result now follows from \Cref{prop:simple} by a Markov's inequality argument on the eigenvalues of $\calK_\ell(\bcalI)$. The second statement follows from the fact that any inequality of the form \Cref{ineq:Cnext} requires the average degree $d$ of the Kikuchi graph to be 1 at the very least (which implies $C_\kappa \binom{k}{k/2} \geq \frac{1}{\ell}$), since the largest eigenvalue of $\Kik_\ell(\calI)$ is at least the largest eigenvalue of any of its submatrices. 
\end{proof}

The second part of \Cref{prop:simple2} guarantees that the quantity $ \braketOP{z^{\odot \ell}}{\bPi_{\geq}}{z^{\odot \ell}}$ is ``large enough'' even if in the future someone manages to establish a tighter bound on $C_\kappa$ in \Cref{thm:alice}, for example by removing the $\ln n$ term in \Cref{ineq:C} (which, just as \cite{WAM19}, we expect is an artifact of the matrix Chernoff bound). 

\subsection{Guiding states} \label{sec:guiding_state_math}

The central ingredient of our super-quadratic quantum speedup is a guiding state that has improved overlap with the leading eigenspace of the Kikuchi graph. 
As shown in \Cref{prop:simple2}, the planted solution $z$ is a good certificate for the non-emptiness of this space in the planted case. 
Of course, we do not know $z$, but we do know that it satisfies roughly a $\left(\frac{1}{2} + \frac{1}{2}\rho\right)$ fraction of the $k$XOR equations. 
This motivates a guiding state whose entries are given by the ``right-hand sides'' of these equations. 

For some intuition, consider a $z$-planted $k$XOR instance $\calI$ with $m$ constraints $\calC_i = (S_i,b_i)$. The central idea is to prepare $\ell/k$ copies of the (approximately normalized) state 
\begin{equation}
   \ket{\gamma(\calI)} = \frac{1}{\sqrt{m}} \sum_i b_i \ket{S_i} \in \C^{n^k}, 
\end{equation} 
since the resulting state has improved overlap with the planted assignment $\ket{z^{\otimes \ell}} \in \C^{n^\ell}$: 
\begin{equation}
    | \braket{z^{\otimes \ell}}{\gamma(\calI)^{\otimes \ell/k}}|^2 = \left( \braket{z^{\otimes k}}{\gamma(\calI)} \right)^{2\ell/k} = \left( \adv_{\calI}(z)^2 \frac{m}{n^k}\right)^{\ell/k}.
\end{equation} 
If $m \approx n^{k/2}$ and $\adv_{\calI}(z)$ is constant, which is the natural setting of the Planted Noisy $k$XOR problem, this overlap is $| \braket{z^{\otimes \ell}}{\gamma(\calI)^{\otimes \ell/k}}|^2 \approx n^{-\ell/2}$, a quadratic improvement over the expected overlap with a random vector. 
Our actual guiding state (approximately) corresponds to a symmetrized and normalized version of $\ket{\gamma(\calI)}^{\otimes \ell/k}$. We now describe our construction more formally.

Let us start by introducing notation on how to partition a set of size $\ell$ into subsets of size $k$ (which will correspond to the ``left-hand sides'' of the $k$XOR instance):
\begin{notation}
\label{not:partition}
    Fix $k$ and $\ell = c k$ for $c \in \N^+$.
    Then for $T \in \binom{[n]}{\ell}$, we write $\Part_k(T)$ for the collection of all $\{S_1, \dots, S_c\} \subset \binom{[n]}{k}$ with $S_1 \sqcup \cdots \sqcup S_c = T$.  
    We write $\Part_k(\ell) = \abs{\Part_k(T)} = \frac{1}{c!}\binom{\ell}{k, \dots, k}$.
\end{notation}
\begin{notation}
    In the preceding setting, if $H \in \R^{\binom{[n]}{k}}$, we write $H^{\circledast c} \in \R^{\binom{[n]}{\ell}}$ for the vector whose $T$-coordinate is
    \begin{equation}    \label{eqn:Hs}
        H^{\circledast c}_{T} = \sum_{\{S_1, \dots, S_c\} \in \Part_k(T)} \prod_{j=1}^c H_{S_j}.
    \end{equation}
\end{notation}
We now define our guiding state in the general setting of a degree-$k$ polynomial $\calA$.
\begin{definition}
\label{def:guiding_state}
    Let $\calA = \sum_{S \in \binom{[n]}{k}} a_S X^S$ be a degree-$k$ polynomial and let $\ell = ck$ for $c \in \N^+$.
    We define the associated \emph{guiding vector} to be
    \begin{equation}
    \ket{\Gamma^\ell(\calA)} 
        = \frac{1}{\chi} \sum_{T \in \binom{[n]}{\ell}} \sum_{\{S_1, \dots, S_c\} \in \Part_k(T)}  \parens*{\prod_{j=1}^c a_{S_j}} \ket{T}, \qquad \chi \coloneqq \chi_{n,k,\ell} = \sqrt{\tbinom{n}{\ell}} \cdot \sqrt{\Part_k(\ell)}.
    \end{equation}
    We remark that $\ket{\Gamma^\ell(\calA)}$ need not be a unit vector (but will approximately be so in our setting; see \Cref{lem:mc}).
\end{definition}
This state corresponds to a symmetrized version of the (not necessarily normalized) state $\ket{\gamma(\calA)}^{\otimes c}$ for $ \ket{\gamma(\calA)} = \binom{n}{k}^{-1/2} \sum_{S \in \binom{[n]}{k}} a_S \ket{S}.$ In particular, if $\ell = k$, no symmetrization is necessary and $\ket{\Gamma^k(\calA)} = \ket{\gamma(\calA)}$. 

For the remainder of this subsection, we take the following standing normalization assumption:
\begin{equation}\label{eq:assumption}
\begin{aligned}
    &\bcalA \text{ is an $\calH$-random $z$-planted degree-$k$ $n$-variate polynomial (\Cref{def:nono}),}\\ 
    &\text{where $\calH$ has mean $\mu \geq 0$ and variance $1$.}
\end{aligned}
\end{equation}

\begin{example} \label{eg:skelly}
    For a $k$XOR instance $\bcalI \sim \wt{\calP}^z_{n,k}(m,\rho)$, the polynomial $\bcalA= \frac{1}{\sqrt{q}}\sum_S B_{\bcalI}(S) X^S$ is $\calH$-random $z$-planted for $\calH = \frac{1}{\sqrt{q}}\mathrm{Skel}((\frac12 + \frac12\rho)q,(\frac12 - \frac12\rho)q)$ and $q = \frac{m}{\binom{n}{k}}$. $\bcalA$ has mean $\mu = \rho \sqrt{q}$ and variance~$1$. 
\end{example}
Thus 
\begin{equation}     
    \ket{\Gamma^\ell(\bcalA)} = \frac{1}{\chi} \sum_{T \in \binom{[n]}{\ell}} z^T \bH_T^{\circledast c} \ket{T}; \quad \text{hence,} \quad 
    \E\bracks*{\ket{\Gamma^\ell(\bcalA)}}  
    = \sqrt{\Part_k(\ell)} \cdot \mu^c  \ket{z^{\odot \ell}}. \label{eqn:cha}
\end{equation}
Thus in expectation we have $\braket{z^{\odot \ell}}{\Gamma^\ell(\bcalA)} \approx \mu^c$ (ignoring $\sqrt{\Part_k(\ell)}$, which is a constant if~$\ell$ is). In the $k$XOR setting, the $\mu$ specified by the natural value of $m \approx n^{k/2}$ corresponds to an overlap of $\braket{z^{\odot \ell}}{\Gamma^\ell(\bcalA)} \approx n^{-\ell/4}$, a quadratic improvement over a random vector. 
But rather than the overlap between $\ket{\Gamma^\ell(\bcalA)}$ and $\ket{z^{\odot \ell}}$, we will actually need to analyze a slightly subtler quantity in our application: how much the guiding vector overlaps with the Kikuchi graph's top eigenspaces.
However, we will know from \Cref{prop:simple2} that there will (likely) be a vector of the form $\ket{v} = \Pi_{\geq} \ket{z^{\odot \ell}}$ in these eigenspaces that is well aligned with $\ket{z^{\odot \ell}}$, and hence hopefully also with~$\ket{\Gamma^\ell(\bcalA)}$.

The following theorem quantifies the above intuition using the second moment method.
\begin{theorem} \label{thm:guide}
    Let $k$ be even, let $\ell = ck$ for $c \in \N^+$, let $n \geq k\ell$, and let $\ket{v} \in \R^{\binom{[n]}{\ell}}$ be a unit vector.
    Then 
    \begin{equation}
        \braket{v}{\Gamma^\ell(\bcalA)} \geq \tfrac12   \sqrt{\Part_k(\ell)} \cdot \mu^c \cdot \braket{v}{z^{\odot \ell}}
    \end{equation}
    except with probability at most
    \begin{equation}
        \frac{O(\ell/k)^2}{\binom{n}{k} \cdot \mu^2 \cdot \braket{v}{z^{\odot \ell}}^2}.
    \end{equation}
\end{theorem}
\begin{proof}
    From \Cref{eqn:cha} we have
    \begin{equation}
        \E\bracks*{\braket{v}{\Gamma^\ell(\bcalA)}} =   \sqrt{\Part_k(\ell)} \cdot \mu^c \cdot \braket{v}{z^{\odot \ell}},
    \end{equation}
    and thus by Chebyshev's inequality it is enough to establish 
    \begin{equation}    \label[ineq]{ineq:var}
        \Var\bracks*{\braket{v}{\Gamma^\ell(\bcalA)}} \leq 2.04(\ell/k)^2 \cdot \Part_k(\ell) \cdot\frac{\tbinom{n}{\ell - k}}{\tbinom{\ell}{k}\tbinom{n}{\ell}} \cdot \mu^{2c-2},
    \end{equation}
    as the above can be bounded by $\Part_k(\ell) \cdot\frac{O(\ell/k)^2}{\tbinom{n}{k}} \cdot \mu^{2c-2}$ using $n \geq k\ell$.
    
    To establish \Cref{ineq:var}, we write $\ket{v} = \sum_{T} w_T z^T \ket{T}$ for some numbers $w_T$ satisfying $\sum_T w_T^2 = \braket{v}{v} = 1$.
    Thus $\braket{v}{\Gamma^\ell(\bcalA)} = \frac{1}{\chi} \sum_T w_T  \bH_T^{\circledast c}$ 
    by \Cref{eqn:cha}, where we use that $z$ is Boolean, and so Cauchy--Schwarz implies
    \begin{equation}
        \Var\bracks*{\braket{v}{\Gamma^\ell(\bcalA)}} = \frac{1}{\chi^2} \sum_{T,U \in \binom{[n]}{\ell}} w_T w_U \Cov[\bH_T^{\circledast c},\bH_U^{\circledast c}] 
        \leq \frac{1}{\chi^2} \sum_T w_T^2 \sum_U \abs{\Cov[\bH_T^{\circledast c},\bH_U^{\circledast c}]}.
    \end{equation}
    Thus the first inequality in \Cref{ineq:var} follows from \Cref{lem:mb} below (whose second hypothesis may be assumed satisfied, as otherwise the probability bound in this theorem exceeds~$1$).
\end{proof}
\begin{lemma}   \label{lem:ma}
    Assume $\mu \leq .1/\sqrt{c}$ and let $(\bH_S)_{S}$ be i.i.d.~according to~$\calH$.
    If $S_1, \dots, S_a, S'_1, \dots, S'_b$, $S''_1, \dots, S''_b \in \binom{[n]}{k}$ are distinct, with $a+b = c$, then 
    \begin{equation}    \label{eqn:0}
        \Cov[\bH_{S_1} \cdots \bH_{S_a} \bH_{S'_1} \cdots \bH_{S'_b}, \bH_{S_1} \cdots \bH_{S_a} \bH_{S''_1} \cdots \bH_{S''_b}] = 0 \quad \text{if $a = 0$;}
    \end{equation}
    and in general the covariance lies between $0$ and
    \begin{equation}    \label[ineq]{ineq:E}
        \E[\bH_{S_1} \cdots \bH_{S_a} \bH_{S'_1} \cdots \bH_{S'_b} \cdot \bH_{S_1} \cdots \bH_{S_a} \bH_{S''_1} \cdots \bH_{S''_b}]
        \leq \exp(.01)\mu^{2c-2a}.
    \end{equation}
\end{lemma}
\begin{proof}
    \Cref{eqn:0} holds because the two random variables are independent.
    Otherwise, the covariance is
    \begin{align}
        {} &\phantom{=} \Var[\bH_{S_1} \cdots \bH_{S_a}] \E[\bH_{S'_1}] \cdots \E[\bH_{S'_b}] \E[\bH_{S''_1}] \cdots \E[\bH_{S''_b}] && \\
        &= ((1+\mu^2)^a- \mu^{2a}) \mu^{2b} 
        = \parens*{\mu^{-2a}(1+\mu^2)^a-1}\mu^{2c} && \text{(which is nonnegative, and $0$ if $a = 0$)} \\
        & \leq (1+\mu^2)^a\mu^{2c-2a}  && \text{(dropping the $-1$)}\\
        & \leq \exp(.01)\mu^{2c-2a} ,
   \end{align}
   where the last inequality used $\mu^2 \leq .01/c \leq .01/a$. 
   One can also see that when the $-1$ was dropped, the resulting quantity in fact equals the expectation in \Cref{ineq:E}.
\end{proof}
\begin{lemma}   \label{lem:mb}
    In the setting of \Cref{lem:ma}, assume that $k\ell \leq n$ and $\mu \geq \ell/\sqrt{\binom{n}{k}}$.
    Then for any $T \in \binom{[n]}{\ell}$, 
    \begin{equation}
        \sum_{U \in \binom{[n]}{\ell}} \abs{\Cov[\bH^{\circledast c}_T, \bH^{\circledast c}_U]} \leq 2.04(\ell/k)^2 \cdot \Part_k(\ell)^2  \cdot\frac{\tbinom{n}{\ell - k}}{\tbinom{\ell}{k}} \cdot \mu^{2c-2}.
    \end{equation}
\end{lemma}
\begin{proof}
    Using \Cref{eqn:Hs}, we have
    \begin{equation}    \label[ineq]{ineq:upme}
        \sum_{U \in \binom{[n]}{\ell}} \abs{\Cov[\bH^{\circledast c}_T, \bH^{\circledast c}_U]} 
        \leq 
        \sum_{\calS = \{S_1, \dots, S_c\} \in \Part_k(T)} 
        \sum_{\substack{\calS' = \{S'_1, \dots, S'_c\} \\
                                  S'_j\text{'s pairwise disjoint}}}
            \abs{\Cov[\bH_{S_1} \cdots \bH_{S_c}, \bH_{S'_1} \cdots \bH_{S'_c}]}.
     \end{equation}
    Fix any $\calS$ in the outer summation.
    We then stratify the inner summation according to $a \coloneqq \abs{\calS \cap \calS'}$.
    By \Cref{lem:ma}, we can ignore the $a = 0$ case;
    and, for a given $1 \leq a \leq c$, the number of possibilities for $\calS'$ (given $\calS$) is at most
    \begin{equation}
        f(a) \coloneqq \tbinom{c}{a} \Part_k(\ell - ak) \tbinom{n}{\ell - ak}.
    \end{equation}
    Thus  we can upper-bound \Cref{ineq:upme} by
    \begin{equation} \label{eqn:blah}
        1.02 \cdot \Part_k(\ell) \cdot \mu^{2c} \cdot  \sum_{a=1}^c g(a), \qquad g(a) \coloneqq f(a) \mu^{-2a}
    \end{equation}
    (where we used $\exp(.01) \leq 1.02$).
    Note that
    \begin{equation}    
        \frac{g(a+1)}{g(a)} = \mu^{-2} \cdot \frac{k! (c-a)^2/(a+1)}{(n+1 - k(c-a)) \cdots (n+k - k(c-a))} \leq 1/2,
    \end{equation}
    where the last inequality used $k\ell \leq n$ and $\mu \geq \ell/\sqrt{\binom{n}{k}}$.
    Thus the sum in \Cref{eqn:blah} is upper-bounded by 
    \begin{equation}
        2g(1) = (2\ell/k)\cdot \Part_k(\ell-k)  \cdot \binom{n}{\ell - k} \cdot \mu^{-2} = 2(\ell/k)^2\cdot \Part_k(\ell)  \cdot \frac{\tbinom{n}{\ell - k}}{\tbinom{\ell}{k}} \cdot \mu^{-2},
    \end{equation}
    completing the proof.
    %
\end{proof}

Finally, we show that the vector $\ket{\Gamma^\ell(\bcalA)}$ is approximately normalized. It is very likely that a stronger bound can be proved, but this bound suffices for our purposes.

\begin{lemma}   \label{lem:mc}
    In the setting of \Cref{lem:mb} and \Cref{eqn:cha}, assuming we also have $\mu \leq .1/\ell^{k/2}$,
    \begin{equation}
        \E\bracks*{\braket{\Gamma^\ell(\bcalA)}{\Gamma^\ell(\bcalA)}} \leq 1.0202.
    \end{equation}
\end{lemma}
\begin{proof}
    Similar to \Cref{ineq:upme}, the expectation equals
    \begin{align}
        \frac{1}{\chi^2} \sum_{T} \E\bracks*{(\bH_T^{\circledast c})^2} &= 
        \avg_{T} \avg_{\calS \in \Part_k(T)} \sum_{\calS' \in \Part_k(T)} \E[\bH_{S_1} \cdots \bH_{S_c} \cdot \bH_{S'_1} \cdots \bH_{S'_c}] \\
        &= \sum_{\calS' \in \Part_k(T_0)} \E[\bH_{S_1} \cdots \bH_{S_c} \cdot \bH_{S'_1} \cdots \bH_{S'_c}],
    \end{align}
    the last equality holding, by symmetry, for any fixed $T_0$ and $\calS \in \Part_k(T_0)$. 
    We again stratify the choices of $\calS'$ according to $a \coloneqq \abs{S \cap S'}$, the number of possibilities being at most $\binom{c}{a} \Part_k(\ell - ak)$.
    Thus using \Cref{lem:ma} again, the above quantity is at most
    \begin{equation}    \label{eqn:yoyo}
        \exp(.01) \cdot \mu^{2c} \cdot \sum_{a=0}^c h(a), \qquad h(a) \coloneqq \binom{c}{a} \Part_k(\ell - ak) \mu^{-2a}.
    \end{equation}
    Now (writing $b = c - a$)
    \begin{equation}
        \frac{h(a)}{h(a-1)} = \mu^{-2} \cdot \frac{(k-1)!(b+1)}{a(kb-(k-1)) \cdots (kb-1)},
    \end{equation}
    which is minimized at $a = c$, where it has minimum value 
    \begin{equation}
        \mu^{-2} \cdot \frac{(k-1)! c}{(\ell - 1) \cdots (\ell- (k-1))} = \mu^{-2} \cdot \frac{\ell}{k} \cdot \frac{1}{\tbinom{\ell-1}{k-1}}.
    \end{equation}
    As $\mu \leq .1/\ell^{k/2}$, the above is certainly at least~$200$, meaning that the sum in \Cref{eqn:yoyo} is at most $1.01h(c) = 1.01\mu^{-2c}$.
    The lemma follows (using $1.01\exp(.01) \leq 1.0202$).
\end{proof}

\subsection{Analysis for the quantum case}  \label{sec:quantum-analysis}

We now combine \Cref{prop:simple2}, \Cref{thm:guide}, and \Cref{lem:mc} to establish a lower bound on the overlap between our guiding state and the leading eigenspace of the Kikuchi graph of a $k$XOR instance.
\begin{theorem} \label{thm:kXOR_analysis}\label{thm:bob}
    Let $k$ be even, let $\ell = c k$ for $c \in \N^+$, let $n \geq k\ell$, and let $\hat{\bcalI} \sim \wt{\calP}^z_{n,k}(\hat{m},\rho)$.
    Write $m = (1-\zeta) \hat{m}$ (for $0 < \zeta < 1$), and introduce the notation $d, \hat{\gamma}, \gamma, \eps$ from \Cref{prop:simple2}.
    Finally, assume
    $
        \ell \leq (\tfrac{.01}{\zeta \rho^2})^{1/k} \cdot (\binom{n}{k}/m)^{1/k}$.
    
    Suppose that $\hat{\bcalI}$ is randomly partitioned into $\bcalI \sqcup \bcalI_{\textnormal{guide}}$ by independently placing each constraint into $\bcalI$ with probability $1-\zeta$ and into $\bcalI_{\textnormal{guide}}$ with probability $\zeta$.
    Let $\ket{\bGamma}$ be the unit vector (``guiding state'') in the direction of $\ket{\Gamma^\ell(\bcalI_{\textnormal{guide}})}$.
    Then (for any $0 < \nu \leq .99$), except with probability at most 
    \begin{equation}
        \mathrm{\textsc{Fail}}_1 + \frac{O(\ell/k)^2 \ell \ln n}{\zeta \eps \rho^3} \cdot \hat{m}^{-1} + \nu,
    \end{equation}
    there is a unit vector $\ket{\bv}$ in the span of $\calK_\ell(\bcalI)$'s eigenspaces of eigenvalue at least $(1-\gamma) \rho d$ such that 
    \begin{equation}
        \braket{\bv}{\bGamma}^2 \geq \xi \cdot (\hat{m}/\tbinom{n}{k})^{\ell/k}, \qquad \text{where } \xi = \Part_k(\ell) \cdot \frac{\rho\eps \nu}{200 \ell \ln n} \cdot (\rho^2 \zeta)^{\ell/k}. 
    \end{equation}
\end{theorem}
\begin{proof}
    By Poisson splitting, we have $\bcalI \sim \wt{\calP}^z_{n,k}(m,\rho)$ and $\bcalI_{\textnormal{guide}} \sim \wt{\calP}^z_{n,k}(\zeta \hat{m},\rho)$ and they are independent.  
    Using \Cref{prop:simple2} (including its notation), and letting $\ket{\bv}$ be the unit vector in the direction of $\bPi_{\geq} \ket{z^{\odot \ell}}$, we conclude that except with probability at most~$\mathrm{\textsc{Fail}}_1$,
    \begin{equation}
        \braket{\bv}{z^{\odot \ell}}^2 \geq \frac{\rho\eps}{49 \ell \ln n}.
    \end{equation}
    Fix any  outcome of $\bcalI$ for which this indeed occurs.  
    Then $\bcalI_{\textnormal{guide}}$ is still distributed as $\wt{\calP}^z_{n,k}(\zeta \hat{m},\rho)$ and thus (per \Cref{eg:skell}) if we view it as a degree-$k$ polynomial, it is $\calH$-noisy $z$-planted for $\calH$ being $\mathrm{Skel}((\half + \half \rho) q, (\half -\half \rho) q)$, where $q = \zeta \hat{m}/\binom{n}{k}$.
    Note that $\calH$ has mean $\rho q$ and variance~$q$.
    Thus if we write $\bcalA = q^{-1/2} \cdot \bcalI_{\textnormal{guide}}$, then $\bcalA$ is an $\calH'$-noisy $z$-planted degree-$k$ polynomial for an $\calH'$ with mean $\mu \coloneqq \rho \sqrt{q}$ and variance~$1$.
    We may now apply \Cref{thm:guide} to conclude that 
    \begin{equation}    \label[ineq]{ineq:alex}
        \braket{\bv}{\Gamma^\ell(\bcalA)}^2 \geq \Part_k(\ell) \cdot \frac{\rho\eps}{196 \ell \ln n} \cdot (\rho^2 q)^{c} 
    \end{equation}
    except with probability at most
    \begin{equation}
        \frac{O(\ell/k)^2 \ell \ln n}{\zeta \eps \rho^3} \cdot \hat{m}^{-1}.
    \end{equation}
    Moreover, our assumed upper bound on~$\ell$ means the hypothesis of \Cref{lem:mc} is satisfied, so we obtain
    \begin{equation}    \label[ineq]{ineq:mark}
        \E[\braket{\Gamma^\ell(\bcalA)}{\Gamma^\ell(\bcalA)}] \leq 1.0202 \quad\implies\quad
        \braket{\Gamma^\ell(\bcalA)}{\Gamma^\ell(\bcalA)}\leq \frac{1.0202}{\nu} \quad \text{except with probability at most~$\nu$.}
    \end{equation}
    Since $\ket{\Gamma^\ell(\bcalA)}$ is in the direction of $\ket{\Gamma^\ell(\bcalI_{\textnormal{guide}})}$ and hence $\ket{\bGamma}$, we may combine \Cref{ineq:alex,ineq:mark} to complete the proof.
\end{proof}

\paragraph{Discussion.}  Let us illustrate how \Cref{thm:bob} may be used, comparing to the discussion towards the end of \Cref{sec:noisy-planted}.  
Recall we saw there that Alice could use \Cref{thm:alice} to show the following for $\bcalI \sim \wt{\calR}_{n,4}(\Delta n)$ (with high probability):
\begin{equation}    \label[ineq]{ineq:banks}
    \text{Provided } \Delta \geq \Delta_0 \coloneqq 12.5 \cdot n \ln n, \quad \lambdamax{\calK_{32}(\bcalI)} \leq .198 \cdot \delta_{32,n,4} \cdot m.
\end{equation}
Based on additional simpler considerations, this yielded an $\wt{O}(n^{32})$ time, $\wt{O}(n^{32})$ space algorithm for distinguishing $\wt{\calR}_{n,4}(\Delta_0 n)$ from $\wt{\calP}^z_{n,4}(\Delta_0 n,0.2)$.
Observe also that we could adjust the constants minutely to obtain essentially the same conclusion; e.g., 
\begin{equation}    \label[ineq]{ineq:myI}
    \text{Provided } \Delta \geq (1-\zeta) \cdot \Delta_0, \quad \lambdamax{\calK_{32}(\bcalI)} \leq .1985 \cdot \delta_{32,n,4} \cdot \Delta_0 n, \quad \text{where } \zeta \coloneqq .003
\end{equation}
(with the same failure probability). 

Now suppose Bob applies \Cref{thm:bob} in the setting where $\hat{\bcalI}$ is drawn from either $\wt{\calR}_{n,4}(\Delta_0 n)$ or $\wt{\calP}^z_{n,4}(\Delta_0 n, 0.2)$ (with $n$ sufficiently large).
On one hand, he can use \Cref{ineq:myI} in the case of  $\wt{\calR}_{n,4}(\Delta_0 n)$.
On the other hand, in the case of $\wt{\calP}^z_{32,n,4}(\Delta_0 n)$, \Cref{thm:bob} (with $\nu = n^{-.1}$, say) tells him that with high probability:
\begin{gather}
    \text{there exists } \ket{\bv} \text{ in the span of $\calK_{32}(\bcalI)$'s eigenspaces of eigenvalue at least } .199 \cdot \delta_{32,n,4} \cdot \Delta_0 n \nonumber\\
    \text{such that } \braket{\bv}{\bGamma}^2 \geq  \wt{\Omega}(n^{-16.1}).
\end{gather}
As we will show in \Cref{sec:quantumalgorithm}, Bob can use a quantum algorithm to certify this (with high probability) in time (gate complexity)
\begin{equation}
    \wt{O}(\Delta_0 n) \cdot \sqrt{\wt{O}(n^{16.1})} = \wt{O}(n^{10.1})
\end{equation}
and space $\O(\log n)$ (qubits).
Here the speedup is not exactly quartic; rather than a power of~$4$ we have a power of $\frac{32}{10.1} \approx 3.17$. 
This is because we are using a ``small''~$\ell$ in our tradeoff of $\wt{O}(n^{\ell})$ classical time versus $\wt{O}(n^{\ell/4 + k/2})$ quantum time. 


To conclude, let us note that our quantum algorithms solves a slightly more general problem than \Cref{prob:planted_noisy_kXOR}. The natural way to solve the Planted Noisy $k$XOR decision problem is to develop a \emph{certification} algorithm for either the random case or the planted case. Indeed, our quantum algorithm certifies that the Kikuchi matrix of a $k$XOR instance has an eigenvalue above the cutoff threshold: For the Kikuchi matrix associated with an instance drawn from $\bcalI \sim \calP^z_{n,k}(m,\rho)$, the quantum algorithm outputs ``planted'' with probability $1-o(1)$ over the distribution $\calP^z_{n,k}(m,\rho)$, whereas for any (even adversarially chosen) instance where the largest eigenvalue of the Kikuchi matrix is below the cutoff threshold, the algorithm outputs ``random'' with near certainty.\footnote{Since quantum algorithms are generally probabilistic, there is still a failure probability over the \emph{internal} randomness of the algorithm, but this can be made arbitrarily close to $0$ (e.g., exponentially small). This is unlike the failure probability in the planted case, where there is some chance of failure due to being given a bad input, and this probability cannot be driven down to $0$.} Combined with an Alice Theorem such as \Cref{thm:alice}, this certification algorithm implies the quantum algorithm for the Planted Noisy $k$XOR problem that we have discussed above.

Let us also note that our quantum speedup also applies to the slightly more general decision problem of distinguishing two Planted Noisy $k$XOR instances with different noise rates $\rho_1 \not = \rho_2$, since the only property of the random $k$XOR instance that our techniques exploit is the small spectral norm of its associated Kikuchi matrix. 
\section{Correctness of the quantum algorithm for Tensor PCA}
\label{sec:proof3} \label{sec:TensorPCA}\label{sec:tensorPCA}
The methods discussed above also apply to other planted inference problems, such as {Tensor PCA} \cite{richard2014statistical, perry2016statistical, hopkins2015tensor}, which was the problem studied by Hastings~\cite{Has20}. In this section, we describe our quantum algorithm for Tensor PCA, which is inspired by Hastings's algorithm, but is simpler to implement and analyze.
In particular, we are able to prove correctness of our algorithm using a standard second-moment-method calculation, without relying on trickier results about low-degree polynomials, Gaussian anticoncentration, and quantum field theory. 
As in the $k$XOR case, our quantum algorithm achieves a nearly quartic speedup over the best known classical algorithm. 
The algorithm and analysis in this section are very similar to that in the previous section and we only describe the key differences.

\subsection{Basic definitions}

The decision version of Tensor PCA concerns the problem of detecting 
a rank-one spike hidden in a Gaussian random tensor \cite{richard2014statistical}.

\begin{definition}
Let $\widetilde{\G} \in\left(\mathbb{R}^n\right)^{\otimes k}$ 
be a Gaussian $k$-tensor with each entry chosen i.i.d.\ from the standard normal distribution $\mathcal{N}(0,1)$. 
A \emph{symmetrized Gaussian $k$-Tensor} is a tensor of the form \begin{equation}
    \boldsymbol{G}:=\frac{1}{\sqrt{k !}} \sum_{\pi \in S_k} \widetilde{\boldsymbol{G}}^\pi,
\end{equation} where $S_k$ is the symmetric group on $k$ elements and $
\widetilde{\G}_{i_1, \ldots, i_p}^\pi:=\widetilde{\G}_{i_{\pi(1)}, \ldots, i_{\pi(p)}}$.
     \end{definition}

\begin{definition}
\label{def:tensor_spiked}
    A \emph{Spiked Noisy $k$-tensor} in dimension $n$ is a $k$th-order $n \times n \times \dots \times n$  tensor of the form \begin{equation}
        \T = \beta z^{\otimes k} + \G
    \end{equation}  where $\G$ is a symmetrized Gaussian $k$-tensor, $\beta \geq 0$ 
    is the Signal-to-Noise ratio (SNR) and $ z \in \mathbb{R}$ 
     is a fixed planted spike, normalized such that $\norm{z} = \sqrt{n}$.
\end{definition}
Given a Spiked Noisy $k$-Tensor $\T$, there are several closely related natural algorithmic tasks: to find~$z$; to find an assignment 
having good correlation with~$z$; or, to detect whether $\T$ was drawn from the ``purely random'' distribution with $\beta = 0$ or the ``planted'' distribution with $\beta = \beta^* > 0$. For simplicity, we focus on this last decision variant of Tensor PCA.

The hardness of the above tasks is
determined by the Signal-to-Noise ratio $\beta$. 
Information theoretically, the decision variant of Tensor PCA is possible for $\beta \gg n^{(1-k)/2}$, but efficient (polynomial-time) algorithms are known only for $\beta \gg n^{-k/4}$. Since this problem is only solvable for $\beta \gg n^{(1-k)/2}$, throughout this section we will assume that $\binom{n}{k}\beta^2 = \Omega(n)$. 

As before, we will typically fix $n$, an \emph{even} value of~$k$, and an integer $\ell \geq k/2$. We think of $k$ as an absolute constant, with $n$ and possibly also $\ell$ asymptotically growing. Following \cite{WAM19}, we consider Boolean spikes $z \in \{\pm1\}^n$ for simplicity, but we discuss more general spikes (such as the Gaussian random spikes analyzed in \cite{Has20}) in \Cref{sec:other_priors}. 

By construction, the entry of $\T$ indexed by a tuple $(i_1,\dots, i_k)$ only depends on the set $S = \{i_1, \dots, i_k)$, which makes it natural to view the Spiked Noisy $k$-Tensor as an $n$-variate degree-$k$ polynomial \begin{equation}
    \widetilde{A}_{\T}(X) = \sum_{S: |S| \leq k} \T_S X^S.
\end{equation} This polynomial is set up such that $\widetilde{A}_{\T}(z)$ is large for the planted spike $z$. Since the terms in $\widetilde{A}_{\T}(X)$ with degree less than $k$ correspond to only a $o(1)$ fraction of the entries of $\T$, it is common \cite{richard2014statistical, WAM19} to ignore them and to define Tensor PCA to only include the homogeneous degree-$k$ part of $\T$. This leads to a homogeneous $n$-variate degree-$k$ polynomial \begin{equation}
     A_{\T}(X) = \sum_{S \in \binom{[n]}{k}} \T_S X^S
\end{equation} as in \Cref{sec:xor-defs}, which means that $A_{\T} (X)$ is precisely a $\calN(\beta,1)$-noisy, $z$-planted random degree-$k$ polynomial as defined in \Cref{def:nono} (cf. also \Cref{ex:tensorPCA}).

\subsection{Algorithms}
As for the Planted Noisy $k$XOR problem, the best known classical algorithm for the decision variant of Spiked Noisy Tensor PCA is based on the Kikuchi Method. Our quantum algorithm for Tensor PCA builds on the same ``quantization'' of the Kikuchi method outlined in \Cref{sec:proof2}, which applies a quantum algorithm for the Sparse Guided Hamiltonian problem to a suitable guiding state. As such, the propositions and proofs in the following section are often very similar (although not exactly identical) with corresponding results in our previous discussion of the Planted Noisy $k$XOR problem, via the mapping $m \mapsto \binom{n}{k}$ and $\rho \mapsto \beta$. For ease of exposition, we state our results in less generality than the previous section. For example, we will fix some of the free parameters (e.g., $\hat{\gamma}, \zeta, \epsilon$ in \Cref{thm:kXOR_analysis}) in agreement with what a reasonable quantum algorithm would choose. As a consequence, we only prove that our algorithm achieves a nearly quartic quantum advantage compared to the classical algorithm  derived from a specific ``Alice-theorem'' (given in \Cref{thm:tensor_bounds}), and not for any possible classical algorithm based on future hypothetically improved ``Alice-theorems'' as we did in \Cref{sec:proof2}. Both simplifications are done solely to keep the statements of our results in this section simple and informative, and can be remedied by following the same steps as in \Cref{sec:proof2}. 

We now describe classical and quantum algorithms for Tensor PCA based on the Kikuchi method. Following \Cref{sec:Kikuchi_matchings}, the $n$-variate degree-$k$ polynomial $A_{\T} (X)$ defines a Kikuchi graph as follows.
\begin{definition}  \label{def:kikuch}
    Any Spiked Noisy $k$-Tensor $\T$ in $n$ dimensions is naturally associated with a $\tbinom{n}{\ell}$-vertex weighted graph. This is the $\ell$-th order \emph{Kikuchi graph} of $\T$, with (weighted) adjacency matrix 
    \begin{equation}    \label{eqn:kiku}
        \Kik_\ell(\T) = \sum_{S \in \tbinom{[n]}{k}} \T_S \cdot \Kik_\ell(S),
    \end{equation}
    where 
    \begin{equation}
        \Kik_\ell(S) = 
        \sum_{T\Delta U = S} (\ketbra{T}{U} + \ketbra{U}{T})
    \end{equation}
    is the adjacency matrix of the Kikuchi matching $\Kik^{\ell,n,k}_S$ introduced in \Cref{def:Kikuchi_matching}.
\end{definition}
The $\ell$-th order (weighted) Kikuchi graph of $\T$ has $\binom{n}{\ell}$ vertices and unlike the $k$XOR case, the Kikuchi graph is $d_\ell$-regular for $d_\ell = \tbinom{n-\ell}{k/2} \tbinom{\ell}{k/2}$.\footnote{We consider the Kikuchi graph $d_\ell$ regular even in the unlikely case that some edges have weight 0.}

For a Spiked Noisy $k$-Tensor $\T = \beta z^{\otimes k} + \G$, the Kikuchi matrix decomposes into 
\begin{equation}
    \Kik_\ell(\T) = \beta \Kik_\ell(z^{\otimes k}) + \Kik_\ell(\G).
\end{equation} To establish correctness of the Kikuchi Method for Tensor PCA, we wish to show that for a suitable choice of $\ell$, $\lambdamax{\Kik_\ell(\T)} $ is significantly greater than $\lambdamax{\Kik_\ell(\G)}$.
\begin{proposition}
\label{prop:tensor_lb}
    Let $\T = \beta z^{\otimes k} + \G$ be a Spiked Noisy $k$-Tensor. The largest eigenvalue of $\Kik_\ell(\T)$ is lower bounded by \begin{equation}
        \lambdamax{\Kik_\ell(\T)} \geq \bra{z^{\odot \ell}} \Kik_\ell(\T) \ket{z^{\odot \ell}} \geq (1-\gamma) \cdot \beta d_\ell
    \end{equation} except with probability at most \begin{equation}
        \frac{1}{ \gamma \beta \sqrt{2 \pi \binom{n}{k}} } \exp \left(-\frac{\gamma^2 \lambda^2 \binom{n}{k}}{2}\right).
    \end{equation}
\end{proposition}
\begin{proof} We have 
    \begin{equation}
        \bra{z^{\odot \ell}} \Kik_\ell(\T) \ket{z^{\odot \ell}} = \bra{z^{\odot \ell}} \Kik_\ell(\beta z^{\otimes k}) \ket{z^{\odot \ell}}+ \bra{z^{\odot \ell}} \Kik_\ell(\G) \ket{z^{\odot \ell}}= \beta d_\ell +  \bra{z^{\odot \ell}} \Kik_\ell(\G) \ket{z^{\odot \ell}},
    \end{equation} where the last term \begin{equation}
        \bra{z^{\odot \ell}} \Kik_\ell(\G) \ket{z^{\odot \ell}} = \frac{d_\ell}{\binom{n}{k}} \sum_{S \in \binom{[n]}{k}} x^S \G_S
    \end{equation} is a Gaussian random variable with mean zero and variance $\frac{d_\ell^2}{\binom{n}{k}}$.  
\end{proof}

\begin{proposition} \label{prop:tensor_ub}
   Let $\G$ be a symmetrized Gaussian $k$-tensor in $n$ dimensions. Then for any $\eps > 0$, except with probability at most $2\binom{n}{\ell}^{-\eps}$, 
    we have
   \begin{equation}
        \lambdamax{\Kik_\ell(\G)}  \leq \sqrt{d_\ell}\cdot \sqrt{2(1+\eps) \ell \ln n}.
    \end{equation}
\end{proposition}
\begin{proof}
    The non-diagonal entries of $\G$ are independent Gaussian random variables.
    Thus matrix Chernoff (Khintchine) bounds
    (see, e.g.,~\cite{Tro12}) tell us that
    \begin{equation}
        \Pr[\opnorm{\bcalG} \geq \lambda_0] \leq 
        2N \cdot \exp\parens*{-\frac{\lambda_0^2}{2\opnorm{\E[\Sigma^2]}}},
    \qquad\text{where}\qquad
        \Sigma^2 \coloneqq \sum_{S \in \binom{[n]}{k}}\left(\G_S \Kik_\ell(S) \right)^2.
    \end{equation}
    The matrix $\Kik_\ell(S)^2$ is diagonal, with its diagonal being the $0$-$1$ indicator of the vertices in the matching.  
    Hence $\Sigma^2$ is diagonal, with the $S$-th diagonal entry equal to the degree of vertex~$S$ in $\Kik_\ell$.
    Thus $\opnorm{\Sigma^2} = d_\ell$, and the result follows.
\end{proof}
Combining the lower bound in \Cref{prop:tensor_lb} with the upper bound in \Cref{prop:tensor_ub} yields the following Theorem. 
\begin{theorem} \label{thm:tensor_bounds}
    (Refinement of Theorem 3.3 in \cite{WAM19}.)
    Let $\ell,n,k \in \N^+$ with $k$  even, and $\ell \geq k/2$.  
    Let $\eps > 0$, and assume that $\ell$ satisfies 
    \begin{equation}    
     \ell^{(k-2)/2}  \geq C \cdot \frac{n^{k/2}}{\lambda^2 \binom{n}{k}}\cdot \ln n, \quad \text{where } C = \frac{2(1+\eps)}{(1-\gamma)^2 \binom{k}{k/2}} .
    \end{equation}
    Then for a Spiked Noisy k-tensor $\T = \beta z^{\otimes k} + \G$ and any constant $\gamma > 0$, except with probability at most $o(1)$, we have
    \begin{equation}
    \lambdamax{\calK_{\ell}(\G)}\leq (1- \gamma) \beta d_\ell \quad \text{ and } \quad \lambdamax{\calK_{\ell}(\T)} \geq (1-\gamma/2) \beta d_\ell.
    \end{equation}
\end{theorem}
\begin{proof}
    The upper bound on $\lambdamax{\calK_{\ell}(\G)}$ follows from \Cref{prop:tensor_ub} by substituting the choice of $\ell$. The lower bound follows from \Cref{prop:tensor_lb} using the additional fact that $\binom{n}{k} \lambda^2 = \Omega(n)$ whenever $\lambda$ is above the information-theoretic limit. 
\end{proof}

\subsection{Guiding state}
\label{sec:tensor_guide}
We now construct a guiding state that has up to quadratically increased overlap with the leading eigenspace of the Kikuchi matrix. Let us first establish a lower bound on the overlap of the planted 
assignment $z$ with the leading eigenspace.
\begin{proposition}[Tensor PCA version of \Cref{prop:simple2}] \label{prop:leading_tensorPCA}
    In the setting of \Cref{thm:tensor_bounds}, let $\bPi_{\geq}$ denote the projector onto the eigenspaces of $\calK_{\ell}(\T)$ of eigenvalue at least $(1-\gamma) \beta d_\ell$.
    Then we have 
    \begin{equation}
        \braketOP{z^{\odot \ell}}{\bPi_{\geq}}{z^{\odot \ell}} \geq \frac{ \gamma}{2}.
    \end{equation}
    except with probability at most $o(1)$.  
\end{proposition}
\begin{proof}
    We have the upper bound $\lambdamax{\Kik_\ell(\T)} \leq \beta d_\ell + \lambdamax{\Kik_\ell(\G)} \leq (2-\gamma)\beta d_\ell$ with probability $1-o(1)$. The result now follows from \Cref{prop:tensor_lb} by a Markov's inequality argument on the eigenvalues of $\calK_\ell(\bcalI)$.
\end{proof}
Now let us define the guiding state. Recall the definition of $\Part_k(V)$ from \Cref{not:partition}.
\begin{notation} \label{fact:skel2}
    In the preceding setting, for a tensor $\T \in \C^{n^k}$, we write $\T^{\circledast c} \in \C^{\binom{n}{\ell}}$ for the vector whose $V$-coordinate is
    \begin{equation}
        \T^{\circledast c}_{V} = \sum_{\{S_1, \dots, S_c\} \in \Part_k(V)} \prod_{j=1}^c \T_{S_j}.
    \end{equation}
\end{notation}
\begin{definition}
    Let $\T$ be a Spiked Noisy $k$-Tensor on $n$ variables and let $\ell = ck$ for $c \in \N^+$.
    We define the associated \emph{guiding state} to be 
\begin{equation}
    \ket{\Gamma^\ell(\T)} = \frac{1}{\sqrt{\binom{n}{\ell} |\Part_k(\ell)|}}\sum_{S \in \binom{[n]}{\ell}} \T_S^{\circledast c} \ket{S}.
\end{equation}
\end{definition} 
Note that this agrees with \Cref{def:guiding_state} for the $\calN(\lambda,1)$-noisy, $z$-planted random degree-$k$ polynomial associated with $\T$. We can thus import \Cref{thm:guide} to argue that this guiding state has improved overlap with the high energy space of $\Kik_\ell(\T)$. To do so, we first need to split the Spiked Noisy $k$-Tensor $\T$ into two independent instances, similar to our splitting of constraints in \Cref{sec:proof2}. 
\begin{lemma}
\label{lem:tensor_split}
    Let $\T = \beta z^{\otimes k} + \G$ be a Spiked Noisy $k$-Tensor and let $\textbf{H}$ be a noise tensor sampled from the same distribution as $\G$. Let $\zeta \not = 0$. Then the two Spiked Noisy $k$-Tensors $\T^+ = \T + \zeta \textbf{H}$ and $\T^- = \T - \frac{1}{\zeta}\textbf{H}$ have independent noise. 
\end{lemma}
\begin{proof}
    This follows from \begin{equation}
        \Cov[\T^+,\T^-] = \Var[\G] - \Var[\textbf{H}] = 0
    \end{equation}
    and the fact that jointly Gaussian random variables are independent if and only if they are uncorrelated.
\end{proof}
Note that $\T^{\pm}$ corresponds to a Spiked Noisy $k$-Tensor with SNR $\lambda^{\pm} = \frac{\lambda}{\sqrt{1 + \zeta^{\pm 2}}}$. We now prove our main theorem. 
\begin{theorem}[Tensor PCA version of \Cref{thm:kXOR_analysis}]
\label{thm:Tensor_analysis}
    Let $k\geq 4$ be even, let $\ell = c k$ for $c \in \N^+$, and $\beta \leq 0.1/\ell^{k/2}$. Let $\T = \beta z^{\otimes k} + \G$ be a Spiked Noisy $k$-Tensor and assume that $\ell$ satisfies the lower bound in \Cref{thm:tensor_bounds}. Partition $\T$ into $\T^+$ and $\T^-$ as in \Cref{lem:tensor_split} for $\zeta = 1/\ln{n}$. Let $\ket{\bGamma(\T^-)}$ and $\Kik_\ell(\T^+)$ be the corresponding guiding state and Kikuchi matrix, respectively. 

    Let $\ket{\bGamma}$ be the unit vector (``guiding state'') in the direction of $\ket{\bGamma(\T^-)}$.
    Then for any constant $\gamma > 0$, with probability at least $1-o(1)$,
    there is a unit vector $\ket{\bv}$ in the span of $\calK_\ell(\T^+)$'s eigenspaces of eigenvalue at least $(1-\gamma) d_\ell \lambda$ such that 
    \begin{equation}
        \braket{\bv}{\bGamma}^2 \geq \lambda^{2c} \xi, \qquad \text{where } \xi = \Part_k(\ell) \cdot \frac{\gamma}{10} \cdot \left(\frac{1}{1 + \ln^2 n}\right)^{c} \cdot \frac{1}{\ln n}. 
    \end{equation}
\end{theorem}
\begin{proof} 
    Using \Cref{prop:leading_tensorPCA} and letting $\ket{\bv}$ be the unit vector in the direction of $\bPi_{\geq} \ket{z^{\odot \ell}}$, we conclude that except with probability at most $o(1)$,
    \begin{equation}
    \label{eq:tbh}
        |\braket{\bv}{z^{\odot \ell}}|^2\geq \frac{\gamma}{2}.
    \end{equation}
    Note that since the SNR $\lambda^+ = \lambda\left(1+\sqrt{\frac{1}{\ln n}}\right)^{-\frac{1}{2}} $ of $\T^+$ is slightly reduced, \Cref{thm:tensor_bounds} in the proof of \Cref{prop:leading_tensorPCA} has to be called with slightly smaller $\gamma^+ = \gamma(1-\frac{1}{2\gamma \ln n}) = \gamma(1-o(1))$, but this still leaves the failure probability at $o(1)$. Fix any  outcome of $\T^+$ for which \Cref{eq:tbh} holds.  
    Since the noise in $\T^-$ is independent of $\T^+$, we can still view $\T^-$ as a $\calN(\lambda,1+\ln^2 n)$-noisy $z$-planted degree-$k$ polynomial. Equivalently, after rescaling, it is a $\calN(\lambda^-,1)$-noisy $z$-planted polynomial, which we call $\mathbf{\calA}$. Here, $\T^-$ has reduced SNR $\lambda^- = \frac{\lambda}{\sqrt{1+\ln^2 n}}$.
    We may now apply \Cref{thm:guide} to conclude that 
    \begin{equation}    \label[ineq]{ineq:alex2}
        \braket{\bv}{\Gamma^\ell(\bcalA)}^2 \geq \Part_k(\ell) \cdot \frac{\gamma}{8} \cdot \left(\frac{\lambda^2 }{1 + \ln{n}^2}\right)^{c} 
    \end{equation}
    except with probability at most
    \begin{equation}
        \frac{O(\ell/k)^2 (1+ \ln^2n)}{ \lambda^2 \gamma} \cdot \binom{n}{k}^{-1} = o(1),
    \end{equation} where we have once again used $\binom{n}{k}\lambda^2 = \Omega(n).$
    Moreover, our assumed upper bound on $\ell$ satisfies the hypothesis of \Cref{lem:mc}, such that we obtain
    \begin{equation}    \label[ineq]{ineq:mark2}
        \E[\braket{\Gamma^\ell(\bcalA)}{\Gamma^\ell(\bcalA)}] \leq 1.0202\quad\implies\quad
        \braket{\Gamma^\ell(\bcalA)}{\Gamma^\ell(\bcalA)}\leq \frac{1.0202}{\nu} \quad \text{except with probability at most~$\nu$.}
    \end{equation}
    Since $\ket{\Gamma^\ell(\bcalA)}$ is in the direction of $\ket{\Gamma^\ell(\bcalI_{\textnormal{guide}})}$ and hence $\ket{\bGamma}$, we may combine \Cref{ineq:alex2,ineq:mark2} for $\nu = \ln^{-1} n$ to complete the proof.
\end{proof}
In the statement of this theorem, we have made the choice $\zeta = 1/\ln n$, which attributes most of the signal to the Kikuchi matrix and an $o(1)$-fraction to the guiding state. We make this choice because reducing the signal given to the Kikuchi matrix forces us to increase $\ell$ to obtain the same guarantees as in \Cref{thm:tensor_bounds}, and thus comes at an exponential cost in the runtime. On the other hand, reducing the signal in the guiding state reduces the overlap in \Cref{thm:guide}, which enters only multiplicatively into the runtime. 
\subsection{General spikes in Tensor PCA}
\label{sec:other_priors}
For simplicity, we have so far focused on Boolean spikes $z \in \{\pm1\}^n$\, following \cite{WAM19}. However, the methods discussed above also apply to spikes drawn from other, more general distributions. Let us illustrate this by considering a \emph{random} spike, with entries $\bz_i \sim \calN(0,1)$ chosen i.i.d.\ from the standard normal distribution. To establish that the Kikuchi method solves the decision variant of Spiked Noisy Tensor PCA for this prior, we need to re-prove the upper and lower bounds on the largest eigenvalue of the Kikuchi matrix in \Cref{thm:tensor_bounds}. The upper bound is independent of the spike, but the lower bound in \Cref{prop:tensor_lb} relied on the fact that for Boolean spikes, $\bz_i^2 = 1$. We now show that this assumption is not crucial and that a similar bound holds for Gaussian spikes. 

\begin{proposition}
\label{prop:tensor_lb_gauss}
    Let $\T = \beta \bz^{\otimes k} + \G$ be a Spiked Noisy $k$-Tensor, 
    $\bz_i \sim \calN(0,1),$ $\ell = o(\sqrt{n})$, $\binom{n}{k}\beta^2 = \Omega(n)$, and $\gamma > 0$ a constant. The largest eigenvalue of $\Kik_\ell(\T)$ is lower bounded by \begin{equation}
        \lambdamax{\Kik_\ell(\T)} \geq \bra{\bz^{\odot \ell}} \Kik_\ell(\T) \ket{\bz^{\odot \ell}} \geq (1-\gamma) \cdot \beta d_\ell
    \end{equation} except with probability at most $o(1)$.
\end{proposition}
\begin{proof}
We need to compute the moments of \begin{equation}
    \bra{\bz^{\odot \ell}} \Kik_\ell(\T) \ket{\bz^{\odot \ell}} = \beta \bra{\bz^{\odot \ell}} \Kik_\ell(\bz^{\otimes k}) \ket{\bz^{\odot \ell}}+ \bra{\bz^{\odot \ell}} \Kik_\ell(\G) \ket{\bz^{\odot \ell}} 
\end{equation} over the distributions $\bz_i \sim \calN(0,1)$ and $\G_S \sim \calN(0,1)$.
The first moment is $\beta d_\ell$. The second moment is (for sets $S_1,S_2,T_1,T_2 \in \binom{[n]}{\ell}$) \begin{align}
    &\E  \left|\bra{\bz^{\odot \ell}} \Kik_\ell(\T) \ket{\bz^{\odot \ell}}\right|^2   \\ &= \E\left[\frac{\beta^2}{{\binom{n}{\ell}}^2} \sum_{|S_1 \Delta T_1| = k, |S_2 \Delta T_2| = k} \prod_{i \in S_1 \cup T_1} \bz_i^2 \prod_{j \in S_2 \cup T_2} \bz_j^2 + \sum_{\substack{S_1\Delta T_2 = S_2 \Delta T_2 \\ |S_1\Delta T_2| = k }} \prod_{i \in S_1 \cap T_1} \bz_i^2 \prod_{j \in S_2 \cap T_2} \bz_j^2 \G_{S_1 \Delta T_1}^2\right] \\ &=\E\left[\frac{\beta^2}{{\binom{n}{\ell}}^2} \sum_{|S_1 \Delta T_1| = k, |S_2 \Delta T_2| = k} 3^{|(S_1 \cup T_1)\cap (S_2 \cup T_2)|} + \sum_{\substack{S_1\Delta T_2 = S_2 \Delta T_2 \\ |S_1\Delta T_2| = k }} 3^{|(S_1 \cap T_1)\cap (S_2 \cap T_2)|} \right] \\ &= 
\frac{\beta^2 d_\ell}{{\binom{n}{\ell}}} \sum_{a = 0}^{\ell + k/2} 3^a \left[ \binom{\ell + k/2}{a}\binom{n-\ell-k/2}{\ell+k/2-a}\binom{\ell+k/2}{k}\binom{k}{k/2}\right] \\ & \qquad \qquad+ \frac{d_\ell}{{\binom{n}{\ell}}} \sum_{b = 0}^{\ell - k/2} 3^b \left[ \binom{\ell - k/2}{b}\binom{n-\ell-k/2}{\ell-k/2-b}\binom{k}{k/2}\right]
\\ &= 
\beta^2 d_\ell^2 \sum_{a = 0}^{\ell + k/2} 3^a \left[ \frac{\binom{\ell + k/2}{a}\binom{n-\ell-k/2}{\ell+k/2-a}}{\binom{n}{\ell + k/2}}\right] + d_\ell^2 \sum_{b = 0}^{\ell - k/2} 3^b \left[ \frac{\binom{\ell - k/2}{b}\binom{n-\ell-k/2}{\ell-k/2-b}}{\binom{\ell+k/2}{k}\binom{n}{\ell + k/2}}\right]
\\ &= 
\beta^2 d_\ell^2 \sum_{a = 0}^{\ell + k/2} 3^a \left[ \frac{a!}{n^a} \binom{\ell+k/2}{a}^2 (1-o(1))\right] + \frac{d_\ell^2}{\binom{n}{k}} \sum_{b = 0}^{\ell - k/2} 3^b \left[ \frac{b!}{n^b} \binom{\ell-k/2}{b}^2 (1-o(1))\right]
\\ &= 
\beta^2 d_\ell^2 (1+o(1)).
\end{align}
Let us explain each step in the above calculation.
\begin{itemize}
    \item In step 1, we use that the odd moments of $x_i$ vanish.
    \item In step 2, we use that the second moment of a standard Gaussian random variable is 1 and the fourth moment is $3$.
    \item In step 3, we stratify the sum according to the intersection cardinalities $a = {|(S_1 \cup T_1)\cap (S_2 \cup T_2)|}$, $b = {|(S_1 \cap T_1)\cap (S_2 \cap T_2)|}$, and compute the corresponding binomial counting factors. 
    \item In step 4, we use the binomial identities \begin{equation}
    \binom{n}{\ell}d_\ell = \binom{n}{\ell}\binom{\ell}{k/2}\binom{n-\ell}{k/2} = \binom{n}{\ell+k/2}\binom{\ell+k/2}{k}\binom{k}{k/2} = \binom{n}{k}\binom{k}{k/2}\binom{n-k}{\ell-k/2}.
\end{equation}
\item In step 5, we expand the binomial coefficients at large $n$ and use $\frac{l^2}{n} = o(1)$ by assumption.
\item In step 6, we notice that all but the summand with $a = 0$ are of magnitude $o(1) \cdot \beta^2 d_\ell^2$ (which, for the second sum, uses the assumption $\binom{n}{k}\beta^2 = \Omega(n)$). 
\end{itemize}
The above computation of moments shows that $\Var[\bra{\bz^{\odot \ell}} \Kik_\ell(\T) \ket{\bz^{\odot \ell}}] = o(1) \cdot \E[\bra{\bz^{\odot \ell}} \Kik_\ell(\T) \ket{\bz^{\odot \ell}}]^2$, and the claim follows. 
\end{proof}

\Cref{prop:tensor_lb_gauss} implies that the Kikuchi method achieves detection for Gaussian priors. However, the correctness (and quartic speedup) of our quantum algorithm is more intricate and relies on \Cref{thm:Tensor_analysis}, so we have to prove a corresponding statement for Gaussian priors. Note that the proof of \Cref{thm:Tensor_analysis} mostly relies only on the first and second moments of the spike, which are equivalent for Boolean and Gaussian spikes. The only non-trivial exceptions are \Cref{prop:leading_tensorPCA} due to its dependency on \Cref{prop:tensor_lb}, and \Cref{thm:guide}, which rely on the fourth moments of the spike.
We have already shown in \Cref{prop:tensor_lb_gauss} that a corresponding statement of  \Cref{prop:leading_tensorPCA} holds for Gaussian priors. Essentially the same fourth-moment calculation shows that \Cref{thm:guide} also holds for Gaussian priors (we do not repeat this calculation here), which finally establishes that our quantum algorithm also achieves a (nearly) quartic speedup for detecting spikes from Gaussian priors.

\section{Implementation of the quantum algorithm }
\label{sec:quantumalgorithm}
In this section we formally prove the upper bound on the quantum complexity of solving the Planted Noisy $k$XOR problem.

We start with some preliminaries on Hamiltonian simulation, then define the Guided Sparse Hamiltonian problem and upper-bound its complexity. 
We then prove some results on state preparation that will be used to create the guiding state efficiently. 
Finally, we present our algorithm for the Planted Noisy $k$XOR problem that uses all these tools.

\subsection{Hamiltonian simulation preliminaries}
\label{sec:hamsim_prelim}
This section summarizes the tools we need from the Hamiltonian simulation literature. Readers familiar with Hamiltonian simulation can safely skip this.

\begin{definition}[Sparse Hamiltonian]
A \emph{Hamiltonian} $H$ on $N$ qubits is a Hermitian matrix of dimension $2^N$. We say $H$ is \emph{$s$-sparse} if $H$ has at most $s$ nonzero entries in any row or column. 
\end{definition} 

Recall that we use $\norm{M}$ to denote the operator norm or largest singular value of $M$, and we define $\maxnorm{M} = \max_{i,j}|M_{ij}|$.

\begin{definition}[Hamiltonian simulation problem]
    In the \emph{Hamiltonian simulation problem}, we are given\footnote{We'll talk about \emph{how} it's given shortly.} a Hamiltonian $H$, a time $t$, and an allowed error $\epsilon$, and the goal is to construct a quantum circuit that implements a unitary $U$ such that $\norm{U-\exp(-\i Ht)}\leq \epsilon$.
    We will assume without loss of generality that $\maxnorm{H} \leq 1$ since this can be achieved by rescaling $t$.
\end{definition}

Although the Hamiltonian is of size $2^N$, we would like to exploit its sparsity and solve the problem faster than time $2^N$, and hence we need a more efficient description of the Hamiltonian than merely writing down all its nonzero entries explicitly. 
The best algorithms for the Hamiltonian simulation problem construct a quantum circuit of size 
$\poly(N,s,t,\log(1/\epsilon))$ when the nonzero entries of the input Hamiltonian can be efficiently computed.

We assume the matrix $H$ is provided via two oracles, $O_H$, the adjacency matrix oracle, and $O_F$, the adjacency list oracle.

\begin{definition}[Adjacency matrix oracle]\label{def:OH}
    For a Hamiltonian $H$ with $\maxnorm{H}\leq 1$, the oracle $O_H$ gives the $(i,j)$ entry of the Hamiltonian to $b$ bits of precision: For all $i,j, \in [2^N]$ and $z\in\{0,1\}^b$, $O_H$ acts as
    \begin{equation}
    O_H|i,j,z\rangle = |i,j,z \oplus \mathrm{round}(H_{ij})\rangle,
\end{equation}
where above $\mathrm{round}(H_{ij})$ denotes the $b$ binary digits of a number that is within $\pm 2^{-b}$ of $H_{ij}$. 
\end{definition}

Since our goal is to solve the Hamiltonian simulation problem to error $\epsilon$, it is sufficient to represent each entry of $H$ to $b=O(\log(st/\epsilon))$ bits of precision. 

\begin{lemma}
    Let $H$ be an $s$-sparse Hermitian matrix with $\maxnorm{H}\leq 1$, and let $H'$ be the matrix $H$ but with every nonzero entry $H_{ij}$ rounded to $b$ binary digits. Then for any $t$, $\norm{\exp(-\i Ht)-\exp(-\i H't)} \leq \eps$ if $b=O(\log(st/\eps))$.
\end{lemma}

\begin{proof}
To upper bound $\norm{\exp(-\i Ht)-\exp(-\i H't)}$, we can use the following folklore fact 
that for any Hermitian matrices $A$ and $B$, $\norm{\exp(\i A)-\exp(\i B)} \leq \norm{A-B}$.
Two proofs can be found in \cite[Appendix A]{CGJ19} and \cite[Lemma 3.2(c)]{HKOT23}, and we provide another: For any $r>0$, we have $\norm{\exp(\i A)-\exp(\i B)} = \norm{\exp(\i A/r)^r-\exp(\i B/r)^r} \leq r\norm{\exp(\i A/r)-\exp(\i B/r)}$, where the last inequality uses the subadditivity of error in unitary approximations. By Taylor expanding the exponential, we have $r\norm{\exp(\i A/r)-\exp(\i B/r)} \leq \norm{A-B} + O(1/r)$. Since $\norm{\exp(\i A)-\exp(\i B)} \leq \norm{A-B} + O(1/r)$ holds for every choice of $r$, we must have $\norm{\exp(\i A)-\exp(\i B)} \leq \norm{A-B}$.

So we know that $\norm{\exp(-\i Ht)-\exp(-\i H't)} \leq \norm{(H-H')t}$. If $E=H-H'$, since $E$ is also $s$-sparse (since $H$ and $H'$ only differ on the nonzero entries), we know that $\norm{E} \leq s\maxnorm{E}$. Thus $\norm{\exp(-\i Ht)-\exp(-\i H't)} \leq st\maxnorm{E} \leq st 2^{-b}$. Choosing $b=O(\log(st/\epsilon))$ makes this error much smaller than $\epsilon$.
\end{proof}

The second oracle for $H$ specifies the locations of the nonzero entries in any row of $H$ via a quantum implementation of an adjacency list. 
The adjacency list of a row $i \in [2^N]$ is an list of distinct indices $j_1,\ldots,j_s$, such that all the nonzero matrix entries of row $i$ appear in this list (and they appear exactly once since the list has distinct elements). 
If a given row $i$ does not have $s$ nonzero entries, some of the entries $j_k$ can point to zero entries in the Hamiltonian, but we still require all the indices $j_k$ to be distinct. (E.g., if row $i$ is all zeros, we still require the adjacency list to output $s$ distinct indices.)

One easy way to enforce this is to have a function $f$ that on input $i$ (the label of the row) and $k$ (the $k$th potentially nonzero entry in row $i$) outputs an index $j$ and a bit $b$ indicating whether or not this is a distinct index ($b=0$) or if it has run out of nonzero entries to specify ($b=1$). When it outputs $b=1$, it should simply output $k$ as the $k$th entry in the adjacency list. This ensures that all the outputs $f(i,k)$ are distinct for different values of $k$.

This is equivalent to doubling the matrix dimension to $2^{N+1}$ and considering the Hamiltonian $|0\rangle\langle 0|\otimes H = \left(\begin{smallmatrix} H & 0 \\ 0 & 0 \end{smallmatrix}\right)$. Then we can view the output $(0,k)$ as representing $k$, and $(1,k)$ as representing $k+N$, which is in the second half of the matrix which is all zeros.

\begin{definition}[Adjacency list oracle]\label{def:OF}
    For a $s$-sparse $N$-qubit Hamiltonian $H$, let $f:[2^N]\times [s] \to \{0,1\} \times [2^N]$ be a function that satisfies:
    \begin{enumerate}[itemsep=0pt,topsep=4pt]
        \item $\forall i \in [2^N], \forall k\neq k' \in [s]: f(i,k)\neq f(i,k')$.
        \item $\forall i,j \in [2^N]$, if $H_{ij} \neq 0$, then there exists a (unique) $k\in[s]$ such that $f(i,k)=(0,j)$.
        \item If for any $i \in [2^N]$ and $k \in [s]$ we have $f(i,k)=(1,k')$, then we require that $k=k'$. 
    \end{enumerate}     
    The \emph{adjacency list oracle} for $H$, call it~$O_F$, implements any function $f$ satisfying these requirements: For any $i \in [2^N], k \in [s]$:
    \begin{equation}
        O_F |i,k\rangle = |i,f(i,k)\rangle.    
    \end{equation}    
\end{definition}
Note that this oracle replaces $k$ in the second register with $f(i,k)$, which is a reversible map. Now that we have defined how $H$ is given, we can state the complexity of Hamiltonian simulation.

\begin{theorem}[Hamiltonian simulation]\label{thm:HamSim}
    Let $H$ be a $s$-sparse Hamiltonian on $N$ qubits with $\maxnorm{H}\leq 1$ and $t$ be such that $st \geq 1$. Then we can approximate the unitary $\exp(-\i Ht)$ with error at most $\epsilon$  using $Q=O(st + \log(1/\epsilon))$ queries to $O_H$ and $O_F$, gate complexity\footnote{By ``gate complexity'', we mean the total number of (arbitrary) 1- and 2-qubit gates used by the quantum algorithm. These gates can be further represented using a finite universal gate set with a logarithmic overhead.} $O(Q(N+\log^{2.5}(st/\epsilon)))$, and $O(N)+\O(\log(st/\epsilon))$ qubits.
\end{theorem}

The query and gate complexity bounds follow from combining Lemma~48 and Corollary~62 of Ref.~\cite{GSLW19}, which block-encodes the Hamiltonian using these oracles and then performs quantum signal processing or a quantum singular value transform to approximately implement $\exp(-\i Ht)\otimes \Id$. For our application, we can also use older techniques based on linear combination of unitaries, which gives an upper bound of $Q = O(st \log(st/\epsilon)/\log\log(st/\epsilon))$~\cite[Theorem 1]{BCK15} and the gate complexity bound stated above. 

The space usage of these algorithms has two contributions. 
First, these algorithms use $O(N)$ space, which is required since $N$ qubits are needed to represent the state, and the adjacency matrix oracle itself acts on over $2N$ qubits. The additional space used by the algorithm is dominated by the complexity of computing the $\arcsin$ function to $\log(st/\epsilon)$ bits of precision. 
As in these papers, it can be computed using the Taylor series, which only needs space $\O(\log(st/\epsilon))$. 
Alternately, one may use more sophisticated methods~\cite{BZ10} that use space $\O(\log(st/\epsilon))$ as well.

Finally, to use these algorithms, we can upper bound the cost of the quantum oracles $O_H$ and $O_F$ using the \emph{classical} cost of computing certain functions.

\begin{fact}\label{fact:classicaloracles}
    The quantum gate complexity of $O_H$ is upper-bounded by (up to constants) the classical gate complexity of computing the function $(i,j)\mapsto H_{ij}$. The quantum gate complexity of $O_F$ is upper bounded by the larger of the classical gate complexities of computing $(i,k) \mapsto f(i,k)$ and $(i,f(i,k)) \mapsto k$.
\end{fact}

The first upper bound follows by simply taking the classical circuit and making it reversible, which computes $H_{ij}$ in an additional register, as needed by $O_H$. If we do the same thing with the classical circuit that maps $(i,k) \mapsto f(i,k)$, then we will go from $|i,k\rangle$ to $|i,k,f(i,k)\rangle$. We then need the inverse map to erase $k$, which will give us $|i,f(i,k)\rangle$.

\subsection{Guided Sparse Hamiltonian problem}
\label{sec:guidedsparseHamSim}

We can now formally define the Guided Sparse Hamiltonian problem.
For any Hamiltonian $H$ and parameter~$\lambda$, we define $\Pi_{\geq \lambda}(H)$ to be the projector on the eigenvectors of $H$ with eigenvalue $\geq \lambda$.

\begin{definition}[Guided Sparse Hamiltonian problem]\label{def:guidedsparse}
    In the \emph{Guided Sparse Hamiltonian problem} we are given the following as input:
    \begin{enumerate}[]
    \item{An $s$-sparse Hamiltonian $H$ on $N$ qubits with $\maxnorm{H}\leq 1$ specified via the oracles $O_H$ (\Cref{def:OH}) and $O_F$ (\Cref{def:OF}).}
    \item{A quantum circuit that uses $G$ gates and maps $\ket{0^N}\ket{0^A}$ to $\ket{\Psi}\ket{0^A}$.}
    \item{Parameters $\lambda  \in [-\Lambda,\Lambda]$, $\alpha \in (0,1)$, and $\gamma \in (0,1]$.}
\end{enumerate}
Output: 
\begin{itemize}[]
    \item YES if $\norm{\Pi_{\geq \lambda }(H)\ket{\Psi}} \geq \gamma$ (which also implies $\norm{H} \geq \lambda$).
    \item NO if $\norm{H}\leq (1-\alpha)\lambda $.
\end{itemize} 
\end{definition}

\begin{theorem}\label{thm:guidedsparse}
The Guided Sparse Hamiltonian problem can be solved with high probability by a quantum algorithm that uses
\begin{align}
Q =&~ \O\left(s/(\gamma \alpha \lambda)\right) \textrm{queries to the oracles for $H$},\\ 
&~\O\left(G/\gamma + \polylog(Q)/\gamma + QN\right) \textrm{ gates, and}\\ 
&~A+O(N)+\O(\log Q) \textrm{ qubits}.    
\end{align}
\end{theorem}

\begin{proof}
    Let $U=\exp(\i Ht)$ for $t = \frac{\pi}{2s}$.
    Thus $U$ and $H$ have the same eigenvectors, with eigenvalue $h$ of $H$ getting mapped to eigenvalue $\exp(\i ht)$ (or eigenphase $ht$) of $U$.
    Since $\norm{H}\leq s\maxnorm{H}\leq s$, any two distinct eigenvalues of $H$ will map to distinct eigenvalues of $U$. 
    Our strategy will be to use Hamiltonian simulation to simulate $U$, then use phase estimation to distinguish the two ranges of eigenvalues of $H$, and finally use amplitude amplification to boost the low success probability due to the small projection $\gamma$.
    
    We can simulate $U$ or $U^{-1}$ (since $U^{-1}$ always has the same gate complexity as $U$) to error $\eps_\mathrm{\scriptscriptstyle HS}$ using \Cref{thm:HamSim}, which uses 
    \begin{align}
    Q_\mathrm{HS} &= O(\log(1/\eps_\mathrm{\scriptscriptstyle HS})) \text{~~queries to the oracles for $H$, and}   \\
    G_\mathrm{HS} &= O(Q_\mathrm{HS}(N+\log^{2.5}(1/\eps_\mathrm{\scriptscriptstyle HS})) \text{~~additional gates.}
    \end{align}
    Note that the assumption that $st \geq 1$ in \Cref{thm:HamSim} is satisfied by our choice of $t$. 
    We will choose $\eps_\mathrm{\scriptscriptstyle HS}$ later so that the sum of all the errors due to approximate implementations of $U$ and $U^{-1}$ is less than a small constant.

    We now use phase estimation to distinguish eigenphases larger than $\lambda t$ from eigenphases smaller than $(1-\alpha)\lambda t$, whose difference is 
    \begin{equation}
        \eps_\mathrm{\scriptscriptstyle PE} = \alpha\lambda t.
    \end{equation} 
    Standard phase estimation combined with success probability amplification by majority voting implements a unitary $U_\mathrm{PE}$ whose behavior on an eigenstate $\ket{\varphi}$ with $U\ket{\varphi}=e^{i\varphi}\ket{\varphi}$ is
    \begin{equation} 
        U_\mathrm{PE} \ket{\varphi} = \ket{\varphi} \sum_{\tilde \varphi} \beta_{\tilde \varphi}\ket{\tilde {\varphi}},
    \end{equation}
    where $\tilde \varphi$ represents a phase written with $\Theta(\log(1/\eps_\mathrm{\scriptscriptstyle PE}))$ bits of precision. We'll say the calculated phase is incorrect if $|\varphi - \tilde \varphi|\geq \eps_\mathrm{\scriptscriptstyle PE}/2$. The probability of an incorrect phase in the second register is $\delta_\mathrm{\scriptscriptstyle PE} = \sum_{\tilde \varphi: |\varphi - \tilde \varphi|\geq\eps_\mathrm{\scriptscriptstyle PE}/2} |\beta_{\tilde \varphi}|^2$. The cost of $U_\mathrm{PE}$ is 
    \begin{align}
    Q_\mathrm{PE} &= O((1/\eps_\mathrm{\scriptscriptstyle PE})\log(1/\delta_\mathrm{\scriptscriptstyle PE})) \text{~~uses of $U$ and $U^{-1}$ and} \\
    G_\mathrm{PE} &= \polylog\left(\frac{1}{\eps_\mathrm{\scriptscriptstyle PE}} \frac{1}{\delta_\mathrm{\scriptscriptstyle PE}}\right) \text{~~additional gates.}
    \end{align}

    Finally, since we only have a state with projection at least $\gamma$ with the space of eigenvectors with eigenphase larger than $\lambda t$ in the YES case, the probability of measuring such a phase if we measure the phase register of $U_\mathrm{PE}\ket{\Psi}$ is at least $\gamma^2(1-\delta_\mathrm{\scriptscriptstyle PE})$. On the other hand, in the NO case if we measure the phase register of $U_\mathrm{PE}\ket{\Psi}$, we will see a phase $\geq \lambda t$ with probability at most $\delta_\mathrm{\scriptscriptstyle PE}$. To have this probability be much smaller than that in the YES case, we choose $\delta_\mathrm{\scriptscriptstyle PE}=\gamma^3$, say, which means 
    \begin{equation}
        \log(1/\delta_\mathrm{\scriptscriptstyle PE})=\Theta(\log(1/\gamma)).
    \end{equation}

    If we were to simply measure the state and repeat $O(1/\gamma^2)$ times, in the YES case we would find a phase $\geq \lambda t$ and in the NO case we would not. Instead, we can use amplitude amplification to decide between the two cases with only a $O(1/\gamma)$ overhead. Amplitude amplification uses
    \begin{align}
        Q_\mathrm{AA} &= O(1/\gamma)  \text{~~uses of $U_\mathrm{PE}$ and} \\
        G_\mathrm{AA} &= O(G/\gamma) \text{~~additional gates,}
    \end{align}
    where the additional gates account for the cost of reflecting about $\ket{\Psi}$ $O(1/\gamma)$ times, which can be done with cost $O(G)$.

    Now the total number of uses of $U$ and $U^{-1}$ is $Q_\mathrm{PE}Q_\mathrm{AA} = O(\log(1/\delta_\mathrm{\scriptscriptstyle PE})/(\gamma \eps_\mathrm{\scriptscriptstyle PE})) = \O(1/(\gamma \alpha \lambda t))$. This means we can choose 
    \begin{equation}
    1/\eps_\mathrm{\scriptscriptstyle HS} = \widetilde\Theta\left(1/(\gamma \alpha \lambda t)\right)    
    \end{equation}
    to ensure the the error in Hamiltonian simulation contributes negligibly to the overall error.

    Now that all the parameters are chosen, we only need to sum the costs. Let's begin with the number of queries to the oracles for $H$. This will be
    \begin{align}
        Q = Q_\mathrm{HS} Q_\mathrm{PE} Q_\mathrm{AA} 
        &= O\left( \log(1/\eps_\mathrm{\scriptscriptstyle HS}) (1/\eps_\mathrm{\scriptscriptstyle PE})\log(1/\delta_\mathrm{\scriptscriptstyle PE}) (1/\gamma) \right) \\
        & = O\left(  \log\left( {1}/{\eps_\mathrm{\scriptscriptstyle HS}} \right) \frac{1}{\alpha \lambda t} \frac{\log(1/\gamma)}{\gamma} \right) \\
        & = O\left( \frac{1}{\gamma \alpha \lambda t }  \log\left(\frac{1}{\gamma \alpha \lambda t} \right) \log\left(\frac 1 \gamma\right) \right) 
        = \O\left( \frac{1}{\gamma \alpha \lambda t} \right) 
        = \O\left( \frac{s}{\gamma \alpha \lambda} \right).
    \end{align}
    The total number of gates used is
    \begin{align}
        G_\mathrm{AA} + Q_\mathrm{AA}G_\mathrm{PE} + Q_\mathrm{AA}Q_\mathrm{PE}G_\mathrm{HS}
        & = O\left( \frac{G}{\gamma} + \frac{G_\mathrm{PE}}{\gamma} + Q_\mathrm{AA}Q_\mathrm{PE}Q_\mathrm{HS}\left(N+\log^{2.5}(1/\eps_\mathrm{\scriptscriptstyle HS})\right)\right)\\
         &= O\left( \frac{G}{\gamma} + \frac{\polylog(1/(\eps_\mathrm{\scriptscriptstyle PE} \delta_\mathrm{\scriptscriptstyle PE}))}{\gamma} + Q(N+\log^{2.5}Q)\right) \\
         &= \O \left(G/\gamma + \polylog(Q)/\gamma + QN\right).
    \end{align}
    Lastly, the quantum algorithm uses $O(N) + \O(\log(st/\eps_\mathrm{\scriptscriptstyle HS}))=O(N)+\O(\log Q)$ qubits for Hamiltonian simulation. Phase estimation uses an additional $O(\log(1/\eps_\mathrm{\scriptscriptstyle PE})+\log(1/\delta_\mathrm{\scriptscriptstyle PE})) = O(\log Q)$ qubits. Amplitude amplification uses $A+O(1)$ additional qubits to prepare and reflect about $\ket{\Psi}$.
\end{proof}

Note that we can also solve the Guided Sparse Hamiltonian problem without using Hamiltonian simulation as a subroutine by implementing a different function of the Hamiltonian other than $\exp(-\i Ht)$ for $t=\frac{\pi}{2s}$.
A natural choice would be $\exp(-\i \arcsin(Ht))$, which can be implemented using a quantum walk~\cite{Chi09}, and will lead to a slightly more efficient algorithm.
Indeed, many Hamiltonian simulation algorithms (such as \cite{BCK15}) use this quantum walk as a subroutine.

\subsection{State preparation preliminaries}
\label{sec:state_prep_prelim}

Let us start with the complexity of preparing an arbitrary quantum state on $N$ qubits with $S$ nonzero entries. 

\begin{lemma}[Sparse state preparation]\label{lem:sparsestate}
    There is an efficient algorithm that takes as input the description of an $N$-qubit state $\ket{\psi}$ with $S$ nonzero entries, and outputs a quantum circuit with $O(NS)$ gates that prepares this state starting from $\ket{0^N}$ using $O(1)$ ancilla qubits.
\end{lemma}

There are many state preparation methods that achieve this complexity. For concreteness, we can use the method of~\cite{MIC21}. We also provide a proof sketch for the reader's convenience.

\begin{proof}[Proof sketch]
    Let the set of nonzero entries of $\ket{\psi}$ be $\mathcal{S}=\{x_1,\ldots,x_S\}$, where each $x_i \in \{0,1\}^N$. Thus
    \begin{equation}
        \ket{\psi}=\sum_{x \in \mathcal{S}} \alpha(x)\ket{x} = \sum_{i \in  [S]} \alpha(x_i)\ket{x_i}.
    \end{equation}
    From this, define the state $\ket{\phi}$ on $\ceil{\log S}$ qubits as follows:
    \begin{equation}
        \ket{\phi} = \sum_{i \in  [S]} \alpha(x_i) \ket{i}.
    \end{equation}
    This is simply an arbitrary state on dimension $S$. It is known that we can create an arbitrary state on dimension $S$ using $O(S)$ CNOT gates and single qubit rotations~\cite{SBM06} with no ancillas. All that remains is to replace each $\ket{i}$ with $\ket{x_i}$. Let's add enough initial qubits to have each $\ket{i}$ be a state on $N$ qubits (instead of $\ceil{\log S}$ qubits). Additionally, let's add a final $\ket{0}$ to all the $\ket{i}$, and a final $\ket{1}$ to all the $\ket{x_i}$, so  that 
    $\mathrm{span} \{\ket{i}\ket{0}\}$ has no intersection with $\mathrm{span}\{\ket{x_i}\ket{1}\}$. Now all we need is a unitary that swaps the basis states $\ket{i}\ket{0}$ and $\ket{x_i}\ket{1}$, and leaves all the other basis states unchanged. Swapping two classical basis states can be done with $O(N)$ gates and $O(1)$ ancillas. Then we repeat this $S$ times for each $i$.
\end{proof}

Next we introduce a primitive that can be viewed as a probabilistic erasure of a register. It erases a register at the cost of decreased amplitude on the state of interest with the rest of the norm going to some garbage state.

\begin{lemma}\label{lem:registererasure}
    Suppose we are given a quantum state 
    \begin{equation}
        |\psi\rangle = \sum_{x \in \{0,1\}^n}\sum_{y\in\{0,1\}^m} \alpha_{xy} |x\rangle|y\rangle,
    \end{equation}
    and our goal is to prepare the unnormalized state 
    \begin{equation}
        |\widehat{\phi}\rangle = \sum_{x \in \{0,1\}^n}\sum_{y\in\{0,1\}^m} \alpha_{xy} |x\rangle.
    \end{equation} 
    Then the quantum circuit $\Id \otimes H^m$, which has $m$ gates, maps 
    \begin{equation}
    |\psi\rangle \mapsto \frac{1}{\sqrt{2^{m}}}|\widehat\phi\rangle|0^m\rangle + \ket{\perp},
    \end{equation}
    where $\ket{\perp}$ is an unnormalized quantum state that satisfies $(\Id \otimes |0^m\rangle\langle0^m|)\ket{\perp} = 0$. 
\end{lemma}
\begin{proof}
    Applying the Hadamard transform to the second register of $\ket{\psi}$, we get
    \begin{align}
        (\Id \otimes H^m) |\psi\rangle 
        &= \frac{1}{\sqrt{2^m}} \sum_{x \in \{0,1\}^n} \sum_{y\in\{0,1\}^m} \sum_{z\in\{0,1\}^m} \alpha_{xy} (-1)^{y\cdot z}|x\rangle|z\rangle         \\
        &= \frac{1}{\sqrt{2^m}} \sum_{x \in \{0,1\}^n}\sum_{y\in\{0,1\}^m} \alpha_{xy} |x\rangle|0^m\rangle + \ket{\perp},\\
        &= \frac{1}{\sqrt{2^m}} \ket{\widehat\phi}\ket{0^m} + \ket{\perp},
    \end{align}
    where $\ket{\perp}$ is an unnormalized quantum state that satisfies $(\Id \otimes |0^m\rangle\langle0^m|)\ket{\perp} = 0$. 
\end{proof}

Note that if we wanted to prepare the normalized version of $\ket{\widehat\phi}$, we could measure the second register and if we see $0^m$, then the first register is the state we want. The probability of seeing $0^m$ is $\langle\widehat\phi|\widehat\phi\rangle/2^m$ and hence by repeating this $O(2^m/\langle\widehat\phi|\widehat\phi\rangle)$ times we will get one copy of the state with high probability.

We now show how we can unitarily map a sorted list and some permutation of that list, to the same sorted list but with only the permutation stored in another register.

\begin{lemma}\label{lem:permutationlabel}
    For any $n \geq \ell \geq 1$, let $x_1<x_2<\ldots<x_{\ell} \in [n]$ be $\ell$ sorted numbers, and let $\pi:[\ell]\to[\ell]$ be a permutation. Then there is a quantum circuit with $O(\ell^2 \log n)$ gates that performs the map
    \begin{equation}
        \ket{x_1,\ldots,x_{\ell}}\ket{x_{\pi(1)},\ldots,x_{\pi(\ell)}} \mapsto 
        \ket{x_1,\ldots,x_{\ell}}\ket{\pi(1),\ldots,\pi(\ell)}.
    \end{equation}
\end{lemma}
\begin{proof}
    We prove this by observing there is a classical algorithm with cost $O(\ell^2 \log n)$ that can compute  $(\pi(1),\ldots,\pi(\ell))$ given $(x_1,\ldots,x_{\ell})$ and $(\ket{x_{\pi(1)},\ldots,x_{\pi(\ell)}})$, and another classical algorithm that can compute $(\ket{x_{\pi(1)},\ldots,x_{\pi(\ell)}})$ given $(x_1,\ldots,x_{\ell})$ and $(\pi(1),\ldots,\pi(\ell))$. 
    
    The first classical algorithm starts with $x_{\pi(1)}$ and searches for its location in the sorted list, and once found, writes this value $\pi(1)$ down. This operation require $\ell$ comparisons of numbers of size $[n]$, giving total cost $O(\ell \log n)$. Since this has to be done $\ell$ times, we get a total cost of $O(\ell^2 \log n)$. The second classical algorithm is similar.
\end{proof}

Given these lemmas we can now establish the following, which essentially replaces a string of distinct elements stored in a register with its sorted version.

\begin{lemma}\label{lem:guide_sort}
    Let $n \geq \ell \geq 1$ and let $X$ be the set of tuples $(x_1,\ldots,x_\ell)\in[n]^\ell$ with all distinct elements. Let
    \begin{equation}
        \ket{\psi} = \sum_{(x_1,\ldots,x_\ell)\in X} \alpha(x_1,\ldots,x_\ell) \ket{x_1,\ldots,x_\ell}.
    \end{equation}
    For any $(x_1,\ldots,x_\ell) \in X$, let $\mathrm{sort}(x_1,\ldots,x_\ell)$ be the string $(y_1,\ldots,y_\ell) \in X$ such that $y$ is the sorted version of $(x_1,\ldots,x_\ell)$, i.e., that $\{x_1,\ldots,x_\ell\} = \{y_1,\ldots,y_\ell\}$ and $y_1<\cdots<y_\ell$. Then let $\ket{\widehat \phi}$ be the unnormalized state obtained by replacing each basis state in in $\ket{\psi}$ with its sorted version:
    \begin{equation}
        \ket{\widehat \phi} = \sum_{(y_1,\ldots,y_\ell)\in X} \sum_{\substack{(x_1,\ldots,x_\ell): \\\mathrm{sort}(x_1,\ldots,x_\ell) = (y_1,\ldots,y_\ell)}} \alpha(x_1,\ldots,x_\ell) |y_1,\ldots,y_\ell\rangle.
    \end{equation}
    Then there is a quantum circuit with $O(\ell^2 \log n)$ gates that maps any state $\ket{\psi}$ to
    \begin{equation}
        \frac{1}{\ell^{\ell/2}} \ket{\widehat \phi}\ket{0^{\ell \log \ell}}\ket{0} + \ket{\perp}\ket{1},
    \end{equation}
    where $\ket{\perp}$ is an unnormalized state.
\end{lemma}

\begin{proof}
    First, we can use any optimal comparison-based sorting algorithm to map $\ket{\psi}$ to 
    \begin{equation}
        \sum_{(x_1,\ldots,x_\ell)\in X} \alpha(x_1,\ldots,x_\ell) \ket{x_1,\ldots,x_\ell} \ket{\mathrm{sort}(x_1,\ldots,x_\ell)},
    \end{equation}
    using $O(\ell \log \ell \log n)$ gates. We can rewrite this state as
    \begin{equation}
        \sum_{(y_1,\ldots,y_\ell)\in X} \sum_{\substack{(x_1,\ldots,x_\ell): \\\mathrm{sort}(x_1,\ldots,x_\ell) = (y_1,\ldots,y_\ell)}} \alpha(x_1,\ldots,x_\ell) \ket{x_1,\ldots,x_\ell}|y_1,\ldots,y_\ell\rangle.
    \end{equation}
    Now $(y_1,\ldots,y_\ell)$ is a sorted list, and $(x_1,\ldots,x_\ell)$ is some permutation of it, so let $\pi_{xy}:[\ell]\to[\ell]$ denote this permutation. Using \Cref{lem:permutationlabel}, there is a circuit of $O(\ell^2 \log n)$ gates that maps this to
    \begin{equation}
        \sum_{(y_1,\ldots,y_\ell)\in X} \sum_{\substack{(x_1,\ldots,x_\ell): \\\mathrm{sort}(x_1,\ldots,x_\ell) = (y_1,\ldots,y_\ell)}} \alpha(x_1,\ldots,x_\ell) \ket{\pi_{xy}(1),\ldots,\pi_{xy}(\ell)}|y_1,\ldots,y_\ell\rangle.
    \end{equation}    
    The first register now has dimension $\ell^\ell$. Now using \Cref{lem:registererasure} with $m=\ell \log \ell$, we can apply a unitary with $O(\ell \log \ell)$ gates that maps this to 
    \begin{equation}
        \frac{1}{\ell^{\ell/2}} \ket{\widehat \phi}\ket{0^{\ell \log \ell}} + \ket{\perp},
    \end{equation}
    where $\ket{\perp}$ is an unnormalized state with  $\langle0^{\ell \log \ell}\ket{\perp} = 0$. With an additional $O(\ell \log \ell)$ gates we can compute the logical OR of the last $\ell \log \ell$ bits into a new register, which gives us
    \begin{equation}
        \frac{1}{\ell^{\ell/2}} \ket{\widehat \phi}\ket{0^{\ell \log \ell}}\ket{0} + \ket{\perp}\ket{1},
    \end{equation}
    since $\langle0^{\ell \log \ell}\ket{\perp} = 0$.
\end{proof}

Finally, we can also project a state into the space with all distinct elements quite efficiently:

\begin{lemma}\label{lem:guide_hom}
    Let $n \geq \ell \geq 1$ and let
    \begin{equation}
        \ket{\psi} = \sum_{(x_1,\ldots,x_\ell)\in [n]^\ell} \alpha(x_1,\ldots,x_\ell) \ket{x_1,\ldots,x_\ell}.
    \end{equation}
    Let $X \subseteq [n]^\ell$ be the set of $\ell$-tuples that are all distinct and let $\ket{\psi_\mathrm{d}}$ be the projection of $\ket{\psi}$ into basis states in $X$:
    \begin{equation}
        \ket{\psi_\mathrm{d}} = \frac{1}{\Psi_\mathrm{d}} \sum_{(x_1,\ldots,x_\ell) \in X} \alpha(x_1,\ldots,x_\ell) |x_1,\ldots,x_\ell\rangle,
    \end{equation}
    where $\Psi_\mathrm{d}$ is the normalization constant that makes this state have norm $1$.
    Then there is a quantum circuit with $O(\ell \log \ell \log n)$ gates that maps any state $\ket{\psi}$ to
    \begin{equation}
        \Psi_\mathrm{d} \ket{\psi_\mathrm{d}}\ket{0} + \ket{\perp}\ket{1},
    \end{equation}
    where $\ket{\perp}$ is an unnormalized state.
\end{lemma}
\begin{proof}
    We can use any optimal comparison-based sorting algorithm to write down a sorted version of $x_1,\ldots,x_\ell$ in a second register using $O(\ell \log \ell \log n)$ gates. 
    Then we can check if any adjacent pair of elements is equal and store these bits in a third register of size $\ell-1$ with cost $O(\ell \log n)$. Finally, we can compute the logical OR of these $\ell-1$ bits in a fourth register at cost $O(\ell)$. We can then uncompute registers 2 and 3. Now register 4 contains 0 if and only if the string in register 1 has all distinct elements.
\end{proof}

\subsection{Algorithm for Planted Noisy \texorpdfstring{$k$}{k}XOR}\label{sec:quantumalgkXOR}

To implement the quantum algorithm, we need to represent the abstract vector space that the Kikuchi matrix acts on concretely. The Kikuchi matrix acts on a space of  dimension $\binom{n}{\ell}$ spanned by states $\ket{S}$, where $S \in \binom{[n]}{\ell}$. For our implementation, we work with $\ell$ qudits each of local dimension $n$ (which can further be represented as qubits in the standard way). We represent the basis state $\ket{S}$ by $\ket{s_1,\ldots,s_\ell}$, where $s_1 < \cdots < s_\ell$ is the sorted list of the elements of $S$. In other words, although the space of $\ell$ qudits has dimension $n^\ell$, we only use $\binom{n}{\ell}$ basis states of this space corresponding to $\ell$ distinct numbers sorted in ascending order.

\subsubsection{Preparing the guiding state}\label{sec:guidingstateprep}

First let us recall the setup.
Our quantum algorithm (discussed in \Cref{sec:quantum-analysis}) will draw a $k$XOR instance $\bcalI \coloneqq \bcalI_{\mathrm{guide}} \sim \widetilde{\calP}^{z}_{n,k}(\tilde m, \rho)$ for $\tilde m = \zeta \hat{m}$.
Throughout this section we will assume $\ell^2 \log n \ll n \leq \tilde m \leq O(n^{k/2})$, as these are the only parameter settings under which our algorithm is used
.
We may then consider (see \Cref{eg:skelly}) the degree-$k$ polynomial $\bcalA$ whose $S$-coefficient is $\frac{1}{\sqrt{\tilde m}} B_{\bcalI}(S)$, where $\tilde q = \tilde m/\binom{n}{k}$ and where the random variables $B_{\bcalI}(S)$ are independent, with $B_{\bcalI}(S) x^S$ distributed as $\mathrm{Skel}((\half + \half \rho)\tilde q, (\half - \half \rho)\tilde q)$.
If we were in the simplest case of $\ell = k$ (i.e., $c = 1$), the guiding state we wish to prepare would be the one parallel to 
\begin{equation} \label{eqn:babyguide}
    \ket{\Gamma^k(\bcalA)} 
        = \frac{1}{\sqrt{\tilde m}} \sum_{S \in \binom{[n]}{k}} B_{\bcalI}(S)  \ket{S}. 
\end{equation}
For general $\ell = ck$, we wish to prepare the following unit-length guiding state:
\begin{equation}\label{eq:guidingstate}
    \ket{\bPsi} \propto \ket{\Gamma^\ell(\bcalA)} \propto 
        \sum_{T \in \binom{[n]}{\ell}} \sum_{\{S_1, \dots, S_c\} \in \Part_k(T)}  \parens*{\prod_{j=1}^c B_{\bcalI}(S) } \ket{T}.
\end{equation}

We now show:
\begin{theorem}\label{thm:guidingstate}
    There is an efficient algorithm that takes as input~$\bcalI$ and (except with probability at most $O(\tilde{m}^{-1/2})$ over the outcome of~$\bcalI$) produces  a quantum circuit with $O(\ell \tilde m \log n)$ gates that prepares the state
    \begin{equation}
    \beta \ket{\bPsi} \ket{0^{\ell \log \ell+3}} + \ket{\perp}\ket{1}, 
\end{equation}
where $\beta \geq \Omega(1/ \ell^{\ell/2})$.
\end{theorem}
\newcommand{\Bad}{\textbf{\textsc{Bad}}}
\newcommand{\Good}{\textbf{\textsc{Good}}}
\begin{proof}
    We first argue that with high probability, the vector in \Cref{eqn:babyguide} is ``almost'' a normalized state with all nonzero amplitudes $\pm 1/\sqrt{\tilde m}$.
    Let us define $\Bad = \{S : \abs{B_{\bcalI}(S)} \geq 2\}$.
    Then
    \begin{equation}
        \E\bracks*{\sum_{S \in \Bad} B_{\bcalI}(S)^2} = \binom{n}{k} \sum_{h= 2}^{\infty} \Pr[\abs{\bH} = h] \cdot h^2,
    \end{equation}
    where $\bH$ denotes a single $\mathrm{Skel}((\half + \half \rho)\tilde q, (\half - \half \rho)\tilde q)$ random variable.  Thinking of $\bH$ as a difference of independent Poisson random variables, the event $\abs{\bH} = h$ implies that at least one of these $\mathrm{Poi}((\half \pm \half \rho) \tilde q)$ random variables is at least~$h$; i.e., 
    \begin{equation}
        \Pr[\abs{\bH} = h] \leq \Pr[\mathrm{Poi}((\half + \half \rho) \tilde q) \geq h] + \Pr[\mathrm{Poi}((\half - \half \rho) \tilde q) \geq h] \leq 2 \Pr[\mathrm{Poi}(\tilde q) \geq h].
    \end{equation}
    Recalling that $\tilde q \leq O(n^{-k/2}) \ll 1$, we have for $h \geq 2$ that $\Pr[\mathrm{Poi}(\tilde q) \geq h] \leq O(\Pr[\mathrm{Poi}(\tilde q) = h]) \leq O(\tilde{q}^h)$. 
    Thus
    \begin{equation}
        \E\bracks*{\sum_{S \in \Bad} B_{\bcalI}(S)^2} \leq \binom{n}{k} \sum_{h= 2}^{\infty} O(\tilde{q}^h) \cdot h^2 \leq \binom{n}{k} 
 \cdot O(\tilde{q}^2) = O(\tilde{m}^2 / \tbinom{n}{k}) \leq O(1),
    \end{equation}
    using our assumption $\tilde{m} \leq O(n^{k/2})$.
    Thus by Markov's inequality, except with probability at most~$\tilde{m}^{-1/2}$ (say) we have
    \begin{equation} \label[ineq]{ineq:bbb1}
        \sum_{S \in \Bad} B_{\bcalI}(S)^2 \leq O(\sqrt{\tilde{m}}).
    \end{equation}
    On the other hand, it is easy to confirm that for $\Good \coloneqq \{S : \abs{B_{\bcalI}(S)} = 1\}$ and $\ol{\bm} \coloneqq \abs{\Good}$,
    \begin{equation}    \label[ineq]{ineq:bbb2}
        \tilde m / 2 \leq \ol{\bm} \leq 2 \tilde m
    \end{equation}
    except with probability exponentially small in~$\tilde m$. 
    These failures only total $O(\tilde{m}^{-1/2})$ in probability, and we will henceforth assume that \Cref{ineq:bbb1,ineq:bbb2} hold.
    
    Let us write $\ket{\bphi}$ for the unit vector in the direction of $\ket{\Gamma^k(\bcalA)}$, so we have
    \begin{equation}    \label{eqn:phigood}
        \ket{\bphi} = \kappa \sum_{S \in \Good} B_{\bcalI}(S) \ket{S} + \kappa\sum_{S \in \Bad} B_{\bcalI}(S) \ket{S} 
    \end{equation}
    for some constant~$\kappa$.
    By \Cref{lem:sparsestate}, there is an efficient algorithm that, given the input, produces a quantum circuit of $O(\tilde m \cdot \log n)$ gates that prepares~$\ket{\bphi}$.
    Moreover, if we reexpress \Cref{eqn:phigood} as
    \begin{equation}
        \ket{\bphi} = \sqrt{1-\eps} \ket{\bphi_{\text{good}}} + \sqrt{\eps} \ket{\bphi_{\text{bad}}}, 
    \end{equation}
    then \Cref{ineq:bbb1,ineq:bbb2} imply $\eps \leq O(1/\tilde{m})$.  

    The next step (recalling $c = \ell/k$) is to define
    \begin{equation}    \label{eqn:yoyoyo}
        \ket{\bPhi} = \ket{\bphi}^{\otimes c} \propto \sum_{S_1, \dots, S_c} B_{\bcalI}(S_1) \cdots B_{\bcalI}(S_c) \ket{S_1} \cdots \ket{S_c}.
    \end{equation}
    Obviously --- and crucially --- we can also produce a quantum circuit of $c \cdot O(\tilde{m} \log n)$ gates that prepares $\ket{\bPhi}$, simply by repeating $c$ times the circuit for $\ket{\bphi}$.

    In preparation for using \Cref{lem:guide_hom}, we'd like to estimate the fractional $\ell^2$-mass of coefficients $B_{\bcalI}(S_1) \cdots B_{\bcalI}(S_c)$ in \Cref{eqn:yoyoyo} on kets $\ket{S_1} \cdots \ket{S_c}$ with $S_1, \dots, S_c$ disjoint.
    Let us call this fractional $\ell^2$-mass $f \in [0,1]$.
    The first step is to observe that the fractional $\ell^2$-mass with $S_1, \dots, S_c$ all falling in $\Good$ is $(1-\eps)^c$.  
    Thus $f \geq (1-\eps)^c f_1 \geq (1-O(\ell/\tilde{m})) f_1$, where $f_1$ is the fraction of tuples $(S_1, \dots, S_c) \in \Good^c$ with $S_1, \dots, S_c$ disjoint (note that $(B_{\bcalI}(S_1)\cdots B_{\bcalI}(S_c))^2 = 1$ for every tuple in $\Good^c$).
    
    Note that if we condition on $\overline{\bm} = m$ for any~$\tilde{m}/2 \leq m \leq 2\tilde{m}$, then $\Good$ is simply distributed as $m$ random sets drawn independently from $\binom{[n]}{k}$ without replacement. 
    It now follows from \Cref{lem:alemma} (below) that $f_1 \geq 1-O(\ell^2\frac{\log n}{n})$; hence $f \geq 1-O(\ell^2\frac{\log n}{n}) \geq .99$.

    We may now apply \Cref{lem:guide_hom}. The resulting additional circuit complexity is negligible compared to the $O(\tilde m \cdot \log n)$ gates we have so far; we thus get a circuit that prepares a state proportional to
    \begin{multline}\label{eqn:hey}
    \sum_{\substack{S_1,\ldots,S_c \in \binom{[n]}{k} \\ \text{all~$S_j$~disjoint}}}  \left(\prod_{j=1}^c B_{\calI}(S_j) \right) \ket{S_1}\cdots\ket{S_c} \ket{0} + \sum_{\substack{S_1,\ldots,S_c \in \binom{[n]}{k} \\ \text{not all~$S_j$~disjoint}}}  \left(\prod_{j=1}^c B_{\calI}(S_j) \right) \ket{S_1}\cdots\ket{S_c} \ket{1} \\
    \eqqcolon \ket{\wt{\bPhi}_{\text{good}}} + \ket{\wt{\bPhi}_{\text{bad}}}, \qquad \text{where }
    \braket{\wt{\bPhi}_{\text{good}}}{\wt{\bPhi}_{\text{good}}} \geq 99 \braket{\wt{\bPhi}_{\text{bad}}}{\wt{\bPhi}_{\text{bad}}}.
    \end{multline}
    Note that 
    \begin{equation}
        \ket{\wt{\bPhi}_{\text{good}}} = \sum_{T \in \binom{[n]}{\ell}} \sum_{\{S_1, \dots, S_c\} \in \Part_k(T)}  \left(\prod_{j=1}^c B_{\calI}(S_j) \right) \ket{S_1}\cdots\ket{S_c},
    \end{equation}
    which is \emph{almost} parallel to the desired guiding state~$\ket{\bPsi}$, except that we would like to replace  $\ket{S_1}\cdots\ket{S_c}$ with its sorted version $\ket{T}$. We can now apply the unitary from \Cref{lem:guide_sort} to~$\ket{\bPsi}$ (on the first $\ell$ qudits; i.e., not acting on the final qubit), which uses $O(\ell^2 \log n) \leq O(\ell \tilde{m} \log n)$ gates to map the state in \Cref{eqn:hey} to a state of the form
\begin{equation}
    \beta \ket{\bPsi}\ket{0^{\ell \log \ell}}\ket{0}\ket{0} +\ket{\perp_2}\ket{1}\ket{0} + \ket{\perp_3}\ket{1}, 
\end{equation}
where $\beta \geq.99/\ell^{\ell/2}$ and $\ket{\perp_2}$ and $\ket{\perp_2}$ are unnormalized states. Computing the logical or of the final~2 qubits into a new final qubit completes the proof.
\end{proof}

\begin{lemma}   \label{lem:alemma}
    Let $\bS_1, \dots, \bS_m$ be drawn independently, without replacement, from $\binom{[n]}{k}$.  
    Let $\bj_1, \dots, \bj_c$ be drawn independently, with replacement, from~$[m]$.  
    Then
    \begin{equation}
        \Pr[\bS_{\bj_1}, \dots, \bS_{\bj_c} \text{ disjoint}] \geq 1 - O\parens*{\ell^2 \frac{\log n}{n}}.
    \end{equation}
\end{lemma}
\begin{proof}
    Thinking of $\bS_1, \dots, \bS_m$ as forming a $k$-uniform $m$-hyperedge hypergraph $\bcalH$ on~$[n]$, the expected degree of any vertex is~$km/n$, and it is well known that except with probability at most~$1/n$ we have $\text{maxdegree}(\bcalH) \leq O(\frac{m\log n}{n})$. 
    Assuming this occurs, each hyperedge in $\bcalH$ touches at most $k \cdot O(\frac{m\log n}{n})$ other hyperedges, so the probability that two hyperedges selected at random (with replacement) overlap is at most
    \begin{equation}
        \frac{k \cdot O(\tfrac{m\log n}{n})}{m} = O\parens*{\frac{\log n}{n}}.
    \end{equation}
    Thus the probability that, among $c$ randomly selected edges, any two overlap is at most $\binom{c}{2} \cdot O\parens*{\frac{\log n}{n}} = O\parens*{\ell^2 \frac{\log n}{n}}$, as desired.
\end{proof}

\subsubsection{Simulating the Kikuchi Hamiltonian}\label{sec:KikuchiHamSim}

To simulate the Hamiltonian corresponding to the Kikuchi matrix for a $k$XOR instance $\calI$, all we need is to show that the quantum oracles $O_H$ and $O_F$ can be efficiently implemented.

\begin{theorem}\label{thm:Hamiltonianoracle}
     Let $\bcalI \sim \wt{\calP}^z_{n,k}(m,\rho)$. Then the corresponding Kikuchi matrix $\calK_\ell(\bcalI)$ has sparsity $O(\ell \log n)$ and number of constraints $\bm=O(m)$, except with probability at most $n^{-\Omega(\ell)}$. When this holds, the quantum oracles $O_H$ (\Cref{def:OH}) and $O_F$ (\Cref{def:OF}) can be implemented with $\O(m \ell \log n)$ gates and $O(\ell \log n)$ qubits.
\end{theorem}

\begin{proof}
For $\bcalI \sim \wt{\calP}^z_{n,k}(m,\rho)$, we know the average degree of the Kikuchi matrix (\Cref{def:Kikuchi_matching}) is $\delta_{\ell,n,k}m$. Using \Cref{thm:alice}, this is at most $O(\ell \log n)$. Even if we use a hypothetical improved version of  \Cref{thm:alice}, this can only decrease the average degree. Furthermore, from \Cref{prop:md}, we know that the maximum degree, or sparsity, of this matrix is $O(\ell \log n)$ except with probability at most $n^{-\Omega(\ell)}$. The number of constraints $\bm$ is drawn from $\mathrm{Poi}(m)$ and is $O(m)$, except with probability exponentially small in $m$, which is smaller than $n^{-\Omega(\ell)}$.

For convenience, let the number of distinct constraints in our instance be $\bar{m} = O(m)$ and let these constraints correspond to the scopes $U_j$ for $j \in [\bar{m}]$ and have right hand sides $B_{\bcalI}(U_j)$. Then there is a circuit of $\O(m \log n)$ gates that outputs $U_j$ on input $j$.\footnote{This can be constructed by brute force enumeration over $j$. For example, we start with $\ket{j,0\ldots 0}$ and our goal is for the circuit to output $\ket{j,U_j}$. One way to do this is to check if $j=1$, and conditioned on that, write $U_1$ to the second register, and so on for each $j$. The circuits to check this condition and write some fixed string to another register are efficient and of size ${\textstyle\widetilde{O}}(\log m \log n)$ and will be repeated $m$ times, yielding the stated bound.} Similarly, there is a circuit of ${\textstyle\widetilde{O}}(m \log n)$ gates that outputs $B_{\bcalI}(U_j)$ on input $U_j$.

We can upper bound the complexity of the oracle $O_H$ by the classical complexity of computing the function that computes a particular entry of the Kikuchi matrix using \Cref{fact:classicaloracles}. Specifically, for $S,T \in \binom{[n]}{\ell}$, we need to upper bound the complexity of the map that accepts as input sorted lists (of size $O(\ell \log n)$) corresponding to $S$ and $T$ and outputs $B_{\bcalI}(S\Delta T)$. The symmetric difference $S\Delta T$ can be computed using $\O(\ell \log n)$ gates. Then we can use the circuit of $\O(m \log n)$ gates that computes $B_{\bcalI}(S\Delta T)$ given $S \Delta T$. This uses $\O(m\log n)$ gates and $O(\ell \log n)$ space.

For the oracle $O_F$, we need to choose a function $f$ such that $f(i,k)$ is the $k$th  nonzero entry of the Kikuchi matrix in row $i$ for some ordering of the nonzero entries. 
A row $i$ of the matrix represents a set $S$. 
For any row set $S$, the entry $T = S \Delta U_j$ is a potential nonzero entry if $|S \Delta U_j|=\ell$, and by assumption there are only $O(\ell \log n)$ such entries. We'll say the $k$th nonzero entry of row $S$ is the $k$th smallest index $j$ for which $|S \Delta U_j|=\ell$.
As before, to use \Cref{fact:classicaloracles}, we  need to compute the classical gate complexities of computing $(i,k) \mapsto f(i,k)$ and $(i,f(i,k)) \mapsto k$. 

Consider the map $(i,k) \mapsto f(i,k)$, or $(S,k)\mapsto f(S,k)$ since each row of the Hamiltonian is indexed by a set $S$. To compute the index $j$ of the $k$th smallest $j$ for which $|S \Delta U_j|=\ell$, we will loop through all the $U_j$ in order, and check if a given $U_j$ satisfies $|S \Delta U_j|=\ell$, and increment a counter each time it does. When this counter reaches $k$, we have found the $k$th nonzero entry in row $S$, $S \Delta U_j$ for the current $j$, and can copy this information to another register. If the counter never reaches $k$, then we return $(1,k)$ as the index of the $k$th nonzero entry in row $S$ as described in \Cref{def:OF}.

More precisely, in addition to registers storing $S$ and $k$, we will have a register storing the loop index $j$, a register storing $U_j$, a counter that holds a number between $0$ and $s$, and an answer register. These registers only use $O(\ell \log n)$ qubits. It remains to show that we can perform the algorithm using $\O(m\ell \log n)$ gates.
Let's say we begin with the loop index register storing $j=1$ and the scope register storing $U_1$. Initializing these registers with these values only uses $O(\ell \log n)$ gates starting from all zeros, since we are simply negating some of the zeros to ones.
Given $U_1$ and $S$, it is easy to compute if $|S \Delta U_1|=\ell$, increment the counter if needed, and uncompute this information, in total using $\O(\ell \log n)$ gates. 
It remains to show how we update the index register from $j=1$ to $j=2$ and from $U_1$ to $U_2$. Since we know we want to go from $j=1$ to $j=2$, we can simply negate exactly those bits in the binary representation of $1$ that make it $2$. Similarly, since we know we want to replace $U_1$ with $U_2$, and these are known strings of length $k \log n$, this uses at most $k \log n$ negation gates to flip the bits that need to be flipped. Thus one step of this loop can be executed with $\O(\ell \log n)$ gates, and since we loop over all scopes, we do this $O(m)$ times, leading to an overall gate complexity of $\O(m\ell \log n)$.

The other function $(i,f(i,k)) \mapsto k$ has the same complexity since we can run essentially the same algorithm. Here we are given a set $S$ and $T$, from which we can compute $U=S \Delta T$, and our goal is to output $k$ for which $T$ is the $k$th nonzero entry in row $S$. We loop through the scopes $U_j$ again, keeping a count of the number of scopes with $|S \Delta U_j|=\ell$ we have seen until we reach $U$ and output the count.
\end{proof}

\subsubsection{Putting it all together}
\label{sec:putting_all_together}

We are now in a position to combine the previous results to arrive at a quantum algorithm for the Planted Noisy $k$XOR problem. Throughout this paper, we have taken great care to allow our quantum algorithm to work with a very general ``Alice Theorem''. This is because it is possible that \Cref{thm:alice}, which achieves the best trade-off between constraint density $\Delta$ and $\ell$ currently known to us, could be improved in the future. For instance, the $\ln{n}$ term in right hand side of \Cref{ineq:C} could be an artifact of the matrix Chernoff bound. 
Since making a rigorous claim that applies to any possible Alice Theorem would require an exceedingly complex theorem statement, we introduce a mild continuity assumption on $C_\kappa$. 
Specifically, we require that the Alice Theorem is ``reasonable'' in the sense that, for all other parameters fixed, a constant relative change in $C_\kappa$ corresponds to a constant relative change of $\kappa$. Note that this is satisfied whenever $C_\kappa$ depends polynomially on $\kappa$, such as in \Cref{thm:alice} and any other Alice Theorem known to us. After stating our theorem, we discuss other forms of reasonable Alice Theorems, and how our quartic quantum speedup applies to them as well. 

We now implement the quantum algorithm implicit in \Cref{thm:kXOR_analysis} for a general ``reasonable'' Alice Theorem and by choosing some specific constants to make the theorem more readable.

\begin{theorem}
    Let $k$ (an even number) and $\rho \in (0,1)$ be constants. Let $\hat{\bcalI}$ be drawn from $\wt{\calP}^z_{n,k}(m,\rho)$ or $\wt{\calR}_{n,k}(m)$. 
    Suppose we have a ``reasonable'' Alice Theorem (e.g., \Cref{thm:alice}) that guarantees that when 
    $\bcalI \sim \wt{\calR}_{n,k}(m)$, 
    if $\Delta =m/n \geq C_\kappa (n/\ell)^{(k-2)/2}$, then 
    with probability $1-o(1)$, we have $\lambdamax{\calK_{\ell}(\bcalI)}\leq \kappa d$. 
    Let $\ell = ck = O(\sqrt{n})$ for $c \in \N^+$ be chosen such that this holds with $\kappa = 0.99 \rho$.

    Then there is a quantum algorithm that uses $\O(n^{\ell/4} m \ell^{O(\ell)}\log^{\ell/2k} n)$ gates to solve the Planted Noisy $k$XOR Problem (\Cref{prob:planted_noisy_kXOR}).
\end{theorem}
\begin{proof}
    We follow the strategy (and notation) laid out in \Cref{thm:kXOR_analysis}. We partition $\hat{\bcalI}$ randomly into $\bcalI \sqcup \bcalI_{\textnormal{guide}}$ by independently placing each constraint into $\bcalI$ with probability $1-\zeta$ and into $\bcalI_{\textnormal{guide}}$ with probability $\zeta$. We choose $\zeta = 1/\ln n$ (see e.g. discussion at the end of \Cref{sec:tensor_guide}). 
    
    Consider $\hat{\bcalI}$ drawn from $\wt{\calP}^z_{n,k}(m,\rho)$. Then if $\bcalB = q^{-1/2} \cdot \bcalI_{\textnormal{guide}}$, where $q = \zeta m/\binom{n}{k}$, we have with probability $1-o(1)$ \begin{equation}  
        \braket{\bv}{\Gamma^\ell(\bcalB)}^2 \geq \Part_k(\ell) \cdot \frac{\rho\eps}{196 \ell \ln n} \cdot (\rho^2 q)^{c} .
    \end{equation}
    By choosing $\eps=0.005$ and simplifying, we get that this is at least 
    $\widetilde{\Omega}(\ell^{c-\ell/2}\zeta^c C_{\kappa}^c/n^{\ell/2})$. From \Cref{thm:guidingstate} (and setting $\nu = 1/\ln n$ in \Cref{thm:kXOR_analysis}), using $O(m\ell^2\log n)$ gates we can prepare a state $\ket{\Psi}$ that has overlap  
    $\gamma^2 = \widetilde{\Omega}(\zeta^c C_{\kappa}^c/(n^{\ell/2}\ell^{2\ell}))$ with the cutoff eigenspace of $\Kik_\ell(\bcalI)$ with eigenvalues larger than $0.995\rho d$. We can now use \Cref{thm:guidedsparse} to decide if the Kikuchi matrix has an eigenvalue in this cutoff space or not. The cost of implementing the Hamiltonian oracles is $\O(m \ell \log n)$  by \Cref{thm:Hamiltonianoracle}, and the sparsity of the Hamiltonian is $O(\ell \log n)$. Applying \Cref{thm:guidedsparse}, we get a quantum algorithm that with probability $1-o(1)$ (over $\wt{\calP}^z_{n,k}(m,\rho)$ and the internal randomness of the algorithm) outputs ``planted'', and has gate complexity 
    $\O\left(m \left( n^{\ell/2}\ell^{O(\ell)}/\zeta^c C_\kappa^c\right)^{{1}/{2}}\right)$, 
    which is $\O\left(n^{\ell/4} m \ell^{O(\ell)}\zeta^{\ell/2k}\right)$.\footnote{Here we used the fact that $C_\kappa = \Omega(1/\ell)$ as discussed after \Cref{thm:alice}.} 

    Now consider $\hat{\bcalI}$ be drawn from the random distribution $\wt{\calR}_{n,k}(m)$. By Poisson-splitting $\bcalI$ is drawn from $\wt{\calR}_{n,k}(m(1-\zeta))$. Invoking the Alice Theorem yields 
    \begin{equation}
        \lambdamax{\calK_{\ell}(\bcalI)}\leq \kappa' d, 
    \end{equation} 
    with probability $1-o(1)$, where $\kappa'$ is such that $C_{\kappa'} \leq (1-\zeta)C_{\kappa}$. By our assumption that the Alice theorem is ``reasonable'', $\kappa' = \kappa (1 + \text{const} \cdot \zeta) < \kappa + \eps$ for large enough $n$. Hence the cutoff eigenspace is empty with high probability, and the quantum algorithm described for $\hat{\bcalI} \sim \wt{\calP}^z_{n,k}(m,\rho)$ outputs ``random'' with probability $1-o(1)$ for $\hat{\bcalI} \sim \wt{\calR}_{n,k}(m)$.
\end{proof}

The necessity for a reasonable Alice Theorem arises from a technical subtlety. The quantum algorithm described above considers a Kikuchi matrix with slightly fewer constraints (i.e.,  $m \mapsto (1-\zeta)m$). For a uniformly random instance, we require that this small change in $m$ increases the upper bound on $\lambdamax{\calK_{\ell}(\bcalI)}$ given by the Alice Theorem only by a small amount, in agreement with the expected behaviour of $\lambdamax{\calK_{\ell}(\bcalI)}$. 
If the continuity condition on the function $C_\kappa \mapsto \kappa$ holds, this change can be accounted for via the gap between the upper bound given by the Alice Theorem and the cutoff threshold $\kappa d$. Alternatively, if the Alice Theorem is such that 
the right-hand side of \Cref{ineq:C} is an inverse polynomial in $\ell$, 
one could account for this increase by running the quantum algorithm with a slightly larger value of $\ell$, e.g., $\ell ' = (1+\zeta) \ell$, to recover the original upper bound $\lambdamax{\calK_{\ell}(\bcalI)}\leq \kappa d$. This approach introduces a negligible overhead of, e.g., $n^{O(\ell \zeta)} = 2^{O(\ell)}$, and thus also achieves a quartic speedup.\footnote{This approach has the additional advantage that it does not depend on the size of the gap between $\kappa d$ and the cutoff threshold, which we have chosen to be constant here for ease of exposition, but could be, for instance, inverse polynomial.} Note that our \Cref{thm:alice} satisfies either of these ``reasonableness'' assumptions.

\bibliography{odonnell-bib}

\begin{thebibliography}{69}%
\makeatletter
\providecommand \@ifxundefined [1]{%
 \@ifx{#1\undefined}
}%
\providecommand \@ifnum [1]{%
 \ifnum #1\expandafter \@firstoftwo
 \else \expandafter \@secondoftwo
 \fi
}%
\providecommand \@ifx [1]{%
 \ifx #1\expandafter \@firstoftwo
 \else \expandafter \@secondoftwo
 \fi
}%
\providecommand \natexlab [1]{#1}%
\providecommand \enquote  [1]{``#1''}%
\providecommand \bibnamefont  [1]{#1}%
\providecommand \bibfnamefont [1]{#1}%
\providecommand \citenamefont [1]{#1}%
\providecommand \href@noop [0]{\@secondoftwo}%
\providecommand \href [0]{\begingroup \@sanitize@url \@href}%
\providecommand \@href[1]{\@@startlink{#1}\@@href}%
\providecommand \@@href[1]{\endgroup#1\@@endlink}%
\providecommand \@sanitize@url [0]{\catcode `\\12\catcode `\$12\catcode `\&12\catcode `\#12\catcode `\^12\catcode `\_12\catcode `\%12\relax}%
\providecommand \@@startlink[1]{}%
\providecommand \@@endlink[0]{}%
\providecommand \url  [0]{\begingroup\@sanitize@url \@url }%
\providecommand \@url [1]{\endgroup\@href {#1}{\urlprefix }}%
\providecommand \urlprefix  [0]{URL }%
\providecommand \Eprint [0]{\href }%
\providecommand \doibase [0]{https://doi.org/}%
\providecommand \selectlanguage [0]{\@gobble}%
\providecommand \bibinfo  [0]{\@secondoftwo}%
\providecommand \bibfield  [0]{\@secondoftwo}%
\providecommand \translation [1]{[#1]}%
\providecommand \BibitemOpen [0]{}%
\providecommand \bibitemStop [0]{}%
\providecommand \bibitemNoStop [0]{.\EOS\space}%
\providecommand \EOS [0]{\spacefactor3000\relax}%
\providecommand \BibitemShut  [1]{\csname bibitem#1\endcsname}%
\let\auto@bib@innerbib\@empty
\bibitem [{\citenamefont {Hastings}(2020)}]{Has20}%
  \BibitemOpen
  \bibfield  {author} {\bibinfo {author} {\bibfnamefont {M.}~\bibnamefont {Hastings}},\ }\bibfield  {title} {\bibinfo {title} {Classical and quantum algorithms for tensor principal component analysis},\ }\href {https://doi.org/10.22331/q-2020-02-27-237} {\bibfield  {journal} {\bibinfo  {journal} {Quantum}\ }\textbf {\bibinfo {volume} {4}},\ \bibinfo {pages} {237} (\bibinfo {year} {2020})}\BibitemShut {NoStop}%
\bibitem [{\citenamefont {Shor}(1999)}]{shor1999polynomial}%
  \BibitemOpen
  \bibfield  {author} {\bibinfo {author} {\bibfnamefont {P.~W.}\ \bibnamefont {Shor}},\ }\bibfield  {title} {\bibinfo {title} {Polynomial-time algorithms for prime factorization and discrete logarithms on a quantum computer},\ }\href@noop {} {\bibfield  {journal} {\bibinfo  {journal} {SIAM review}\ }\textbf {\bibinfo {volume} {41}},\ \bibinfo {pages} {303} (\bibinfo {year} {1999})}\BibitemShut {NoStop}%
\bibitem [{\citenamefont {Grover}(1996)}]{grover1996fast}%
  \BibitemOpen
  \bibfield  {author} {\bibinfo {author} {\bibfnamefont {L.~K.}\ \bibnamefont {Grover}},\ }\bibfield  {title} {\bibinfo {title} {A fast quantum mechanical algorithm for database search},\ }in\ \href@noop {} {\emph {\bibinfo {booktitle} {Proceedings of the twenty-eighth annual ACM symposium on Theory of computing}}}\ (\bibinfo {year} {1996})\ pp.\ \bibinfo {pages} {212--219}\BibitemShut {NoStop}%
\bibitem [{\citenamefont {Babbush}\ \emph {et~al.}(2021)\citenamefont {Babbush}, \citenamefont {McClean}, \citenamefont {Newman}, \citenamefont {Gidney}, \citenamefont {Boixo},\ and\ \citenamefont {Neven}}]{BMN+21}%
  \BibitemOpen
  \bibfield  {author} {\bibinfo {author} {\bibfnamefont {R.}~\bibnamefont {Babbush}}, \bibinfo {author} {\bibfnamefont {J.~R.}\ \bibnamefont {McClean}}, \bibinfo {author} {\bibfnamefont {M.}~\bibnamefont {Newman}}, \bibinfo {author} {\bibfnamefont {C.}~\bibnamefont {Gidney}}, \bibinfo {author} {\bibfnamefont {S.}~\bibnamefont {Boixo}},\ and\ \bibinfo {author} {\bibfnamefont {H.}~\bibnamefont {Neven}},\ }\bibfield  {title} {\bibinfo {title} {Focus beyond quadratic speedups for error-corrected quantum advantage},\ }\href {https://doi.org/10.1103/PRXQuantum.2.010103} {\bibfield  {journal} {\bibinfo  {journal} {PRX Quantum}\ }\textbf {\bibinfo {volume} {2}},\ \bibinfo {pages} {010103} (\bibinfo {year} {2021})}\BibitemShut {NoStop}%
\bibitem [{\citenamefont {Hoefler}\ \emph {et~al.}(2023)\citenamefont {Hoefler}, \citenamefont {H{\"a}ner},\ and\ \citenamefont {Troyer}}]{hoefler2023disentangling}%
  \BibitemOpen
  \bibfield  {author} {\bibinfo {author} {\bibfnamefont {T.}~\bibnamefont {Hoefler}}, \bibinfo {author} {\bibfnamefont {T.}~\bibnamefont {H{\"a}ner}},\ and\ \bibinfo {author} {\bibfnamefont {M.}~\bibnamefont {Troyer}},\ }\bibfield  {title} {\bibinfo {title} {Disentangling hype from practicality: On realistically achieving quantum advantage},\ }\href {https://doi.org/10.1145/3571725} {\bibfield  {journal} {\bibinfo  {journal} {Communications of the ACM}\ }\textbf {\bibinfo {volume} {66}},\ \bibinfo {pages} {82} (\bibinfo {year} {2023})}\BibitemShut {NoStop}%
\bibitem [{\citenamefont {H{\aa}stad}(1984)}]{Has84}%
  \BibitemOpen
  \bibfield  {author} {\bibinfo {author} {\bibfnamefont {J.}~\bibnamefont {H{\aa}stad}},\ }\emph {\bibinfo {title} {An {NP}-complete problem --- some aspects of its solution and some possible applications}},\ \href@noop {} {Master's thesis},\ \bibinfo  {school} {Uppsala University} (\bibinfo {year} {1984})\BibitemShut {NoStop}%
\bibitem [{\citenamefont {Applebaum}\ \emph {et~al.}(2010)\citenamefont {Applebaum}, \citenamefont {Barak},\ and\ \citenamefont {Wigderson}}]{ABW10}%
  \BibitemOpen
  \bibfield  {author} {\bibinfo {author} {\bibfnamefont {B.}~\bibnamefont {Applebaum}}, \bibinfo {author} {\bibfnamefont {B.}~\bibnamefont {Barak}},\ and\ \bibinfo {author} {\bibfnamefont {A.}~\bibnamefont {Wigderson}},\ }\bibfield  {title} {\bibinfo {title} {Public-key cryptography from different assumptions},\ }in\ \href {https://doi.org/10.1145/1806689.1806715} {\emph {\bibinfo {booktitle} {Proceedings of the 42nd Annual ACM Symposium on Theory of Computing}}}\ (\bibinfo {year} {2010})\ pp.\ \bibinfo {pages} {171--180}\BibitemShut {NoStop}%
\bibitem [{\citenamefont {Dao}\ \emph {et~al.}(2023)\citenamefont {Dao}, \citenamefont {Ishai}, \citenamefont {Jain},\ and\ \citenamefont {Lin}}]{DIJL23}%
  \BibitemOpen
  \bibfield  {author} {\bibinfo {author} {\bibfnamefont {Q.}~\bibnamefont {Dao}}, \bibinfo {author} {\bibfnamefont {Y.}~\bibnamefont {Ishai}}, \bibinfo {author} {\bibfnamefont {A.}~\bibnamefont {Jain}},\ and\ \bibinfo {author} {\bibfnamefont {H.}~\bibnamefont {Lin}},\ }\bibfield  {title} {\bibinfo {title} {Multi-party homomorphic secret sharing and sublinear {MPC} from sparse {LPN}},\ }in\ \href {https://doi.org/10.1007/978-3-031-38545-2_11} {\emph {\bibinfo {booktitle} {Proceedings of the 43rd Annual International Conference on the Theory and Applications of Cryptographic Techniques}}}\ (\bibinfo {year} {2023})\ pp.\ \bibinfo {pages} {315--348}\BibitemShut {NoStop}%
\bibitem [{\citenamefont {Dao}\ and\ \citenamefont {Jain}(2024)}]{DJ24}%
  \BibitemOpen
  \bibfield  {author} {\bibinfo {author} {\bibfnamefont {Q.}~\bibnamefont {Dao}}\ and\ \bibinfo {author} {\bibfnamefont {A.}~\bibnamefont {Jain}},\ }\href@noop {} {\emph {\bibinfo {title} {Lossy Cryptography from Code-Based Assumptions}}},\ \bibinfo {type} {Tech. Rep.}\ (\bibinfo  {institution} {Cryptology ePrint Archive},\ \bibinfo {year} {2024})\ \Eprint {https://arxiv.org/abs/2402.03633} {2402.03633} \BibitemShut {NoStop}%
\bibitem [{\citenamefont {Alon}\ \emph {et~al.}(1998)\citenamefont {Alon}, \citenamefont {Krivelevich},\ and\ \citenamefont {Sudakov}}]{alon1998finding}%
  \BibitemOpen
  \bibfield  {author} {\bibinfo {author} {\bibfnamefont {N.}~\bibnamefont {Alon}}, \bibinfo {author} {\bibfnamefont {M.}~\bibnamefont {Krivelevich}},\ and\ \bibinfo {author} {\bibfnamefont {B.}~\bibnamefont {Sudakov}},\ }\bibfield  {title} {\bibinfo {title} {Finding a large hidden clique in a random graph},\ }\href@noop {} {\bibfield  {journal} {\bibinfo  {journal} {Random Structures \& Algorithms}\ }\textbf {\bibinfo {volume} {13}},\ \bibinfo {pages} {457} (\bibinfo {year} {1998})}\BibitemShut {NoStop}%
\bibitem [{\citenamefont {Abbe}(2018)}]{JMLR:v18:16-480}%
  \BibitemOpen
  \bibfield  {author} {\bibinfo {author} {\bibfnamefont {E.}~\bibnamefont {Abbe}},\ }\bibfield  {title} {\bibinfo {title} {Community detection and stochastic block models: Recent developments},\ }\href {http://jmlr.org/papers/v18/16-480.html} {\bibfield  {journal} {\bibinfo  {journal} {Journal of Machine Learning Research}\ }\textbf {\bibinfo {volume} {18}},\ \bibinfo {pages} {1} (\bibinfo {year} {2018})}\BibitemShut {NoStop}%
\bibitem [{\citenamefont {Abbe}\ and\ \citenamefont {Sandon}(2015)}]{abbe2015detection}%
  \BibitemOpen
  \bibfield  {author} {\bibinfo {author} {\bibfnamefont {E.}~\bibnamefont {Abbe}}\ and\ \bibinfo {author} {\bibfnamefont {C.}~\bibnamefont {Sandon}},\ }\bibfield  {title} {\bibinfo {title} {Detection in the stochastic block model with multiple clusters: proof of the achievability conjectures, acyclic bp, and the information-computation gap},\ }\href@noop {} {\bibfield  {journal} {\bibinfo  {journal} {arXiv preprint arXiv:1512.09080}\ } (\bibinfo {year} {2015})}\BibitemShut {NoStop}%
\bibitem [{\citenamefont {Bandeira}\ \emph {et~al.}(2021)\citenamefont {Bandeira}, \citenamefont {Banks}, \citenamefont {Kunisky}, \citenamefont {Moore},\ and\ \citenamefont {Wein}}]{bandeira2021spectral}%
  \BibitemOpen
  \bibfield  {author} {\bibinfo {author} {\bibfnamefont {A.~S.}\ \bibnamefont {Bandeira}}, \bibinfo {author} {\bibfnamefont {J.}~\bibnamefont {Banks}}, \bibinfo {author} {\bibfnamefont {D.}~\bibnamefont {Kunisky}}, \bibinfo {author} {\bibfnamefont {C.}~\bibnamefont {Moore}},\ and\ \bibinfo {author} {\bibfnamefont {A.}~\bibnamefont {Wein}},\ }\bibfield  {title} {\bibinfo {title} {Spectral planting and the hardness of refuting cuts, colorability, and communities in random graphs},\ }in\ \href@noop {} {\emph {\bibinfo {booktitle} {Conference on Learning Theory}}}\ (\bibinfo {organization} {PMLR},\ \bibinfo {year} {2021})\ pp.\ \bibinfo {pages} {410--473}\BibitemShut {NoStop}%
\bibitem [{\citenamefont {Singer}(2011)}]{singer2011angular}%
  \BibitemOpen
  \bibfield  {author} {\bibinfo {author} {\bibfnamefont {A.}~\bibnamefont {Singer}},\ }\bibfield  {title} {\bibinfo {title} {Angular synchronization by eigenvectors and semidefinite programming},\ }\href@noop {} {\bibfield  {journal} {\bibinfo  {journal} {Applied and computational harmonic analysis}\ }\textbf {\bibinfo {volume} {30}},\ \bibinfo {pages} {20} (\bibinfo {year} {2011})}\BibitemShut {NoStop}%
\bibitem [{\citenamefont {Perry}\ \emph {et~al.}(2018)\citenamefont {Perry}, \citenamefont {Wein}, \citenamefont {Bandeira},\ and\ \citenamefont {Moitra}}]{perry2018message}%
  \BibitemOpen
  \bibfield  {author} {\bibinfo {author} {\bibfnamefont {A.}~\bibnamefont {Perry}}, \bibinfo {author} {\bibfnamefont {A.~S.}\ \bibnamefont {Wein}}, \bibinfo {author} {\bibfnamefont {A.~S.}\ \bibnamefont {Bandeira}},\ and\ \bibinfo {author} {\bibfnamefont {A.}~\bibnamefont {Moitra}},\ }\bibfield  {title} {\bibinfo {title} {Message-passing algorithms for synchronization problems over compact groups},\ }\href@noop {} {\bibfield  {journal} {\bibinfo  {journal} {Communications on Pure and Applied Mathematics}\ }\textbf {\bibinfo {volume} {71}},\ \bibinfo {pages} {2275} (\bibinfo {year} {2018})}\BibitemShut {NoStop}%
\bibitem [{\citenamefont {Richard}\ and\ \citenamefont {Montanari}(2014)}]{richard2014statistical}%
  \BibitemOpen
  \bibfield  {author} {\bibinfo {author} {\bibfnamefont {E.}~\bibnamefont {Richard}}\ and\ \bibinfo {author} {\bibfnamefont {A.}~\bibnamefont {Montanari}},\ }\bibfield  {title} {\bibinfo {title} {A statistical model for tensor {PCA}},\ }in\ \href@noop {} {\emph {\bibinfo {booktitle} {Proceedings of the 27th International Conference on Neural Information Processing Systems}}},\ Vol.~\bibinfo {volume} {27}\ (\bibinfo {year} {2014})\ \Eprint {https://arxiv.org/abs/1411.1076} {1411.1076} \BibitemShut {NoStop}%
\bibitem [{\citenamefont {M{\'e}zard}\ \emph {et~al.}(1987)\citenamefont {M{\'e}zard}, \citenamefont {Parisi},\ and\ \citenamefont {Virasoro}}]{mezard1987spin}%
  \BibitemOpen
  \bibfield  {author} {\bibinfo {author} {\bibfnamefont {M.}~\bibnamefont {M{\'e}zard}}, \bibinfo {author} {\bibfnamefont {G.}~\bibnamefont {Parisi}},\ and\ \bibinfo {author} {\bibfnamefont {M.~A.}\ \bibnamefont {Virasoro}},\ }\href@noop {} {\emph {\bibinfo {title} {Spin glass theory and beyond: An Introduction to the Replica Method and Its Applications}}},\ Vol.~\bibinfo {volume} {9}\ (\bibinfo  {publisher} {World Scientific Publishing Company},\ \bibinfo {year} {1987})\BibitemShut {NoStop}%
\bibitem [{\citenamefont {Hastings}(2006)}]{hastings2006community}%
  \BibitemOpen
  \bibfield  {author} {\bibinfo {author} {\bibfnamefont {M.~B.}\ \bibnamefont {Hastings}},\ }\bibfield  {title} {\bibinfo {title} {Community detection as an inference problem},\ }\href@noop {} {\bibfield  {journal} {\bibinfo  {journal} {Physical Review E—Statistical, Nonlinear, and Soft Matter Physics}\ }\textbf {\bibinfo {volume} {74}},\ \bibinfo {pages} {035102} (\bibinfo {year} {2006})}\BibitemShut {NoStop}%
\bibitem [{\citenamefont {Moore}(2017)}]{moore2017computer}%
  \BibitemOpen
  \bibfield  {author} {\bibinfo {author} {\bibfnamefont {C.}~\bibnamefont {Moore}},\ }\bibfield  {title} {\bibinfo {title} {The computer science and physics of community detection: Landscapes, phase transitions, and hardness},\ }\href@noop {} {\bibfield  {journal} {\bibinfo  {journal} {arXiv preprint arXiv:1702.00467}\ } (\bibinfo {year} {2017})}\BibitemShut {NoStop}%
\bibitem [{\citenamefont {Wein}\ \emph {et~al.}(2019)\citenamefont {Wein}, \citenamefont {Alaoui},\ and\ \citenamefont {Moore}}]{WAM19}%
  \BibitemOpen
  \bibfield  {author} {\bibinfo {author} {\bibfnamefont {A.~S.}\ \bibnamefont {Wein}}, \bibinfo {author} {\bibfnamefont {A.~E.}\ \bibnamefont {Alaoui}},\ and\ \bibinfo {author} {\bibfnamefont {C.}~\bibnamefont {Moore}},\ }\bibfield  {title} {\bibinfo {title} {The {K}ikuchi hierarchy and tensor {PCA}},\ }in\ \href {https://doi.org/10.1109/focs.2019.000-2} {\emph {\bibinfo {booktitle} {Proceedings of the 60th Annual IEEE Symposium on Foundations of Computer Science}}}\ (\bibinfo {year} {2019})\ pp.\ \bibinfo {pages} {1446--1468}\BibitemShut {NoStop}%
\bibitem [{\citenamefont {Kikuchi}(1951)}]{PhysRev.81.988}%
  \BibitemOpen
  \bibfield  {author} {\bibinfo {author} {\bibfnamefont {R.}~\bibnamefont {Kikuchi}},\ }\bibfield  {title} {\bibinfo {title} {A theory of cooperative phenomena},\ }\href {https://doi.org/10.1103/PhysRev.81.988} {\bibfield  {journal} {\bibinfo  {journal} {Phys. Rev.}\ }\textbf {\bibinfo {volume} {81}},\ \bibinfo {pages} {988} (\bibinfo {year} {1951})}\BibitemShut {NoStop}%
\bibitem [{\citenamefont {Gharibian}\ and\ \citenamefont {{Le Gall}}(2022)}]{GL22}%
  \BibitemOpen
  \bibfield  {author} {\bibinfo {author} {\bibfnamefont {S.}~\bibnamefont {Gharibian}}\ and\ \bibinfo {author} {\bibfnamefont {F.}~\bibnamefont {{Le Gall}}},\ }\bibfield  {title} {\bibinfo {title} {Dequantizing the quantum singular value transformation: hardness and applications to quantum chemistry and the quantum pcp conjecture},\ }in\ \href {https://doi.org/10.1145/3519935.3519991} {\emph {\bibinfo {booktitle} {Proceedings of the 54th Annual ACM Symposium on Theory of Computing}}}\ (\bibinfo {year} {2022})\ p.\ \bibinfo {pages} {19–32}\BibitemShut {NoStop}%
\bibitem [{\citenamefont {Jain}\ \emph {et~al.}(2021)\citenamefont {Jain}, \citenamefont {Lin},\ and\ \citenamefont {Sahai}}]{JLS21}%
  \BibitemOpen
  \bibfield  {author} {\bibinfo {author} {\bibfnamefont {A.}~\bibnamefont {Jain}}, \bibinfo {author} {\bibfnamefont {H.}~\bibnamefont {Lin}},\ and\ \bibinfo {author} {\bibfnamefont {A.}~\bibnamefont {Sahai}},\ }\bibfield  {title} {\bibinfo {title} {Indistinguishability obfuscation from well-founded assumptions},\ }in\ \href {https://doi.org/10.1145/3406325.3451093} {\emph {\bibinfo {booktitle} {Proceedings of the 53rd Annual ACM Symposium on Theory of Computing}}}\ (\bibinfo {year} {2021})\ pp.\ \bibinfo {pages} {60--73}\BibitemShut {NoStop}%
\bibitem [{\citenamefont {Jain}\ \emph {et~al.}(2022)\citenamefont {Jain}, \citenamefont {Lin},\ and\ \citenamefont {Sahai}}]{JLS22}%
  \BibitemOpen
  \bibfield  {author} {\bibinfo {author} {\bibfnamefont {A.}~\bibnamefont {Jain}}, \bibinfo {author} {\bibfnamefont {H.}~\bibnamefont {Lin}},\ and\ \bibinfo {author} {\bibfnamefont {A.}~\bibnamefont {Sahai}},\ }\bibfield  {title} {\bibinfo {title} {Indistinguishability obfuscation from {LPN} over {$\mathbb{F}_p$}, {DLIN}, and {PRGs} in {$\mathsf{NC}^0$}},\ }in\ \href {https://doi.org/10.1007/978-3-031-06944-4_23} {\emph {\bibinfo {booktitle} {Advances in Cryptology -- EUROCRYPT 2022}}}\ (\bibinfo {year} {2022})\ pp.\ \bibinfo {pages} {670--699}\BibitemShut {NoStop}%
\bibitem [{\citenamefont {Alekhnovich}(2003)}]{Ale03}%
  \BibitemOpen
  \bibfield  {author} {\bibinfo {author} {\bibfnamefont {M.}~\bibnamefont {Alekhnovich}},\ }\bibfield  {title} {\bibinfo {title} {More on average case vs approximation complexity},\ }in\ \href {https://doi.org/10.1109/SFCS.2003.1238204} {\emph {\bibinfo {booktitle} {Proceedings of the 44th Annual IEEE Symposium on Foundations of Computer Science}}}\ (\bibinfo {organization} {IEEE},\ \bibinfo {year} {2003})\ pp.\ \bibinfo {pages} {298--307}\BibitemShut {NoStop}%
\bibitem [{\citenamefont {Feige}(2002)}]{Fei02}%
  \BibitemOpen
  \bibfield  {author} {\bibinfo {author} {\bibfnamefont {U.}~\bibnamefont {Feige}},\ }\bibfield  {title} {\bibinfo {title} {Relations between average case complexity and approximation complexity},\ }in\ \href {https://doi.org/10.1145/509907.509985} {\emph {\bibinfo {booktitle} {Proceedings of the 34th Annual ACM Symposium on Theory of Computing}}}\ (\bibinfo {year} {2002})\ pp.\ \bibinfo {pages} {543--543}\BibitemShut {NoStop}%
\bibitem [{\citenamefont {Goerdt}\ and\ \citenamefont {Krivelevich}(2001)}]{GK01}%
  \BibitemOpen
  \bibfield  {author} {\bibinfo {author} {\bibfnamefont {A.}~\bibnamefont {Goerdt}}\ and\ \bibinfo {author} {\bibfnamefont {M.}~\bibnamefont {Krivelevich}},\ }\bibfield  {title} {\bibinfo {title} {Efficient recognition of random unsatisfiable $k$-{SAT} instances by spectral methods},\ }in\ \href {https://doi.org/10.1007/3-540-44693-1_26} {\emph {\bibinfo {booktitle} {Proceedings of the 18th Annual Symposium on Theoretical Aspects of Computer Science}}}\ (\bibinfo {year} {2001})\ pp.\ \bibinfo {pages} {294--304}\BibitemShut {NoStop}%
\bibitem [{\citenamefont {Goerdt}\ and\ \citenamefont {Jurdzi{\'n}ski}(2002)}]{GJ02}%
  \BibitemOpen
  \bibfield  {author} {\bibinfo {author} {\bibfnamefont {A.}~\bibnamefont {Goerdt}}\ and\ \bibinfo {author} {\bibfnamefont {T.}~\bibnamefont {Jurdzi{\'n}ski}},\ }\bibfield  {title} {\bibinfo {title} {Some results on random unsatisfiable {$k$}-{SAT} instances and approximation algorithms applied to random structures},\ }in\ \href {https://doi.org/10.1007/3-540-45687-2_23} {\emph {\bibinfo {booktitle} {Proceedings of the 27th Annual International Symposium on Mathematical Foundations of Computer Science}}}\ (\bibinfo {year} {2002})\ pp.\ \bibinfo {pages} {280--291}\BibitemShut {NoStop}%
\bibitem [{\citenamefont {Coja{-}Oghlan}\ \emph {et~al.}(2007)\citenamefont {Coja{-}Oghlan}, \citenamefont {Goerdt},\ and\ \citenamefont {Lanka}}]{CGL07}%
  \BibitemOpen
  \bibfield  {author} {\bibinfo {author} {\bibfnamefont {A.}~\bibnamefont {Coja{-}Oghlan}}, \bibinfo {author} {\bibfnamefont {A.}~\bibnamefont {Goerdt}},\ and\ \bibinfo {author} {\bibfnamefont {A.}~\bibnamefont {Lanka}},\ }\bibfield  {title} {\bibinfo {title} {Strong refutation heuristics for random $k$-{SAT}},\ }\href {https://doi.org/10.1137/070699354} {\bibfield  {journal} {\bibinfo  {journal} {Combinatorics, Probability and Computing}\ }\textbf {\bibinfo {volume} {16}},\ \bibinfo {pages} {5} (\bibinfo {year} {2007})}\BibitemShut {NoStop}%
\bibitem [{\citenamefont {Coja{-}Oghlan}\ \emph {et~al.}(2010)\citenamefont {Coja{-}Oghlan}, \citenamefont {Cooper},\ and\ \citenamefont {Frieze}}]{CCF10}%
  \BibitemOpen
  \bibfield  {author} {\bibinfo {author} {\bibfnamefont {A.}~\bibnamefont {Coja{-}Oghlan}}, \bibinfo {author} {\bibfnamefont {C.}~\bibnamefont {Cooper}},\ and\ \bibinfo {author} {\bibfnamefont {A.}~\bibnamefont {Frieze}},\ }\bibfield  {title} {\bibinfo {title} {An efficient sparse regularity concept},\ }\href {https://doi.org/10.1137/080730160} {\bibfield  {journal} {\bibinfo  {journal} {SIAM Journal on Discrete Mathematics}\ }\textbf {\bibinfo {volume} {23}},\ \bibinfo {pages} {2000} (\bibinfo {year} {2010})}\BibitemShut {NoStop}%
\bibitem [{\citenamefont {Allen}\ \emph {et~al.}(2015)\citenamefont {Allen}, \citenamefont {O'Donnell},\ and\ \citenamefont {Witmer}}]{AOW15}%
  \BibitemOpen
  \bibfield  {author} {\bibinfo {author} {\bibfnamefont {S.}~\bibnamefont {Allen}}, \bibinfo {author} {\bibfnamefont {R.}~\bibnamefont {O'Donnell}},\ and\ \bibinfo {author} {\bibfnamefont {D.}~\bibnamefont {Witmer}},\ }\bibfield  {title} {\bibinfo {title} {How to refute a random {CSP}},\ }in\ \href {https://doi.org/10.1109/FOCS.2015.48} {\emph {\bibinfo {booktitle} {Proceedings of the 56th Annual IEEE Symposium on Foundations of Computer Science}}}\ (\bibinfo {organization} {IEEE},\ \bibinfo {year} {2015})\ pp.\ \bibinfo {pages} {689--708}\BibitemShut {NoStop}%
\bibitem [{\citenamefont {Barak}\ and\ \citenamefont {Moitra}(2022)}]{BM22}%
  \BibitemOpen
  \bibfield  {author} {\bibinfo {author} {\bibfnamefont {B.}~\bibnamefont {Barak}}\ and\ \bibinfo {author} {\bibfnamefont {A.}~\bibnamefont {Moitra}},\ }\bibfield  {title} {\bibinfo {title} {Noisy tensor completion via the sum-of-squares hierarchy},\ }\href {https://doi.org/10.1007/s10107-022-01793-9} {\bibfield  {journal} {\bibinfo  {journal} {Mathematical Programming}\ }\textbf {\bibinfo {volume} {193}},\ \bibinfo {pages} {513} (\bibinfo {year} {2022})}\BibitemShut {NoStop}%
\bibitem [{\citenamefont {d'Orsi}\ and\ \citenamefont {Trevisan}(2023)}]{DT23}%
  \BibitemOpen
  \bibfield  {author} {\bibinfo {author} {\bibfnamefont {T.}~\bibnamefont {d'Orsi}}\ and\ \bibinfo {author} {\bibfnamefont {L.}~\bibnamefont {Trevisan}},\ }\bibfield  {title} {\bibinfo {title} {{A Ihara-Bass Formula for Non-Boolean Matrices and Strong Refutations of Random CSPs}},\ }in\ \href {https://doi.org/10.4230/LIPIcs.CCC.2023.27} {\emph {\bibinfo {booktitle} {Proceedings of the 38th Annual Computational Complexity Conference}}},\ \bibinfo {series} {Leibniz International Proceedings in Informatics (LIPIcs)}, Vol.\ \bibinfo {volume} {264}\ (\bibinfo {year} {2023})\ pp.\ \bibinfo {pages} {27:1--27:16}\BibitemShut {NoStop}%
\bibitem [{\citenamefont {Alekhnovich}\ and\ \citenamefont {Razborov}(2001)}]{AR01}%
  \BibitemOpen
  \bibfield  {author} {\bibinfo {author} {\bibfnamefont {M.}~\bibnamefont {Alekhnovich}}\ and\ \bibinfo {author} {\bibfnamefont {A.}~\bibnamefont {Razborov}},\ }\bibfield  {title} {\bibinfo {title} {Lower bounds for polynomial calculus: {N}on-binomial case},\ }in\ \href {https://doi.org/10.1109/sfcs.2001.959893} {\emph {\bibinfo {booktitle} {Proceedings of the 42nd Annual IEEE Symposium on Foundations of Computer Science}}}\ (\bibinfo {year} {2001})\ pp.\ \bibinfo {pages} {190--199}\BibitemShut {NoStop}%
\bibitem [{\citenamefont {Schoenebeck}(2008)}]{Sch08}%
  \BibitemOpen
  \bibfield  {author} {\bibinfo {author} {\bibfnamefont {G.}~\bibnamefont {Schoenebeck}},\ }\bibfield  {title} {\bibinfo {title} {Linear level {L}asserre lower bounds for certain $k$-{CSP}s},\ }in\ \href {https://doi.org/10.1109/focs.2008.74} {\emph {\bibinfo {booktitle} {Proceedings of the 49th Annual IEEE Symposium on Foundations of Computer Science}}}\ (\bibinfo {year} {2008})\ pp.\ \bibinfo {pages} {593--602}\BibitemShut {NoStop}%
\bibitem [{\citenamefont {O'Donnell}\ and\ \citenamefont {Witmer}(2014)}]{OW14}%
  \BibitemOpen
  \bibfield  {author} {\bibinfo {author} {\bibfnamefont {R.}~\bibnamefont {O'Donnell}}\ and\ \bibinfo {author} {\bibfnamefont {D.}~\bibnamefont {Witmer}},\ }\bibfield  {title} {\bibinfo {title} {Goldreich's {PRG}: Evidence for near-optimal polynomial stretch},\ }in\ \href {https://doi.org/10.1109/ccc.2014.9} {\emph {\bibinfo {booktitle} {Proceedings of the 29th Annual Computational Complexity Conference}}}\ (\bibinfo {year} {2014})\ pp.\ \bibinfo {pages} {1--12}\BibitemShut {NoStop}%
\bibitem [{\citenamefont {Mori}\ and\ \citenamefont {Witmer}(2016)}]{MW16}%
  \BibitemOpen
  \bibfield  {author} {\bibinfo {author} {\bibfnamefont {R.}~\bibnamefont {Mori}}\ and\ \bibinfo {author} {\bibfnamefont {D.}~\bibnamefont {Witmer}},\ }\bibfield  {title} {\bibinfo {title} {Lower bounds for {CSP} refutation by {SDP} hierarchies},\ }in\ \href {https://doi.org/10.4230/LIPIcs.APPROX-RANDOM.2016.41} {\emph {\bibinfo {booktitle} {Proceedings of the 20th Annual International Workshop on Randomized Techniques in Computation}}},\ \bibinfo {series} {Leibniz International Proceedings in Informatics (LIPIcs)}, Vol.~\bibinfo {volume} {60}\ (\bibinfo {year} {2016})\ pp.\ \bibinfo {pages} {41:1--41:30}\BibitemShut {NoStop}%
\bibitem [{\citenamefont {Kothari}\ \emph {et~al.}(2017)\citenamefont {Kothari}, \citenamefont {Mori}, \citenamefont {O'Donnell},\ and\ \citenamefont {Witmer}}]{KMOW17}%
  \BibitemOpen
  \bibfield  {author} {\bibinfo {author} {\bibfnamefont {P.}~\bibnamefont {Kothari}}, \bibinfo {author} {\bibfnamefont {R.}~\bibnamefont {Mori}}, \bibinfo {author} {\bibfnamefont {R.}~\bibnamefont {O'Donnell}},\ and\ \bibinfo {author} {\bibfnamefont {D.}~\bibnamefont {Witmer}},\ }\bibfield  {title} {\bibinfo {title} {Sum of squares lower bounds for refuting any {CSP}},\ }in\ \href {https://doi.org/10.1145/3055399.3055485} {\emph {\bibinfo {booktitle} {Proceedings of the 49th Annual ACM Symposium on Theory of Computing}}}\ (\bibinfo {year} {2017})\ pp.\ \bibinfo {pages} {132--145}\BibitemShut {NoStop}%
\bibitem [{\citenamefont {Raghavendra}\ \emph {et~al.}(2017)\citenamefont {Raghavendra}, \citenamefont {Rao},\ and\ \citenamefont {Schramm}}]{RRS17}%
  \BibitemOpen
  \bibfield  {author} {\bibinfo {author} {\bibfnamefont {P.}~\bibnamefont {Raghavendra}}, \bibinfo {author} {\bibfnamefont {S.}~\bibnamefont {Rao}},\ and\ \bibinfo {author} {\bibfnamefont {T.}~\bibnamefont {Schramm}},\ }\bibfield  {title} {\bibinfo {title} {Strongly refuting random {CSP}s below the spectral threshold},\ }in\ \href {https://doi.org/10.1145/3055399.3055417} {\emph {\bibinfo {booktitle} {Proceedings of the 49th Annual ACM Symposium on Theory of Computing}}}\ (\bibinfo {year} {2017})\ pp.\ \bibinfo {pages} {121--131}\BibitemShut {NoStop}%
\bibitem [{\citenamefont {Ahn}(2020)}]{Ahn20}%
  \BibitemOpen
  \bibfield  {author} {\bibinfo {author} {\bibfnamefont {K.}~\bibnamefont {Ahn}},\ }\href@noop {} {\emph {\bibinfo {title} {A simpler strong refutation of random $k$-{XOR}}}},\ \bibinfo {type} {Tech. Rep.}\ (\bibinfo  {institution} {arXiv:2008.03556},\ \bibinfo {year} {2020})\ \Eprint {https://arxiv.org/abs/2008.03556} {2008.03556} \BibitemShut {NoStop}%
\bibitem [{\citenamefont {Kitaev}\ \emph {et~al.}(2002)\citenamefont {Kitaev}, \citenamefont {Shen},\ and\ \citenamefont {Vyalyi}}]{KSV02}%
  \BibitemOpen
  \bibfield  {author} {\bibinfo {author} {\bibfnamefont {A.}~\bibnamefont {Kitaev}}, \bibinfo {author} {\bibfnamefont {A.}~\bibnamefont {Shen}},\ and\ \bibinfo {author} {\bibfnamefont {M.}~\bibnamefont {Vyalyi}},\ }\href {https://books.google.com/books?id=08vZYhafYEAC} {\emph {\bibinfo {title} {Classical and Quantum Computation}}},\ Graduate studies in mathematics\ (\bibinfo  {publisher} {American Mathematical Society},\ \bibinfo {year} {2002})\BibitemShut {NoStop}%
\bibitem [{\citenamefont {Bookatz}(2014)}]{Boo14}%
  \BibitemOpen
  \bibfield  {author} {\bibinfo {author} {\bibfnamefont {A.~D.}\ \bibnamefont {Bookatz}},\ }\bibfield  {title} {\bibinfo {title} {{QMA}-complete problems},\ }\href {https://doi.org/10.26421/QIC14.5-6-1} {\bibfield  {journal} {\bibinfo  {journal} {Quantum Inf. Comput.}\ }\textbf {\bibinfo {volume} {14}},\ \bibinfo {pages} {361} (\bibinfo {year} {2014})}\BibitemShut {NoStop}%
\bibitem [{\citenamefont {Cade}\ \emph {et~al.}(2023)\citenamefont {Cade}, \citenamefont {Folkertsma}, \citenamefont {Gharibian}, \citenamefont {Hayakawa}, \citenamefont {Le~Gall}, \citenamefont {Morimae},\ and\ \citenamefont {Weggemans}}]{CFG+23}%
  \BibitemOpen
  \bibfield  {author} {\bibinfo {author} {\bibfnamefont {C.}~\bibnamefont {Cade}}, \bibinfo {author} {\bibfnamefont {M.}~\bibnamefont {Folkertsma}}, \bibinfo {author} {\bibfnamefont {S.}~\bibnamefont {Gharibian}}, \bibinfo {author} {\bibfnamefont {R.}~\bibnamefont {Hayakawa}}, \bibinfo {author} {\bibfnamefont {F.}~\bibnamefont {Le~Gall}}, \bibinfo {author} {\bibfnamefont {T.}~\bibnamefont {Morimae}},\ and\ \bibinfo {author} {\bibfnamefont {J.}~\bibnamefont {Weggemans}},\ }\bibfield  {title} {\bibinfo {title} {{Improved Hardness Results for the Guided Local Hamiltonian Problem}},\ }in\ \href {https://doi.org/10.4230/LIPIcs.ICALP.2023.32} {\emph {\bibinfo {booktitle} {50th International Colloquium on Automata, Languages, and Programming (ICALP 2023)}}},\ \bibinfo {series} {Leibniz International Proceedings in Informatics (LIPIcs)}, Vol.\ \bibinfo {volume} {261}\ (\bibinfo {year} {2023})\ pp.\ \bibinfo {pages} {32:1--32:19}\BibitemShut {NoStop}%
\bibitem [{\citenamefont {Hastings}(2018)}]{hastings2018short}%
  \BibitemOpen
  \bibfield  {author} {\bibinfo {author} {\bibfnamefont {M.~B.}\ \bibnamefont {Hastings}},\ }\bibfield  {title} {\bibinfo {title} {A short path quantum algorithm for exact optimization},\ }\href@noop {} {\bibfield  {journal} {\bibinfo  {journal} {Quantum}\ }\textbf {\bibinfo {volume} {2}},\ \bibinfo {pages} {78} (\bibinfo {year} {2018})}\BibitemShut {NoStop}%
\bibitem [{\citenamefont {Dalzell}\ \emph {et~al.}(2023)\citenamefont {Dalzell}, \citenamefont {Pancotti}, \citenamefont {Campbell},\ and\ \citenamefont {Brand{\~a}o}}]{Dalzell2023mind}%
  \BibitemOpen
  \bibfield  {author} {\bibinfo {author} {\bibfnamefont {A.~M.}\ \bibnamefont {Dalzell}}, \bibinfo {author} {\bibfnamefont {N.}~\bibnamefont {Pancotti}}, \bibinfo {author} {\bibfnamefont {E.~T.}\ \bibnamefont {Campbell}},\ and\ \bibinfo {author} {\bibfnamefont {F.~G.}\ \bibnamefont {Brand{\~a}o}},\ }\bibfield  {title} {\bibinfo {title} {Mind the gap: Achieving a super-grover quantum speedup by jumping to the end},\ }in\ \href@noop {} {\emph {\bibinfo {booktitle} {Proceedings of the 55th Annual ACM Symposium on Theory of Computing}}}\ (\bibinfo {year} {2023})\ pp.\ \bibinfo {pages} {1131--1144}\BibitemShut {NoStop}%
\bibitem [{\citenamefont {Boulebnane}\ and\ \citenamefont {Montanaro}(2022)}]{Boulebnane2022}%
  \BibitemOpen
  \bibfield  {author} {\bibinfo {author} {\bibfnamefont {S.}~\bibnamefont {Boulebnane}}\ and\ \bibinfo {author} {\bibfnamefont {A.}~\bibnamefont {Montanaro}},\ }\href {https://arxiv.org/abs/2208.06909} {\bibinfo {title} {Solving boolean satisfiability problems with the quantum approximate optimization algorithm}},\ \bibinfo {howpublished} {arXiv preprint arXiv:2208.06909} (\bibinfo {year} {2022})\BibitemShut {NoStop}%
\bibitem [{\citenamefont {Chakrabarti}\ \emph {et~al.}(2024)\citenamefont {Chakrabarti}, \citenamefont {Herman}, \citenamefont {Ozgul}, \citenamefont {Zhu}, \citenamefont {Augustino}, \citenamefont {Hao}, \citenamefont {He}, \citenamefont {Shaydulin},\ and\ \citenamefont {Pistoia}}]{chakrabarti2024generalized}%
  \BibitemOpen
  \bibfield  {author} {\bibinfo {author} {\bibfnamefont {S.}~\bibnamefont {Chakrabarti}}, \bibinfo {author} {\bibfnamefont {D.}~\bibnamefont {Herman}}, \bibinfo {author} {\bibfnamefont {G.}~\bibnamefont {Ozgul}}, \bibinfo {author} {\bibfnamefont {S.}~\bibnamefont {Zhu}}, \bibinfo {author} {\bibfnamefont {B.}~\bibnamefont {Augustino}}, \bibinfo {author} {\bibfnamefont {T.}~\bibnamefont {Hao}}, \bibinfo {author} {\bibfnamefont {Z.}~\bibnamefont {He}}, \bibinfo {author} {\bibfnamefont {R.}~\bibnamefont {Shaydulin}},\ and\ \bibinfo {author} {\bibfnamefont {M.}~\bibnamefont {Pistoia}},\ }\bibfield  {title} {\bibinfo {title} {Generalized short path algorithms: Towards super-quadratic speedup over markov chain search for combinatorial optimization},\ }\href@noop {} {\bibfield  {journal} {\bibinfo  {journal} {arXiv preprint arXiv:2410.23270}\ } (\bibinfo {year} {2024})}\BibitemShut {NoStop}%
\bibitem [{\citenamefont {Kapit}\ \emph {et~al.}(2023)\citenamefont {Kapit}, \citenamefont {Barton}, \citenamefont {Feeney}, \citenamefont {Grattan}, \citenamefont {Patnaik}, \citenamefont {Sagal}, \citenamefont {Carr},\ and\ \citenamefont {Oganesyan}}]{kapit2023approximability}%
  \BibitemOpen
  \bibfield  {author} {\bibinfo {author} {\bibfnamefont {E.}~\bibnamefont {Kapit}}, \bibinfo {author} {\bibfnamefont {B.~A.}\ \bibnamefont {Barton}}, \bibinfo {author} {\bibfnamefont {S.}~\bibnamefont {Feeney}}, \bibinfo {author} {\bibfnamefont {G.}~\bibnamefont {Grattan}}, \bibinfo {author} {\bibfnamefont {P.}~\bibnamefont {Patnaik}}, \bibinfo {author} {\bibfnamefont {J.}~\bibnamefont {Sagal}}, \bibinfo {author} {\bibfnamefont {L.~D.}\ \bibnamefont {Carr}},\ and\ \bibinfo {author} {\bibfnamefont {V.}~\bibnamefont {Oganesyan}},\ }\bibfield  {title} {\bibinfo {title} {On the approximability of random-hypergraph max-3-xorsat problems with quantum algorithms},\ }\href@noop {} {\bibfield  {journal} {\bibinfo  {journal} {arXiv preprint arXiv:2312.06104}\ } (\bibinfo {year} {2023})}\BibitemShut {NoStop}%
\bibitem [{\citenamefont {Zhou}\ \emph {et~al.}(2024)\citenamefont {Zhou}, \citenamefont {Basso},\ and\ \citenamefont {Mei}}]{zhou2024statistical}%
  \BibitemOpen
  \bibfield  {author} {\bibinfo {author} {\bibfnamefont {L.}~\bibnamefont {Zhou}}, \bibinfo {author} {\bibfnamefont {J.}~\bibnamefont {Basso}},\ and\ \bibinfo {author} {\bibfnamefont {S.}~\bibnamefont {Mei}},\ }\bibfield  {title} {\bibinfo {title} {Statistical estimation in the spiked tensor model via the quantum approximate optimization algorithm},\ }\href@noop {} {\bibfield  {journal} {\bibinfo  {journal} {arXiv preprint arXiv:2402.19456}\ } (\bibinfo {year} {2024})}\BibitemShut {NoStop}%
\bibitem [{\citenamefont {Prabhu}(2024)}]{qualtranPR1348}%
  \BibitemOpen
  \bibfield  {author} {\bibinfo {author} {\bibfnamefont {A.}~\bibnamefont {Prabhu}},\ }\href@noop {} {\bibinfo {title} {[wip] algorithm for planted noisy kxor}},\ \bibinfo {howpublished} {\url{https://github.com/quantumlib/Qualtran/pull/1348}} (\bibinfo {year} {2024}),\ \bibinfo {note} {gitHub pull request \#1348, quantumlib/Qualtran}\BibitemShut {NoStop}%
\bibitem [{\citenamefont {Harrigan}\ \emph {et~al.}(2024)\citenamefont {Harrigan}, \citenamefont {Khattar}, \citenamefont {Yuan}, \citenamefont {Peduri}, \citenamefont {Yosri}, \citenamefont {Malone}, \citenamefont {Babbush},\ and\ \citenamefont {Rubin}}]{harrigan2024expressing}%
  \BibitemOpen
  \bibfield  {author} {\bibinfo {author} {\bibfnamefont {M.~P.}\ \bibnamefont {Harrigan}}, \bibinfo {author} {\bibfnamefont {T.}~\bibnamefont {Khattar}}, \bibinfo {author} {\bibfnamefont {C.}~\bibnamefont {Yuan}}, \bibinfo {author} {\bibfnamefont {A.}~\bibnamefont {Peduri}}, \bibinfo {author} {\bibfnamefont {N.}~\bibnamefont {Yosri}}, \bibinfo {author} {\bibfnamefont {F.~D.}\ \bibnamefont {Malone}}, \bibinfo {author} {\bibfnamefont {R.}~\bibnamefont {Babbush}},\ and\ \bibinfo {author} {\bibfnamefont {N.~C.}\ \bibnamefont {Rubin}},\ }\bibfield  {title} {\bibinfo {title} {Expressing and analyzing quantum algorithms with qualtran},\ }\href@noop {} {\bibfield  {journal} {\bibinfo  {journal} {arXiv preprint arXiv:2409.04643}\ } (\bibinfo {year} {2024})}\BibitemShut {NoStop}%
\bibitem [{\citenamefont {Reiher}\ \emph {et~al.}(2017)\citenamefont {Reiher}, \citenamefont {Wiebe}, \citenamefont {Svore}, \citenamefont {Wecker},\ and\ \citenamefont {Troyer}}]{reiher2017elucidating}%
  \BibitemOpen
  \bibfield  {author} {\bibinfo {author} {\bibfnamefont {M.}~\bibnamefont {Reiher}}, \bibinfo {author} {\bibfnamefont {N.}~\bibnamefont {Wiebe}}, \bibinfo {author} {\bibfnamefont {K.~M.}\ \bibnamefont {Svore}}, \bibinfo {author} {\bibfnamefont {D.}~\bibnamefont {Wecker}},\ and\ \bibinfo {author} {\bibfnamefont {M.}~\bibnamefont {Troyer}},\ }\bibfield  {title} {\bibinfo {title} {Elucidating reaction mechanisms on quantum computers},\ }\href@noop {} {\bibfield  {journal} {\bibinfo  {journal} {Proceedings of the national academy of sciences}\ }\textbf {\bibinfo {volume} {114}},\ \bibinfo {pages} {7555} (\bibinfo {year} {2017})}\BibitemShut {NoStop}%
\bibitem [{\citenamefont {G{\"u}nther}\ \emph {et~al.}(2025)\citenamefont {G{\"u}nther}, \citenamefont {Witteveen}, \citenamefont {Schmidhuber}, \citenamefont {Miller}, \citenamefont {Christandl},\ and\ \citenamefont {Harrow}}]{gunther2025phase}%
  \BibitemOpen
  \bibfield  {author} {\bibinfo {author} {\bibfnamefont {J.}~\bibnamefont {G{\"u}nther}}, \bibinfo {author} {\bibfnamefont {F.}~\bibnamefont {Witteveen}}, \bibinfo {author} {\bibfnamefont {A.}~\bibnamefont {Schmidhuber}}, \bibinfo {author} {\bibfnamefont {M.}~\bibnamefont {Miller}}, \bibinfo {author} {\bibfnamefont {M.}~\bibnamefont {Christandl}},\ and\ \bibinfo {author} {\bibfnamefont {A.}~\bibnamefont {Harrow}},\ }\bibfield  {title} {\bibinfo {title} {Phase estimation with partially randomized time evolution},\ }\href@noop {} {\bibfield  {journal} {\bibinfo  {journal} {arXiv preprint arXiv:2503.05647}\ } (\bibinfo {year} {2025})}\BibitemShut {NoStop}%
\bibitem [{\citenamefont {Tropp}(2012)}]{Tro12}%
  \BibitemOpen
  \bibfield  {author} {\bibinfo {author} {\bibfnamefont {J.}~\bibnamefont {Tropp}},\ }\bibfield  {title} {\bibinfo {title} {User-friendly tail bounds for sums of random matrices},\ }\href {https://doi.org/10.1007/s10208-011-9099-z} {\bibfield  {journal} {\bibinfo  {journal} {Foundations of Computational Mathematics. The Journal of the Society for the Foundations of Computational Mathematics}\ }\textbf {\bibinfo {volume} {12}},\ \bibinfo {pages} {389} (\bibinfo {year} {2012})}\BibitemShut {NoStop}%
\bibitem [{\citenamefont {Guruswami}\ \emph {et~al.}(2022)\citenamefont {Guruswami}, \citenamefont {Kothari},\ and\ \citenamefont {Manohar}}]{guruswami2022algorithms}%
  \BibitemOpen
  \bibfield  {author} {\bibinfo {author} {\bibfnamefont {V.}~\bibnamefont {Guruswami}}, \bibinfo {author} {\bibfnamefont {P.~K.}\ \bibnamefont {Kothari}},\ and\ \bibinfo {author} {\bibfnamefont {P.}~\bibnamefont {Manohar}},\ }\bibfield  {title} {\bibinfo {title} {Algorithms and certificates for boolean {CSP} refutation: smoothed is no harder than random},\ }in\ \href {https://doi.org/10.1145/3519935.3519955} {\emph {\bibinfo {booktitle} {Proceedings of the 54th Annual ACM Symposium on Theory of Computing}}}\ (\bibinfo {year} {2022})\ pp.\ \bibinfo {pages} {678--689}\BibitemShut {NoStop}%
\bibitem [{\citenamefont {Hsieh}\ \emph {et~al.}(2023)\citenamefont {Hsieh}, \citenamefont {Kothari},\ and\ \citenamefont {Mohanty}}]{hsieh2023simple}%
  \BibitemOpen
  \bibfield  {author} {\bibinfo {author} {\bibfnamefont {J.-T.}\ \bibnamefont {Hsieh}}, \bibinfo {author} {\bibfnamefont {P.~K.}\ \bibnamefont {Kothari}},\ and\ \bibinfo {author} {\bibfnamefont {S.}~\bibnamefont {Mohanty}},\ }\bibfield  {title} {\bibinfo {title} {A simple and sharper proof of the hypergraph moore bound},\ }in\ \href {https://doi.org/10.1137/1.9781611977554.ch89} {\emph {\bibinfo {booktitle} {Proceedings of the 2023 Annual ACM-SIAM Symposium on Discrete Algorithms (SODA)}}}\ (\bibinfo {organization} {SIAM},\ \bibinfo {year} {2023})\ pp.\ \bibinfo {pages} {2324--2344}\BibitemShut {NoStop}%
\bibitem [{\citenamefont {Alrabiah}\ \emph {et~al.}(2023)\citenamefont {Alrabiah}, \citenamefont {Guruswami}, \citenamefont {Kothari},\ and\ \citenamefont {Manohar}}]{alrabiah2023near}%
  \BibitemOpen
  \bibfield  {author} {\bibinfo {author} {\bibfnamefont {O.}~\bibnamefont {Alrabiah}}, \bibinfo {author} {\bibfnamefont {V.}~\bibnamefont {Guruswami}}, \bibinfo {author} {\bibfnamefont {P.~K.}\ \bibnamefont {Kothari}},\ and\ \bibinfo {author} {\bibfnamefont {P.}~\bibnamefont {Manohar}},\ }\bibfield  {title} {\bibinfo {title} {A near-cubic lower bound for 3-query locally decodable codes from semirandom {CSP} refutation},\ }in\ \href {https://doi.org/10.1145/3564246.3585143} {\emph {\bibinfo {booktitle} {Proceedings of the 55th Annual ACM Symposium on Theory of Computing}}}\ (\bibinfo {year} {2023})\ pp.\ \bibinfo {pages} {1438--1448}\BibitemShut {NoStop}%
\bibitem [{\citenamefont {Hsieh}\ \emph {et~al.}(2024)\citenamefont {Hsieh}, \citenamefont {Kothari}, \citenamefont {Mohanty}, \citenamefont {Correia},\ and\ \citenamefont {Sudakov}}]{hsieh2024small}%
  \BibitemOpen
  \bibfield  {author} {\bibinfo {author} {\bibfnamefont {J.-T.}\ \bibnamefont {Hsieh}}, \bibinfo {author} {\bibfnamefont {P.~K.}\ \bibnamefont {Kothari}}, \bibinfo {author} {\bibfnamefont {S.}~\bibnamefont {Mohanty}}, \bibinfo {author} {\bibfnamefont {D.~M.}\ \bibnamefont {Correia}},\ and\ \bibinfo {author} {\bibfnamefont {B.}~\bibnamefont {Sudakov}},\ }\bibfield  {title} {\bibinfo {title} {Small even covers, locally decodable codes and restricted subgraphs of edge-colored kikuchi graphs},\ }\href@noop {} {\bibfield  {journal} {\bibinfo  {journal} {arXiv:2401.11590}\ } (\bibinfo {year} {2024})},\ \Eprint {https://arxiv.org/abs/2401.11590} {2401.11590} \BibitemShut {NoStop}%
\bibitem [{\citenamefont {Christandl}\ \emph {et~al.}(2007)\citenamefont {Christandl}, \citenamefont {K{\"o}nig}, \citenamefont {Mitchison},\ and\ \citenamefont {Renner}}]{christandl2007one}%
  \BibitemOpen
  \bibfield  {author} {\bibinfo {author} {\bibfnamefont {M.}~\bibnamefont {Christandl}}, \bibinfo {author} {\bibfnamefont {R.}~\bibnamefont {K{\"o}nig}}, \bibinfo {author} {\bibfnamefont {G.}~\bibnamefont {Mitchison}},\ and\ \bibinfo {author} {\bibfnamefont {R.}~\bibnamefont {Renner}},\ }\bibfield  {title} {\bibinfo {title} {One-and-a-half quantum de {F}inetti theorems},\ }\href {https://doi.org/10.1007/s00220-007-0189-3} {\bibfield  {journal} {\bibinfo  {journal} {Communications in mathematical physics}\ }\textbf {\bibinfo {volume} {273}},\ \bibinfo {pages} {473} (\bibinfo {year} {2007})}\BibitemShut {NoStop}%
\bibitem [{\citenamefont {Perry}\ \emph {et~al.}(2020)\citenamefont {Perry}, \citenamefont {Wein},\ and\ \citenamefont {Bandeira}}]{perry2016statistical}%
  \BibitemOpen
  \bibfield  {author} {\bibinfo {author} {\bibfnamefont {A.}~\bibnamefont {Perry}}, \bibinfo {author} {\bibfnamefont {A.~S.}\ \bibnamefont {Wein}},\ and\ \bibinfo {author} {\bibfnamefont {A.~S.}\ \bibnamefont {Bandeira}},\ }\bibfield  {title} {\bibinfo {title} {Statistical limits of spiked tensor models},\ }\href {https://doi.org/10.1214/19-aihp960} {\bibfield  {journal} {\bibinfo  {journal} {Annales de l'Institut Henri Poincaré, Probabilités et Statistiques}\ }\textbf {\bibinfo {volume} {56}},\ \bibinfo {pages} {230 } (\bibinfo {year} {2020})}\BibitemShut {NoStop}%
\bibitem [{\citenamefont {Hopkins}\ \emph {et~al.}(2015)\citenamefont {Hopkins}, \citenamefont {Shi},\ and\ \citenamefont {Steurer}}]{hopkins2015tensor}%
  \BibitemOpen
  \bibfield  {author} {\bibinfo {author} {\bibfnamefont {S.~B.}\ \bibnamefont {Hopkins}}, \bibinfo {author} {\bibfnamefont {J.}~\bibnamefont {Shi}},\ and\ \bibinfo {author} {\bibfnamefont {D.}~\bibnamefont {Steurer}},\ }\bibfield  {title} {\bibinfo {title} {Tensor principal component analysis via sum-of-square proofs},\ }in\ \href@noop {} {\emph {\bibinfo {booktitle} {Conference on Learning Theory}}}\ (\bibinfo {organization} {PMLR},\ \bibinfo {year} {2015})\ pp.\ \bibinfo {pages} {956--1006},\ \Eprint {https://arxiv.org/abs/1507.03269} {1507.03269} \BibitemShut {NoStop}%
\bibitem [{\citenamefont {Chakraborty}\ \emph {et~al.}(2019)\citenamefont {Chakraborty}, \citenamefont {Gily\'{e}n},\ and\ \citenamefont {Jeffery}}]{CGJ19}%
  \BibitemOpen
  \bibfield  {author} {\bibinfo {author} {\bibfnamefont {S.}~\bibnamefont {Chakraborty}}, \bibinfo {author} {\bibfnamefont {A.}~\bibnamefont {Gily\'{e}n}},\ and\ \bibinfo {author} {\bibfnamefont {S.}~\bibnamefont {Jeffery}},\ }\bibfield  {title} {\bibinfo {title} {{The Power of Block-Encoded Matrix Powers: Improved Regression Techniques via Faster Hamiltonian Simulation}},\ }in\ \href {https://doi.org/10.4230/LIPIcs.ICALP.2019.33} {\emph {\bibinfo {booktitle} {46th International Colloquium on Automata, Languages, and Programming (ICALP 2019)}}},\ \bibinfo {series} {Leibniz International Proceedings in Informatics (LIPIcs)}, Vol.\ \bibinfo {volume} {132}\ (\bibinfo {year} {2019})\ pp.\ \bibinfo {pages} {33:1--33:14}\BibitemShut {NoStop}%
\bibitem [{\citenamefont {Haah}\ \emph {et~al.}(2023)\citenamefont {Haah}, \citenamefont {Kothari}, \citenamefont {O'Donnell},\ and\ \citenamefont {Tang}}]{HKOT23}%
  \BibitemOpen
  \bibfield  {author} {\bibinfo {author} {\bibfnamefont {J.}~\bibnamefont {Haah}}, \bibinfo {author} {\bibfnamefont {R.}~\bibnamefont {Kothari}}, \bibinfo {author} {\bibfnamefont {R.}~\bibnamefont {O'Donnell}},\ and\ \bibinfo {author} {\bibfnamefont {E.}~\bibnamefont {Tang}},\ }\bibfield  {title} {\bibinfo {title} {Query-optimal estimation of unitary channels in diamond distance},\ }in\ \href {https://doi.org/10.1109/FOCS57990.2023.00028} {\emph {\bibinfo {booktitle} {2023 IEEE 64th Annual Symposium on Foundations of Computer Science (FOCS)}}}\ (\bibinfo {year} {2023})\ pp.\ \bibinfo {pages} {363--390}\BibitemShut {NoStop}%
\bibitem [{\citenamefont {Gily\'{e}n}\ \emph {et~al.}(2019)\citenamefont {Gily\'{e}n}, \citenamefont {Su}, \citenamefont {Low},\ and\ \citenamefont {Wiebe}}]{GSLW19}%
  \BibitemOpen
  \bibfield  {author} {\bibinfo {author} {\bibfnamefont {A.}~\bibnamefont {Gily\'{e}n}}, \bibinfo {author} {\bibfnamefont {Y.}~\bibnamefont {Su}}, \bibinfo {author} {\bibfnamefont {G.~H.}\ \bibnamefont {Low}},\ and\ \bibinfo {author} {\bibfnamefont {N.}~\bibnamefont {Wiebe}},\ }\bibfield  {title} {\bibinfo {title} {Quantum singular value transformation and beyond: exponential improvements for quantum matrix arithmetics},\ }in\ \href {https://doi.org/10.1145/3313276.3316366} {\emph {\bibinfo {booktitle} {Proceedings of the 51st Annual ACM SIGACT Symposium on Theory of Computing}}},\ \bibinfo {series and number} {STOC 2019}\ (\bibinfo {year} {2019})\ p.\ \bibinfo {pages} {193–204}\BibitemShut {NoStop}%
\bibitem [{\citenamefont {Berry}\ \emph {et~al.}(2015)\citenamefont {Berry}, \citenamefont {Childs},\ and\ \citenamefont {Kothari}}]{BCK15}%
  \BibitemOpen
  \bibfield  {author} {\bibinfo {author} {\bibfnamefont {D.~W.}\ \bibnamefont {Berry}}, \bibinfo {author} {\bibfnamefont {A.~M.}\ \bibnamefont {Childs}},\ and\ \bibinfo {author} {\bibfnamefont {R.}~\bibnamefont {Kothari}},\ }\bibfield  {title} {\bibinfo {title} {Hamiltonian simulation with nearly optimal dependence on all parameters},\ }in\ \href {https://doi.org/10.1109/FOCS.2015.54} {\emph {\bibinfo {booktitle} {2015 IEEE 56th Annual Symposium on Foundations of Computer Science}}}\ (\bibinfo {year} {2015})\ pp.\ \bibinfo {pages} {792--809}\BibitemShut {NoStop}%
\bibitem [{\citenamefont {Brent}\ and\ \citenamefont {Zimmermann}(2010)}]{BZ10}%
  \BibitemOpen
  \bibfield  {author} {\bibinfo {author} {\bibfnamefont {R.~P.}\ \bibnamefont {Brent}}\ and\ \bibinfo {author} {\bibfnamefont {P.}~\bibnamefont {Zimmermann}},\ }\href {https://doi.org/10.1017/cbo9780511921698} {\emph {\bibinfo {title} {Modern Computer Arithmetic}}}\ (\bibinfo  {publisher} {Cambridge University Press},\ \bibinfo {year} {2010})\BibitemShut {NoStop}%
\bibitem [{\citenamefont {Childs}(2009)}]{Chi09}%
  \BibitemOpen
  \bibfield  {author} {\bibinfo {author} {\bibfnamefont {A.~M.}\ \bibnamefont {Childs}},\ }\bibfield  {title} {\bibinfo {title} {On the relationship between continuous- and discrete-time quantum walk},\ }\href {https://doi.org/10.1007/s00220-009-0930-1} {\bibfield  {journal} {\bibinfo  {journal} {Communications in Mathematical Physics}\ }\textbf {\bibinfo {volume} {294}},\ \bibinfo {pages} {581–603} (\bibinfo {year} {2009})}\BibitemShut {NoStop}%
\bibitem [{\citenamefont {Malvetti}\ \emph {et~al.}(2021)\citenamefont {Malvetti}, \citenamefont {Iten},\ and\ \citenamefont {Colbeck}}]{MIC21}%
  \BibitemOpen
  \bibfield  {author} {\bibinfo {author} {\bibfnamefont {E.}~\bibnamefont {Malvetti}}, \bibinfo {author} {\bibfnamefont {R.}~\bibnamefont {Iten}},\ and\ \bibinfo {author} {\bibfnamefont {R.}~\bibnamefont {Colbeck}},\ }\bibfield  {title} {\bibinfo {title} {Quantum {C}ircuits for {S}parse {I}sometries},\ }\href {https://doi.org/10.22331/q-2021-03-15-412} {\bibfield  {journal} {\bibinfo  {journal} {{Quantum}}\ }\textbf {\bibinfo {volume} {5}},\ \bibinfo {pages} {412} (\bibinfo {year} {2021})}\BibitemShut {NoStop}%
\bibitem [{\citenamefont {Shende}\ \emph {et~al.}(2006)\citenamefont {Shende}, \citenamefont {Bullock},\ and\ \citenamefont {Markov}}]{SBM06}%
  \BibitemOpen
  \bibfield  {author} {\bibinfo {author} {\bibfnamefont {V.~V.}\ \bibnamefont {Shende}}, \bibinfo {author} {\bibfnamefont {S.~S.}\ \bibnamefont {Bullock}},\ and\ \bibinfo {author} {\bibfnamefont {I.~L.}\ \bibnamefont {Markov}},\ }\bibfield  {title} {\bibinfo {title} {Synthesis of quantum-logic circuits},\ }\href {https://doi.org/10.1109/TCAD.2005.855930} {\bibfield  {journal} {\bibinfo  {journal} {IEEE Transactions on Computer-Aided Design of Integrated Circuits and Systems}\ }\textbf {\bibinfo {volume} {25}},\ \bibinfo {pages} {1000} (\bibinfo {year} {2006})}\BibitemShut {NoStop}%
\end{thebibliography}%

\end{document}